\documentclass[reprint,aps,prx,twocolumn]{revtex4-2}
\usepackage{graphicx}
\usepackage{bm}
\usepackage{color}
\usepackage{amsmath,amssymb} 
\usepackage{empheq}
\usepackage{ulem}
\usepackage{mathrsfs,comment}
\usepackage{mediabb}
\newcommand{\FRAC}[2]{ {\textstyle \frac{#1}{#2}} }
\newcommand{\dd}{ \bm{d} }
\newcommand{\ee}{ \bm{e} }
\newcommand{\hh}{ \bm{h} }
\newcommand{\kk}{ \bm{k} }
\newcommand{\pp}{ \bm{p} }
\newcommand{\qq}{ \bm{q} }
\newcommand{\rr}{ \bm{r} }
\newcommand{\uuu}{ \bm{u} } 
\newcommand{\vvv}{ \bm{v} } 
\newcommand{\GG}{ \bm{G} }

\newcommand{\QQ}{ \bm{Q} }

\newcommand{\XX}{ \bm{X} }
\newcommand{\bdelta}{ \bm{\delta} }
\newcommand{\bgamma}{ \bm{\gamma} }
\newcommand{\bphi}{ \bm{\phi} }
\newcommand{\bpsi}{ \bm{\psi} }
\newcommand{\btheta}{ \bm{\theta} }
\newcommand{\bzero}{ \bm{0} }
\newcommand{\RE}{{\mathrm{Re} \,}}
\newcommand{\IM}{{\mathrm{Im} \,}}

\newcommand{\hee}{\hat{\bm{e}}}

\newcommand{\msm}[1]{\mspace{-#1mu}}
\newcommand{\msp}[1]{\mspace{#1mu}}
\newcommand{\hs}[1]{\hspace{#1}}

\newcommand{\Sp}[1]{\mspace{-2mu}#1\mspace{-2mu}}

\newcommand{\PLUS}{\msm2 + \msm2}
\newcommand{\MINUS}{\msm2 - \msm2}
\newcommand{\Eq}{\msm2 = \msm2}
\newcommand{\Equiv}{\msm2 \equiv \msm2}
\newcommand{\Neq}{\msm2 \neq \msm2}
\newcommand{\GT}{\msm2 > \msm2}
\newcommand{\LT}{\msm2 < \msm2}
\newcommand{\GE}{\msm2 \ge \msm2}
\newcommand{\LE}{\msm2 \le \msm2}
\newcommand{\To}{\msm2 \to \msm2}
\newcommand{\Sim}{\msm2 \sim \msm2}

\newcommand{\PM}{\msm2 \pm \msm2}

\newcommand{\dprime}{{' \msm2 '}}

\newcommand{\black}{\color{black}}
\newcommand{\blue}{\color{blue}}

\newcommand{\NSpm}{{NS$'$ }}

\allowdisplaybreaks[4]

\newcommand{\Blue}[1]{\textcolor{blue}{#1}}

\begin{document}

\title{Quadrupole partial orders and triple-$\qq$ states on the face-centered cubic lattice}

\author{Kazumasa Hattori}
\author{Takayuki Ishitobi}
\affiliation{%
Department of Physics, 
Tokyo Metropolitan University,
1-1, Minami-osawa, Hachioji, Tokyo 192-0397, Japan}
\author{Hirokazu Tsunetsugu}
\affiliation{%
The Institute for Solid State Physics, 
The University of Tokyo, 
Kashiwanoha 5-1-5, Chiba 277-8581, Japan }

\date{\today}

\begin{abstract}
We study $\Gamma_3$ quadrupole orders in a face-centered cubic lattice. 
The $\Gamma_3$ quadrupole moments under cubic symmetry 
possess a unique cubic invariant in their free energy 
in the uniform ($\qq=\bzero $) sector and the triple-$\qq$ sector 
for the X points $\qq=(2\pi,0,0),(0,2\pi,0)$, 
and $(0,0,2\pi)$. 
Competition between this cubic anisotropy 
and anisotropic quadrupole-quadrupole interactions 
causes a drastic impact on the phase diagram 
both in the ground state and at finite temperatures. 
We show details about the model construction and its properties, 
the phase diagram, and the mechanism of the various triple-$\qq$ 
quadrupole orders reported in our preceding letter 
[J.~Phys.~Soc.~Jpn.~\textbf{90}, 43701 (2021)]. 
By using a mean-field approach, 
we analyze a quadrupole exchange model that consists of 
a crystalline-electric field scheme with 
the ground-state $\Gamma_3$ non-Kramers doublet 
and the excited singlet $\Gamma_1$ state. 
We have found various triple-$\qq$ orders 
in the four-sublattice 
mean-field approximation. 
A few partially ordered phases  
are stabilized in a wide range of parameter space 
and they have a higher transition temperature 
than single-$\qq$ orders. 
With lowering  temperature, 
there occur transitions from these partially ordered phases 
into further symmetry broken phases  
in which previously disordered sites acquire 
nonvanishing quadrupole moments.
The identified phases in the mean-field approximation 
are further analyzed by a phenomenological Landau theory. 
This analysis reveals results 
qualitatively consistent with the mean-field results 
and also shows that 
the cubic invariant plays an important role 
for stabilizing the triple-$\qq$ states. 
The present mechanism 
for the triple-$\qq$ states also takes effect in 
systems with different types of quadrupoles, 
and we discuss its implications for 
recent experiments in a few $f$- and $d$-electron compounds.
\end{abstract}

\pacs{Valid MAC'S appear here}
\maketitle

\section{Introduction}
\label{sec:1}

Strongly correlated electron systems possess a variety of possibilities 
of exotic phenomena \cite{Tokura2000,zhou2017quantum} and also 
innovative devices in future \cite{Fiebig2016-or}. 
 Strong correlations between 
$d$- or $f$-electrons often lead to fascinating and mysterious 
ordered phases \cite{mydosh2014hidden}, 
phase transitions \cite{nasu2014vaporization}, 
and even liquid like phases \cite{zhou2017quantum}. 
 In addition to these correlation effects, 
the importance of spin-orbit (SO) coupling 
has been recognized for materials 
including heavy elements with 5$d$ or 4$d$ 
electrons \cite{kim2009phase,Jackeli2009,takagi2019concept}. 
 These electrons are subject to both the crystalline-electric field (CEF) 
and the SO coupling. 
Various phenomena of multipoles 
emerge in such configurations \cite{Hayami2018-zx}.  
This causes nontrivial properties in their spin, orbital, 
or more general
multipole models \cite{Hayami2018-zx,Yatsushiro2021-ob}. 

Several years ago, 
the present authors studied antiferro quadrupole orders 
in Pr 1-2-20 compounds \cite{Hattori2014,Hattori2016,Ishitobi2019-du},  
in which Pr ions form a diamond sublattice. 
Each Pr$^{3+}$ ion has two $f$-electrons and 
its ground state is a non-Kramers doublet $\Gamma_3$ 
in the cubic environment, which is a consequence of 
the interplay of 
CEF and SO couplings \cite{OnimaruKusunose2016}.  
This $\Gamma_3$ doublet has two active components of 
electric quadrupole O$_{20}$ and O$_{22}$, 
and they form a two-dimensional basis of $E$ irreducible representation 
of the cubic point group $\QQ$=$(Q_u , Q_v)$=$({\rm O}_{20}, \sqrt{3}{\rm O}_{22})$. 
To understand quadrupole orders in this system, 
it is too naive to make predictions based on 
the understanding of an apparently similar 
spin-1/2 model on the same lattice.  
Despite their local degrees of freedom 
being a doublet in common, 
there exists a crucial difference. This is due to their 
opposite parities under time reversal operation: spin 
has odd parity, while electric quadrupole has even parity.  
Its even parity protects the presence of a nonvanishing 
third-order term of quadrupole moments 
in the system's free energy. Its consequences and implications 
were briefly discussed 
in the previous study \cite{Tsunetsugu2021}. 

More recently,  Kusanose \textit{et al.} have measured 
the low-temperature properties of 
$\Gamma_3$ quadrupole moments in another compound 
PrMgNi$_4$ \cite{Kusanose2019,Kusanose2020,Kusanose2022}. 
An important difference is that 
Pr ions form a face-centered cubic (fcc) sublattice there.
In our preceding study \cite{Tsunetsugu2021}, 
we have shown that different lattice structures
open a way to stabilize novel triple-$\qq$ orders 
including partial-ordered states. 
One can understand this by noting 
that the third-order term in the free energy contains 
a following coupling of the moments at 
three wavevectors $\pp_{1,2,3}$, 
\begin{align}
\sim b_{ABC} A(\bm{p}_1) B(\bm{p}_2) C(\bm{p}_3) , 
\end{align}
with $b_{ABC}$ being the third-order coupling 
constant.  A requirement exists such that
$\bm{p}_1+\bm{p}_2+\bm{p}_3=\GG $ should be one of 
the reciprocal lattice vectors.   
The set of $A(\bm{p}_1)$, $B(\bm{p}_2)$, and $C(\bm{p}_3)$ is 
a certain combination of multipole operators 
such that the product $ABC$ remains invariant under 
any point-group symmetry operation.  
As the third-order coupling is nonvanishing $b_{ABC} \ne 0$, 
one naively expects a first-order transition 
occuring at a temperature higher than the second-order one, which is  
determined as a vanishing point of the second-order coefficient 
in the free energy. 
Such a type of third-order term has been discussed 
 in several contexts for nonmagnetic systems including 
the $\gamma$-$\alpha$ transition in Ce \cite{Nikolaev1999}, 
possible low-temperature phases in UPd$_3$ \cite{Walker_1994}, 
and multiple phases in PrV$_2$Al$_{20}$ \cite{Ishitobi2021}. 
As for the simplest cases with $\pp_1=\pp_2=\pp_3=\bf{0}$, 
quadrupole orders in PrTi$_2$Al$_{20}$ have been theoretically 
discussed \cite{Hattori2014,Hattori2016,Lee2018,Freyer2018,Freyer2020}, 
and also experimentally 
explored \cite{Sakai2011, Matsubayashi2012, Taniguchi2019, Kittaka2020}.
A system with $\Gamma_3$ quadrupole moments on an fcc lattice 
is the simplest realization for such triple-$\qq$ physics 
and we will study this in this paper in more detail 
than in the previous report \cite{Tsunetsugu2021}.

This paper is organized as follows. 
In Sec.~\ref{sec:Model}, we will introduce a low-energy effective model 
defined in the space of $\Gamma_3$ 
and $\Gamma_1$ CEF states relevant to the Pr-based systems. 
We will explain the basic properties of exchange interactions 
between the quadrupole moments, and in Sec.~\ref{sec:SSproperty}, 
we will examine the local quadrupole Hamiltonian. 
In Sec.~\ref{sec:MFresults}, we will perform a 
single-site mean-field analysis for a $\Gamma_3$-only model, and find that 
its results turn out insufficient. 
Then, we will proceed to the discussions about the four-site 
mean-field approximations including various triple-$\qq$ orders
 in Sec.~\ref{sec:phase}.  
Sections~\ref{sec:Landau}--\ref{sec:Knegative} are devoted to phenomenological 
analyses of the microscopic mean-field results 
and we will clarify how various triple-$\qq$ orders 
including partially ordered states emerge. 
We will also discuss a possible application 
of the present 
triple-$\qq$ mechanism to other systems in Sec.~\ref{sec:Discussion} 
and summarize this paper in Sec.~\ref{sec:Summary}.

\section{Model}
\label{sec:Model}
In this section, we will introduce a model Hamiltonian 
to be studied in this paper 
for discussing quadrupole orders in 
Pr-based $\Gamma_3$ systems. 
In Sec.~\ref{sec:G3-G1}, 
we will analyze their local Hilbert space and 
construct an effective exchange Hamiltonian 
of $f$-electron quadrupoles 
based on their symmetry property.  
In Sec.~\ref{sec:exJ},  
we will then analyze this exchange Hamiltonian at the classical level 
and carry out a mode analysis to 
identify the leading ordering patterns. 
One should understand that the quadrupole interactions are 
determined by integrating out the degrees of freedom of 
conduction electrons in the real Pr-based compounds, 
and we do not consider those conduction electrons explicitly.  
This is because our aim is to study the physics related to 
symmetry breakings exhibited by the localized $f$-electrons. 
Whether the system is insulating or metallic does not 
alter the qualitative aspect of the symmetry breaking.  
Exceptions are some details at quantum critical points \cite{Sachdev}, 
but they are not the issue of the present study. 
In order to determine the phase diagram 
and identify ordering patterns in each phase, 
the localized electron model is sufficient 
and has been used as a reasonable starting point in the studies  
of, e.g., CeB$_6$~\cite{Shiina1997-ia,Kusunose2005-ti}, 
URu$_2$Si$_2$~\cite{PhysRevB.72.014432}, 
and Pr 1-2-20 systems~\cite{Freyer2018,Freyer2020}. 
Throughout this paper, we analyze fundamental properties 
of quadrupole orders in 
the minimal $\Gamma_3$ model \eqref{eq:defH} 
on the fcc lattice
with the simplest fcc nearest-neighbor interactions 
and clarify their general trends with 
a few important material parameters 
rather than focus on a specific material.

\subsection{$\Gamma_3-\Gamma_1$ model}
\label{sec:G3-G1}

We first explain the local Hilbert space of the $\Gamma_3$ systems 
for discussing quadrupole orders in their ground state systems. 
As discussed in Refs.~\cite{Hattori2014,Hattori2016,Tsunetsugu2021}, 
it is important to include the excited $\Gamma_1$ state 
in order to take into account local anisotropy. 
This is because the quadrupole operators have 
quite large matrix elements 
connecting the $\Gamma_3$ doublet 
to the $\Gamma_1$ state.  
See Eq.~(\ref{eq:QuQv}). 
The other CEF excited states can be safely neglected, 
since they have no quadrupole matrix elements 
with $\Gamma_3$.  
Thus, our local Hilbert space consists of 
the singlet $\Gamma_1 = \{ |s\rangle \}$ 
and the non-Kramers doublet 
$\Gamma_3 = \{ |u\rangle , |v \rangle \}$:
\begin{align}
|s\rangle
&=
\FRAC{1}{\sqrt{12}}
\Bigl[ 
   \sqrt{\FRAC{5}{2}} \, 
   \bigl( |4\rangle+|-4\rangle \bigr)
  +\sqrt{7} \, |0\rangle
\Bigr], 
\\
|u\rangle
&=
\FRAC{1}{\sqrt{12}}
\Bigl[
   \sqrt{\FRAC{7}{2}} \, 
   \bigl( |4\rangle+|-4\rangle \bigr)
  -\sqrt{5}\, |0\rangle
\Bigr], 
\\
|v\rangle
&=
\FRAC{1}{\sqrt{2}} \,  
\bigl( |2\rangle+|-2\rangle \bigr).
\end{align}
Here $|J_z\rangle$ denotes the eigenstate of the $z$-component of the total angular momentum $J_z$ 
in the multiplet of the total angular momentum $J$=4.  
Using these
three basis states
$\{ |s\rangle, |u\rangle,|v\rangle \}$, 
the $\Gamma_3$ quadrupole operators 
$\QQ \equiv (Q_u,Q_v)^{\rm T}$ with T being the transpose are 
represented as \cite{Hattori2014}  
\begin{equation}
Q_u
\equiv 
\left(
\begin{array}{@{\hspace{1pt}}c@{\hspace{5pt}}c@{\hspace{5pt}}c@{\hspace{1pt}}}
0 & \alpha  & 0\\
\alpha & 1 & 	0 \\
0 & 0 & -1
\end{array}
\right) , 
\ \ 
Q_v
\equiv 
\left(
\begin{array}{@{\hspace{1pt}}c@{\hspace{5pt}}c@{\hspace{5pt}}c@{\hspace{1pt}}}
0 & 0  & \alpha\\
0 & 0 & 	-1 \\
\alpha & -1 & 0
\end{array}
\right) , \ \ 
\alpha \equiv \frac{\sqrt{35}}{2} . 
\label{eq:QuQv}
\end{equation}
As noted before, the matrix elements connecting 
$\Gamma_3$ to $\Gamma_1$ are quite large 
$\alpha \sim 3$.  

\begin{figure}[bt]
\begin{center}
\includegraphics[width=0.45\textwidth]{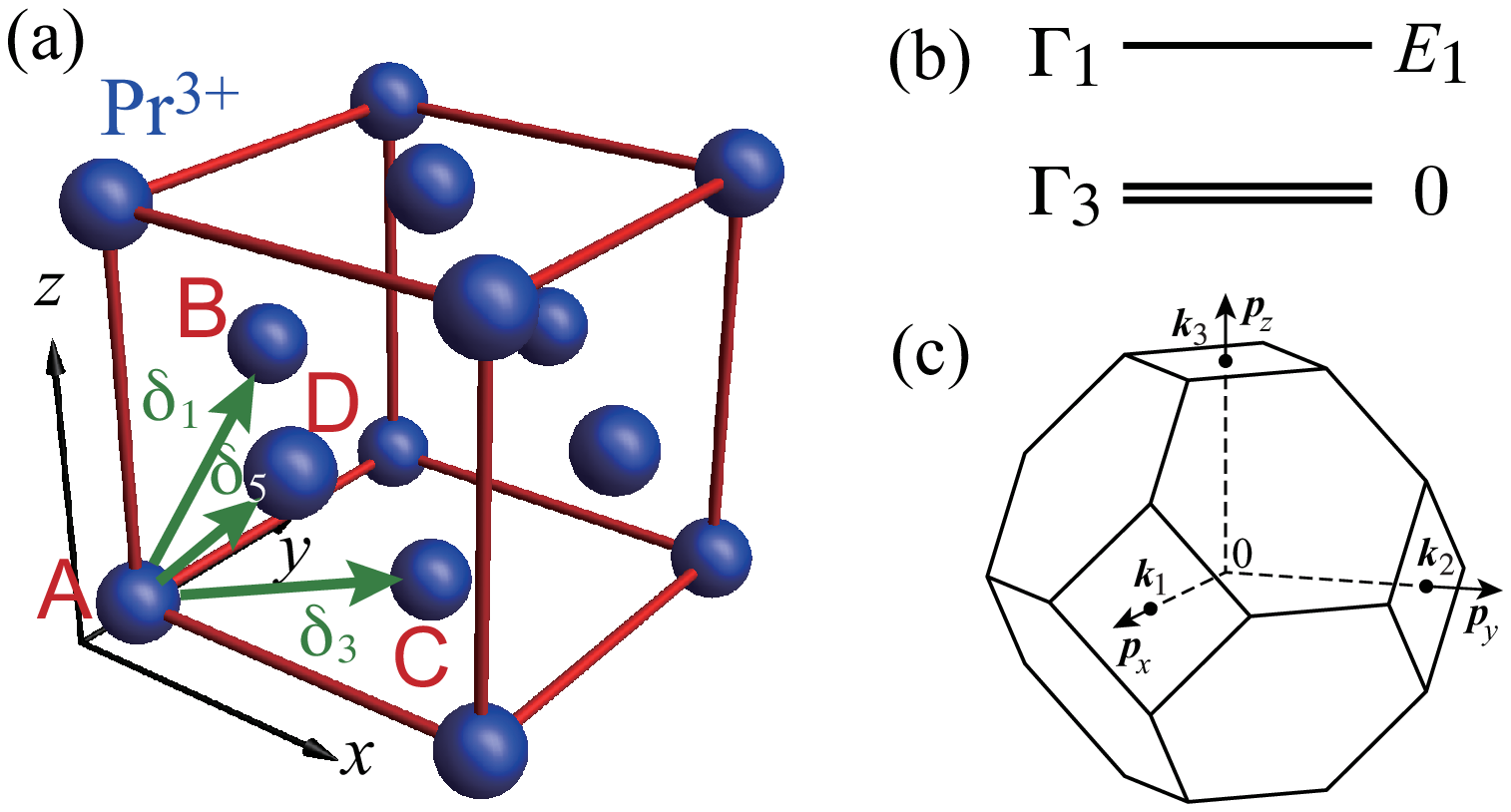}
\vspace{-10pt}
\end{center}
\caption[]{
(a) Cubic unit cell of the 
fcc lattice of Pr ions and its four sublattices 
labeled by A-D. 
$\overrightarrow{\mathrm{AB}}$=$\bdelta _1$, 
$\overrightarrow{\mathrm{AC}}$=$\bdelta _3$, 
and 
$\overrightarrow{\mathrm{AD}}$=$\bdelta _5$.  
(b) 
The relevant CEF states of a Pr$^{3+}$ ion in this paper.
$\Gamma_3$ is the non-Kramers ground-state doublet and 
$\Gamma_1$ is a singlet excited state. 
$| m \rangle$ is the eigenstate $J_z |m \rangle = m |m \rangle$ 
in the $J$=4 multiplet. 
(c) Brillouin zone of the fcc lattice. The three X points are 
$\kk_1$=($2\pi$,0,0), $\kk_2$=(0,$2\pi$,0), and $\kk_3$=(0,0,$2\pi$). 
}
\label{fig1}
\end{figure}

In the fcc lattice, each site has twelve nearest neighbors 
separated by 
$\bdelta _{1,4}$=$(0,\FRAC12 ,\pm \FRAC12 )$, 
$\bdelta _{2,5}$=$(\mp \FRAC12 ,0, \FRAC12 )$, 
$\bdelta _{3,6}$=$(\FRAC12 ,\pm \FRAC12 , 0)$, 
and their counterparts $-\bdelta $'s. 
See Fig.~\ref{fig1}.  
The symmetry analysis in our previous study concluded that 
quadrupole interactions generally have anisotropic 
couplings in addition to isotropic ones \cite{Tsunetsugu2021}. 
The minimal model of quadrupole interactions reads as 
\begin{align}
H_Q &= 
\msm4
\sum_{\langle \rr , \rr ' \rangle}
\QQ (\rr) \cdot \mathsf{J}_{\rr-\rr'}\QQ (\rr ' ),
\label{eq2:HQ}
\\
\mathsf{J}_{\rr-\rr'}&=
  J\mathsf{g}_0+ K \mathsf{g} (\mbox{$\rr$$-$$\rr '$} ),
\label{eq:def_J_delta}
\end{align}
where $\langle \rr, \rr'\rangle$ indicates 
that the sum runs over the nearest-neighbor site pairs 
and $\mathsf{g}_0=\hat{\sigma}_0$. 
Throughout this paper, 
$\hat{\sigma}_0$ denotes the 2$\times$2 identity matrix.  
The anisotropy factor has a form 
represented with the Pauli matrices 
$\hat{\sigma}_1$ and $\hat{\sigma}_3$ 
\begin{equation}
\mathsf{g} (\bdelta  )
= 
 \cos \zeta (\bdelta  ) \, \hat{\sigma}_3
-\sin \zeta (\bdelta  ) \, \hat{\sigma}_1,
\label{eq2:g-mat}
\end{equation}
where $\hat{\sigma}_{1,3}$ operate 
in the $(Q_u, Q_v)$ space.  
The angle parameter $\zeta (\bdelta )$ is defined 
for the bond vector $\bdelta $=$(x,y,z)$ as
\begin{equation}
\zeta (\bdelta  ) = 
\IM \log 
\bigl( e^{i\omega} x^2 + e^{-i\omega} y^2 + z^2 \bigr), \quad 
\omega \equiv \FRAC{2}{3}\pi . 
\label{eq:zeta-delta}
\end{equation}
As for the nearest-neighbor bonds, 
$\zeta (\pm \bdelta_{n}) \!= \! n \omega \MINUS \pi$,  
and 
this leads to the relation 
$\mathsf{g} (\bdelta_{1}) \!+\! \mathsf{g} (\bdelta_{3})
 \!+\! \mathsf{g} (\bdelta_{5})$=$\mathsf{0}$.  
The same type of anisotropic coupling has been used 
in the so-called compass model 
to study orbital orders \cite{Kugel1973}.  
The effective model \eqref{eq2:HQ} was obtained 
with the special value $K$=$J$ by Kubo and Hotta 
starting from a microscopic electron Hamiltonian \cite{Kubo2017}. 
The special $K$ value is due to a simple form 
of their microscopic Hamiltonian, and 
various other types of super-exchange processes 
generate $K$$\ne$$J$.

We also note that, conduction electrons cause 
the RKKY interactions between the localized 
quadrupole moments. 
The RKKY interactions have, in general, 
a longer range beyond the nearest-neighbor distance
used in this paper. 
However, their range remains finite 
at finite temperatures, since electron propagation 
loses long-range coherence due to thermal fluctuations. 
The distance dependence of the RKKY interactions 
is quite complicated 
and one needs to consider
the details of the Fermi surfaces 
to calculate its dependence.  
Before performing that type of elaborate calculation, 
we have to clarify fundamental features of quadrupole orders 
and focus on 
a minimal model in which conduction electron degrees of freedom 
are traced out.

We denote by $E_1$ the $\Gamma_1$ energy level relative to 
the $\Gamma_3$'s value, 
and assume $E_1 > 0$ throughout this paper.  
Then, we define the Hamiltonian of the $\Gamma_3$--$\Gamma_1$ model 
by 
\begin{equation}
H = 
E_1 \sum_{\rr} |s (\rr ) \rangle \langle s (\rr ) | 
+ H_{Q}. 
\label{eq:defH}
\end{equation}
In the following sections, 
we will analyze in detail the exchange interactions $H_Q$ 
and the properties of a single-site Hamiltonian 
under quadrupolar molecular fields.

%
%

%
%
\begin{figure*}[t]
\begin{center}
\includegraphics[width=\textwidth]{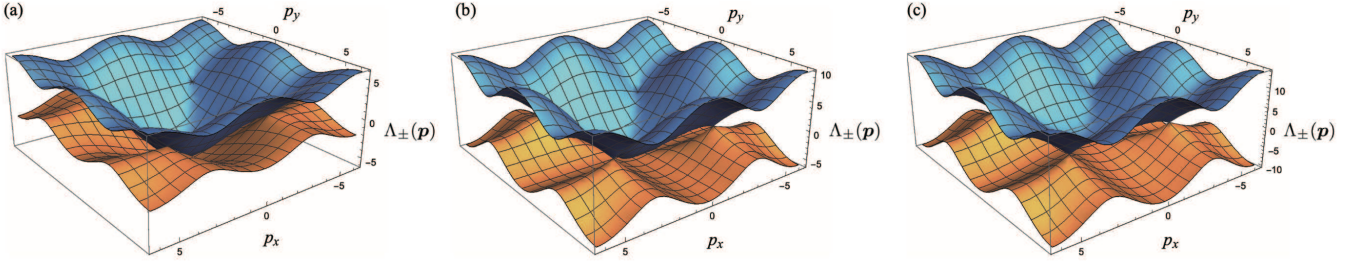}
\end{center}
\vspace{-10pt}
\caption{ The eigenvalues 
$\Lambda _{\pm} (\pp) $ 
in the $p_x$--$p_y$ plane for $J<0$.  
$K/|J|=$ (a) 1.0, (b) 2.0, and (c) 3.0. 
Note that $(2\pi,2\pi,0)\equiv (0,0,2\pi)$ for the fcc lattice}. 
\label{fig2}
\end{figure*}

\subsection{Exchange interactions}\label{sec:exJ}

Let us start with finding a classical ground state based 
on the Fourier mode analysis of $H_Q$.  
In terms of Fourier components 
$\QQ _{\pp} = N^{-1/2} \sum_{\rr} e^{-i \pp \cdot \rr} \QQ (\rr )$ 
($N$ is the number of sites), 
$H_Q$ is represented as 
\begin{subequations} 
\begin{align}
H_Q
&=
\sum_{\pp} 
\QQ _{\pp} \cdot \mathsf{J}  (\pp) \, \QQ _{-\pp},
\\
\mathsf{J}  (\pp) 
&\equiv 
\sum_{n=1}^6
e^{i\pp\cdot \bdelta _n} 
\mathsf{J}_{\bdelta _n}
=
J \gamma_0 (\pp)\hat{\sigma}_0 + 
K \bgamma  (\pp),
\label{eq:def_J_p} 
\end{align}
where the sum $\sum_{\pp}$ 
is taken over the wavevectors in the Brillouin zone (BZ)
shown in Fig.~\ref{fig1}(c).  
The coupling constants are given by 
\begin{align}
\hspace{-10pt}
\gamma_{n=0,1} (\pp)
&\equiv
2 \bigl[
c_{xy} (\pp)  \PLUS 
e^{in \omega} c_{yz} (\pp) \PLUS 
e^{-in\omega} c_{zx} (\pp)
\bigr],  
\\
\bgamma  (\pp)
&= 
- [ \RE \gamma_1 (\pp) ] \, \hat{\sigma}_3 
+ [ \IM \gamma_1 (\pp) ] \, \hat{\sigma}_1  
\nonumber\\
&=
\left[- \FRAC12 \gamma_1 (\pp)
(\hat{\sigma}_3 + i \hat{\sigma}_1 ) \right] 
+ \mbox{h.c.}, 
\end{align} 
and the form factor $c_{ab} (\pp)$ is defined as 
\begin{equation}
c_{ab}(\pp) \equiv 
\cos \bigl( \FRAC12 p_a \bigr)
\cos \bigl( \FRAC12 p_b \bigr). 
\end{equation}
\label{eqs:HamQ}
\end{subequations}
\!\!Note that $\mathsf{J} (\pp) \Eq \mathsf{J} (-\pp)$ 
is a real symmetric matrix, 
and this guarantees the hermicity of $H_Q$.  

%
%
\begin{figure}[tb]
\begin{center}
\includegraphics[width=0.4\textwidth]{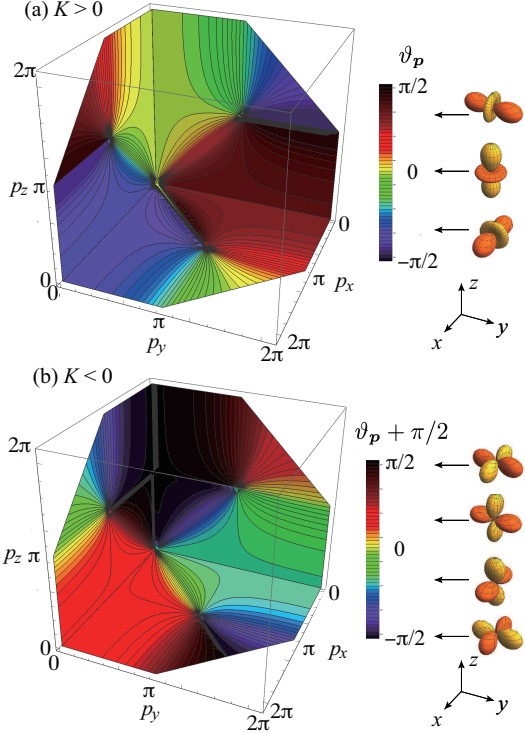}
\end{center}
\vspace{-10pt}
\caption{
$\vartheta_{\pp}$  in the $p_x$--$p_y$, $p_y$--$p_z$, 
and $p_z$--$p_x$ planes. 
Shown are (a) $\vartheta_{\pp}$ for $K>0$ and 
(b) $\vartheta_{\pp} \PLUS \pi/2$ for $K<0$. 
When the angle
$|\vartheta_{\pp}|$ or $|\vartheta_{\pp}+\pi/2|$ 
exceeds $\pi/2$, 
it is shifted by $\pi$ to an equivalent value. 
In (a), the eigenvectors at the X points are those with
$\vartheta_{\kk_1}=\bar{\vartheta}(\kk_1)=-\pi/3$, 
$\vartheta_{\kk_2}=\bar{\vartheta}(\kk_2)=\pi/3$, and 
$\vartheta_{\kk_3}=\bar{\vartheta}(\kk_3)=0$, 
where schematic $u$-type orbital shapes are illustrated. 
In (b), $\vartheta_{\kk_1}=\bar{\vartheta}(\kk_1)=\pi/6$, 
$\vartheta_{\kk_2}=\bar{\vartheta}(\kk_2)=-\pi/6$, and 
$\vartheta_{\kk_3}=\bar{\vartheta}(\kk_3)=\pi/2$ or equivalently $-\pi/2$.
}
\label{fig3}
\end{figure}

A classical ground state is a spiral state, 
and its propagating vector 
$\pp$=$\qq _*$ is the position 
where the coefficient matrix $\mathsf{J}  (\pp)$ 
has the maximally negative eigenvalue.  
The diagonalization of $\mathsf{J} (\pp)$ is straightforward and 
the eigenvalues are 
\begin{equation}
\Lambda _{\mp} (\pp) = 
J \gamma_0 (\pp) \mp K | \gamma_1 (\pp) | . 
\label{eq:EigenValueLambda}
\end{equation}
In terms of the notation introduced for 
a unit vector pointing to the direction of angle $\theta$ 
\begin{equation}
 \hee _\theta \equiv 
(\cos \theta , \sin \theta )^{\rm T} =: \hee  (\theta ) , 
\label{eq:evec}
\end{equation}
the corresponding eigenvectors are written as 
\begin{equation}
\vvv _- (\pp) =  \hee  ( \vartheta_{\pp} ) , 
\quad 
\vvv _+ (\pp) = \hee  (\vartheta_{\pp }+ \pi/2 ) , 
\end{equation}
for $\Lambda_{-} (\pp)$ and $\Lambda_{+} (\pp)$, respectively. 
The angle parameter $\vartheta _{\pp}$ is given 
by the following parametrization: 
\begin{equation}
\gamma _1 (\pp)  = | \gamma_1 (\pp) | \, 
e^{-2i \vartheta_{\pp} } , 
\quad  \mbox{$\vartheta_{\pp}$: real}.
\end{equation}
Figure~\ref{fig2} shows the eigenvalues 
$\Lambda_{\mp} (\pp)$ 
in the $p_x \,$--$\, p_y$ plane 
for $J$$<$0 as an illustrative example. 

We can easily show that the maximally negative eigenvalue is 
\begin{equation}
\Lambda_{\mathrm{min}}=
\left\{
\begin{array}{@{}ll}
-2J-4|K|  & \mbox{for $|K|$$+$$2J$$>$0}
\\[4pt]
\phantom{-}6J  & \mbox{for $2J$$<$$|K|$$<$$-2J$}
\end{array}
\right. . \label{eq:Lambda_min}
\end{equation}
The minimum position is located at 
$\kk_1$=$(2\pi, 0,0)$, $\kk_2$=$(0,2\pi,0)$, and 
$\kk_3$=$(0,0,2\pi)$ when $|K| \PLUS 2J>0$, 
while at $\kk_0$=$(0,0,0)$ otherwise. 
Thus, the propagating wavevector $\qq _*$ is either 
one of $\kk _{1,2,3}$ or $\kk _0$ depending on the parameters.  
Note that $\kk_{1,2,3}$ are the X points in the BZ, i.e., 
the centers of the three pairs of square parts on the BZ surface.  
See Fig.~\ref{fig1}(c).  
When $J$$<$0 and $|K|$$<$$2|J|$, 
one expects a ferro order in the ground state. 

The eigenvector $\uuu  (\qq _*)$ of the maximally negative eigenvalue 
$\Lambda_{\mathrm{min}}$ 
describes the unstable mode that orders 
at a phase transition approaching from the paramagnetic phase side, 
and the local ordered moments are given as 
$\langle \QQ (\rr ) \rangle \Eq 
 \mathrm{Re}\, \uuu  (\qq _*) \, e^{i \qq _* \cdot \rr}$.  
For the ferro ordering, the coefficient matrix 
$\mathsf{J} (\kk _0)$=$6J \hat{\sigma}_0$ 
is isotropic and thus the order parameter $\langle \QQ \rangle$ 
may point to 
any direction in the quadrupole space 
within the level of Fourier mode analysis.  

In the following parts of this paper,  
we concentrate on ``antiferro'' orders with the ordering wavevector 
$\kk_{\ell \ne 0}$ in detail.  
The eigenvector $\uuu  (\qq _*)$ is $\vvv  _- (\qq _*)$ for $K$$>$0, 
and $\vvv  _+ (\qq _*)$ for $K$$<$0, 
and let us represent as
$\uuu  (\qq _*)$=%
$(\cos \bar{\vartheta} (\qq _* ) , 
  \sin \bar{\vartheta} (\qq _* ))^{\rm T}$ for both cases.   
Then, its direction is given as 
\begin{align}
&\hspace{7mm}
\mbox{$K$$>$0} \hspace{7mm}
\mbox{$K$$<$0}
\nonumber
\\
\bar{\vartheta} (\mbox{$\qq _*$=$\kk_\ell$} ) = 
&\left\{
\begin{array}{rrl}
\mbox{$\FRAC76 \pi$$\mp$$\FRAC12 \pi$},   & 
\mbox{$\FRAC23 \pi$$\pm$$\FRAC12 \pi$}   & \ 
\mbox{at $\kk_1$}
\\[6pt]
\mbox{$\FRAC56 \pi$$\pm$$\FRAC12 \pi$},   & 
\mbox{$\FRAC43 \pi$$\pm$$\FRAC12 \pi$}   & \ 
\mbox{at $\kk_2$}.
\\[6pt]
\mbox{$\FRAC12 \pi$$\mp$$\FRAC12 \pi$},   & 
\mbox{$\pi$$\mp$$\FRAC12 \pi$}   & \ 
\mbox{at $\kk_3$}
\end{array}
\right. \label{eq:eigen_theta}
\end{align}
See Fig.~\ref{fig3}. For example, the order with $\qq _*$=$\kk_3$ has 
the order parameter 
$\langle \QQ (\rr ) \rangle$=$(-1)^{2z} \, (Q_u , 0 )^{\rm T}$ for $K$$>$0, 
while $(-1)^{2z} \, (0, Q_v )^{\rm T}$ for $K$$<$0. 
They are known as 
O$_{20}$- and O$_{22}$-type AFQ order, respectively.  
Note that for a more general order parameter, 
the cubic lattice symmetry 
implies that generally six values of the order parameters 
\begin{equation}
\begin{split}
\langle \QQ \rangle = 
&q \msp2 \hee  (0) \pm q' \hee  (\FRAC{\pi}{2}), \ \ \ 
 q \msp2 \hee  (\omega ) \pm q' \hee  (\omega + \FRAC{\pi}{2}), 
\\
&q \msp2 \hee  (-\omega ) \pm q' \hee  (-\omega + \FRAC{\pi}{2}), 
\end{split}
\label{eq:Q6equiv}
\end{equation}
are equivalent.  
They are related to two types of lattice symmetry operations:
one is a $\FRAC12 \pi$ rotation about one of the three principle 
axes $R_i (\pm \FRAC12 \pi)$ ($i \Eq  x, y, z$), 
and the other is an $\omega$ rotation about the axis along [1,1,1] 
direction $R_1 (\pm \omega)$.  
The above six $\langle \QQ \rangle$ values are transformed 
from one to another by these operations.

We emphasize that the phase factor of the eigenvectors 
$\uuu  (\kk_\ell )$ are not relevant 
in this Fourier mode analysis.  
However, they play a role at the stage that 
mode couplings are taken into account.

\section{Single-site property}
\label{sec:SSproperty}
For discussing possible 
phases and their transitions in this system, 
it is important to first understand single-site properties 
under quadrupole molecular field 
$\hh$=$(h_u,h_v)^{\rm T}$. 
It turns out that the wavefunction changes its topological 
character as the field strength grows, 
and this change is reflected in the phase diagrams 
of interacting quadrupoles. 

A single-site Hamiltonian is given as
\begin{equation}
{\mathcal H}_{\rm mf}
=
-\hh \cdot \QQ 
+ E_1|s\rangle\langle s|, 
\label{eq:Hloc}
\end{equation}
where the site label $\rr$ is omitted. 
As for the external fields, 
only those coupled to $\QQ$ are considered here, 
since our concern in this paper is quadrupole orders. 
Since multiplying a positive factor to $\mathcal{H}_{\rm mf}$ 
does not change its ground state, 
we normalize the Hamiltonian  
such that $\hh ^2  \! +\! E_1^2 \Eq  1$ 
and introduce the polar coordinates defined by 
\begin{equation}
(h_u, h_v ) =\sin\Theta ( \cos\Phi, \sin\Phi ), 
\ \ 
E_1=\cos\Theta . 
\end{equation}
The normalized Hamiltonian is now 
represented as 
\begin{align}
&\mspace{-6mu}
{\mathcal H}_{\rm mf} = 
{\mathcal H}_{\rm mf} (\Theta, \Phi ) 
\nonumber\\
&\mspace{-6mu}
=
\left(
\begin{array}%
{@{}r@{\hs{1pt}}r@{\hs{1pt}}r@{\hs{1pt}}r@{\hs{3pt}}r@{\hs{1pt}}r@{}}
\cos\Theta  & &
-\alpha \sin\Theta & \cos\Phi & 
-\alpha \sin\Theta &  \sin\Phi 
\\
-\alpha \sin\Theta & \cos\Phi & 
-\sin\Theta & \cos\Phi & 
\sin\Theta & \sin\Phi
\\
-\alpha \sin\Theta & \sin\Phi  & 
\sin\Theta & \sin\Phi &
\sin\Theta & \cos\Phi 
\end{array}
\right) . 
\label{eq:HamilHloc}
\vspace{-6pt}
\end{align}
Our primary concern is about the case 
of $0 \LT  \Theta \LE  \FRAC12 \pi$, 
and there exists the symmetry 
$
{\mathcal H}_{\rm mf} (\mbox{$\pi$$-$$\Theta$}, \mbox{$\pi$$+$$\Phi$} )
\Eq  
- {\mathcal H}_{\rm mf} (\Theta , \Phi) 
$.  

It is important to understand the symmetry of this Hamiltonian 
that is related to the cubic lattice structure. 
Let $R$ be one of the aforementioned two types of rotation operators. 
Its application transforms 
a wavefunction with an orthogonal matrix as
$ \psi ' \Eq  \mathcal{U}_R \psi$, 
and thus the Hamiltonian is correspondingly transformed as 
$ \mathcal{H}_{\rm mf} ' \Eq  
  \mathcal{U}_R \mathcal{H}_{\rm mf} 
  \mathcal{U}_R^{\rm T} $. 
The transformation matrices 
for $R$=$R_z (\pm \pi/2)$ and $R_1 (\pm \omega)$ 
are 
$\mathcal{U}_R \Eq  \mathcal{M}_z \Eq  \mathcal{M} (0)$ and 
$\mathcal{U} (\pm \omega)$,   
respectively, in terms of the notations 
defined for general $\theta$ by
\begin{subequations}
\begin{align}
\mathcal{M} (\theta ) &= 
\left(
\begin{array}{@{}c@{\hspace{4pt}}|@{\hspace{4pt}}c@{\hspace{5pt}}c@{}}
1 & 0  & 0 \\
\hline
0 & \phantom{-} \cos \theta & -\sin \theta \\
0 & -\sin \theta & -\cos \theta
\end{array}
\right) 
= 1 \oplus \mu (\theta), \ 
\\
\mathcal{U} (\theta ) &= 
\left(
\begin{array}{@{}c@{\hspace{4pt}}|@{\hspace{4pt}}c@{\hspace{5pt}}c@{}}
1 & 0  & 0 \\
\hline
0 & \cos \theta  & -\sin \theta \\
0 & \sin \theta  & \phantom{-}\cos \theta
\end{array}
\right) 
= 1 \oplus U (\theta) , 
\end{align}
\label{eq:MandU}
\end{subequations}
\!\!where $\mu (\theta)$ and $U (\theta)$ are $2\! \times \! 2$ 
orthogonal matrices operating in the $\Gamma_3$ subspace. 
Note that the $\Gamma_1$ space is invariant for any $R$, 
since it is singlet.   
$\mu (\theta)$ is a mirror operator in the $\Gamma_3$ multiplet. 
The other symmetry operations of 
$\theta \Eq \pm \FRAC12 \pi$ rotations can 
be represented as 
$R_x (\theta ) \Eq R_1 ( \omega ) R_z (\theta ) R_1 (-\omega)$ and 
$R_y (\theta ) \Eq R_1 (-\omega ) R_z (\theta ) R_1 (\omega)$. 
The corresponding mirrors are given as  
$\mathcal{M}_{x,y} \Eq  
 \mathcal{U} (\pm \omega) \mathcal{M}_z \mathcal{U} (\mp \omega) \Eq  
 \mathcal{M} (\pm \omega)$.  
Note the relations 
$
[\mathcal{U}(n \omega)]^{\rm T} 
\Eq  
\mathcal{U}(\mbox{$-$$n \omega$})
$
and 
$
\mathcal{M}(n \omega) \mathcal{M}(n' \omega) 
\Eq  
\mathcal{U}((\mbox{$n'$$-$$n$}) \omega)
$.  

Now that the transformation matrices are obtained, we can directly 
show that the transformed Hamiltonian also has the form of 
$\mathcal{H}_{\rm mf} (\Theta , \Phi ')$  
and the new field direction $\Phi '$ is determined.  
The result is summarized as follows:
\begin{subequations} 
\begin{align}
\mathcal{U} \left( n\omega \right) \, 
\mathcal{H}_{\rm mf} (\Theta, \Phi) \, 
[\mathcal{U} \left( n\omega \right)]^{\rm T}
&= 
\mathcal{H}_{\rm mf} \left( \Theta, \mbox{$\Phi$$+$$n\omega$} \right) , 
\\
\mathcal{M} \left( n\omega \right) \, 
\mathcal{H}_{\rm mf} (\Theta, \Phi) \,
\mathcal{M} \left( n\omega \right) 
&= 
\mathcal{H}_{\rm mf} \left( \Theta, \mbox{$-$$\Phi$$-$$n\omega$} \right) , 
\end{align}
\label{eq:Hmf_trans}
\end{subequations}
for $n \Eq  0, \pm1$.  
Note that the transformations change $\Phi$ alone.  
These manifest the equivalence of 
the following six field directions 
\begin{equation}
 \Phi' = \pm ( \Phi + n\omega ), \quad
(n = 0, \pm1) .  
\label{eq:Phi_equiv}
\end{equation}
To be more precise, the Hamiltonians for these $\Phi$'s 
are related to each other through orthogonal 
transformations.  
For the three special directions $\Phi \Eq  n\omega$ ($n \Eq  0, \pm1$), 
$[\mathcal{M} \left( \Phi \right), \, 
\mathcal{H}_{\rm mf} (\Theta, \Phi) \msp2 ] \Eq 0$, 
and therefore the ground state wavefunction should be a simultaneous 
eigenvector of $\mathcal{M} (\Phi)$.  
Finally, in the limit $\Theta$$\to$0, 
the $\Gamma_1$ component of 
the ground state vanishes.  
The Hamiltonian projected to 
the remaining $\Gamma_3$ subspace 
$\mathcal{P}\mathcal{H}_{\rm mf}\mathcal{P}$
has a continuous symmetry 
$
\mathcal{P} \mathcal{H}_{\rm mf} (0,\Phi) \mathcal{P} 
\Eq  
U (\FRAC12 \Phi )  
[ \mathcal{P}  \mathcal{H}_{\rm mf} (0,0) \mathcal{P} ] 
[U (\FRAC12 \Phi )]^{\blue \rm T}
$.  

Let $\langle \QQ \rangle_0 (\Phi)$ be 
the ground-state expectation value calculated for 
$\mathcal{H}_{\rm mf} (\Theta,\Phi)$, and then 
it is also accordingly transformed for the equivalent $\Phi$'s in 
Eq.~\eqref{eq:Phi_equiv}.  
For simplicity, we drop the $\Theta$-dependence for a while.  
The relations \eqref{eq:Hmf_trans} imply 
\begin{subequations} 
\begin{align}
\langle \QQ \rangle_0 \left( \mbox{$\Phi$$+$$n\omega$} \right) &= 
\bigl\langle 
  [\mathcal{U} \left( n\omega \right)]^{\rm T} 
  \QQ \, 
  \mathcal{U} \left( n\omega \right) 
\bigr\rangle_0 ( \Phi ) 
\nonumber\\
&= 
U \left( n\omega \right) \langle \QQ \rangle_0 (\Phi), 
\\
\langle \QQ \rangle_0 \left( \mbox{$2n\omega$$-$$\Phi$} \right) &= 
\bigl\langle 
   \mathcal{M} \left( n\omega \right) 
   \QQ \, 
   \mathcal{M} \left( n\omega \right) 
\bigr\rangle_0 ( \Phi ) 
\nonumber\\
&= 
\mu \left( n\omega \right) \langle \QQ \rangle_0 (\Phi).
\end{align} 
\label{eq:Q_trans}%
\end{subequations}
This means that the quadrupole moment $\langle \QQ \rangle_0$ 
is transformed as a two-dimensional vector  
with the rotation $U(n \omega)$ or mirror $\mu (n \omega)$ 
operation defined in Eqs.~\eqref{eq:MandU}.  
The transformed values are those six listed in Eq.~\eqref{eq:Q6equiv}.  
These symmetries also imply the following properties for the special 
field directions
$\Phi \Eq \FRAC12 \omega \msm2 \times \msm2 \mbox{(integer)}$:
\begin{equation}
 \langle \QQ \rangle_0 (\Phi ) \parallel \hh , \ \ \ 
\frac{\partial}{\partial \Phi} 
\bigl| \langle \QQ \rangle_0 (\Phi) \bigr| = 0, 
\end{equation}

\begin{figure}[t]
\begin{center}
\includegraphics[width=0.5\textwidth]{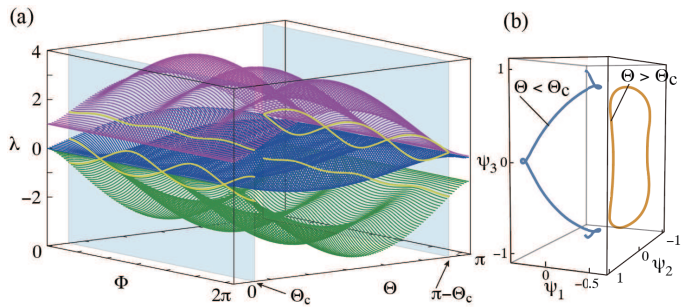}
\end{center}
\caption{
(a) Eigenenergies $\lambda$ of 
${\mathcal H}_{\rm mf} (\Theta, \Phi )$ in 
Eq.~(\ref{eq:HamilHloc}). 
For clarity, the data for $\Theta=\Theta_c$ or 
$\pi-\Theta_c$, 
showing the gap closing, are highlighted. 
(b) Trajectory of the ground state wavefunction 
$\psi=(\psi_1,\psi_2,\psi_3)$ from $\Phi=0$ to $2\pi$. 
The values with 
$\Theta \Eq \cot^{-1} 5 \LT \Theta_c$ 
and $\Theta \Eq \cot^{-1} 3 \GT \Theta_c$ are used as examples. 
$\psi_{i} (\Phi \Eq 2\pi) =-\psi_{i} (0)$ for all $i$'s 
when $\Theta \LT \Theta_c$, 
while $\psi_{i} (2\pi ) \Eq \psi_{i} (0)$ when $\Theta >\Theta_c$. 
The upper and lower end points 
correspond to $\Phi \Eq \pm\pi$ for $\Theta<\Theta_c$.
}
\label{fig4}
\end{figure}

Let us analyze the eigenvalues of 
$\mathcal{H}_{\rm mf} (\Theta,\Phi)$.   
It is enlightening to regard this problem 
as a one-dimensional system 
with three internal states 
where $\Phi$ is its ``wave number'' while $\Theta$ is a control 
parameter \cite{Bernevig}.  
At $\Theta$=$\pi$, the ground state wavefunction 
is trivially a pure singlet $\Gamma_1$ state 
and the excitations are fully gapped for all $\Phi$'s  
as shown in Fig.~\ref{fig4}. 
As $\Theta$ decreases, the gap reduces and closes at 
$\Theta_c 
\Eq \cot^{-1} \FRAC{27}{8}
\! \simeq \! 0.092\pi$ 
and $\Phi \Eq \pi \PLUS n\omega$ ($n \Eq 0, \pm1$).   
The symmetry of the Hamiltonian also implies 
a level crossing between the two excited states at 
$\Theta \Eq \pi \MINUS \Theta_c$.  
With varying $\Theta$, 
the ground state changes its nature drastically 
from a half-integer ``spin'' type ($\Theta \! \sim \! 0$) 
to an integer ``spin'' one ($\Theta \! > \! \Theta_c$). 
This change is characterized by the Berry phase 
factor $(-1)^{2\pi \chi_B}$ 
of the ground state wavefunction 
$\bm{\psi}$
acquired during the adiabatic change in $\Phi$ from 0 to $2\pi$: 
$\bm{\psi}(\mbox{$\Phi$=$2\pi$}) 
 \Eq  
 (-1)^{2\pi \chi_B} \bm{\psi}(0)$ \cite{Haldane1983a,Haldane1983-xf}.  
This factor $(-1)^{2\pi \chi_B}$ is $-1$ for $\Theta<\Theta_c$, 
while $+1$ for $\Theta>\Theta_c$.  
This change comes from the contribution of the three singular points 
at $(\Theta, \Phi) \Eq  (\Theta_c , \mbox{$\pi$$+$$n\omega$})$, 
at which the two low-energy modes exhibit a dispersion of Dirac-cone type.  

%
%

\begin{figure}[t]
\begin{center}
\includegraphics[width=0.5\textwidth]{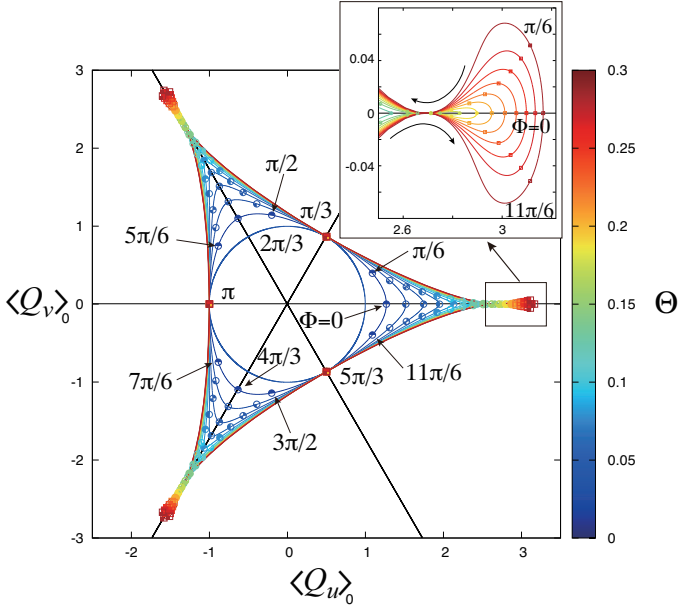}
\end{center}
\vspace{-10pt}
\caption{
Trajectory of $\langle \QQ \rangle_0$ as $\Phi$ varies 
from 0 to $2\pi$ is drawn for several values of $\Theta \LT \Theta_c$. 
The $\Theta$ value is shown by color, 
and positions for typical $\Phi$ values are indicated. 
The unit circle represents the result in
the pseudospin-1/2 limit ($\Theta=0$). 
The magnitude  
$|\langle \QQ \rangle_0|$ increases as $\Theta$ increases. 
Inset is a zoom up near $\langle \QQ \rangle_0 \sim (3,0)^{\rm T}$ where
$\Phi\sim 0$. The arrows indicate the direction 
for increasing $\Phi$. Trajectories show a similar behavior 
for $\Phi \sim 2\pi/3$ and $4\pi/3$.  
}
\label{fig5}
\end{figure}

\begin{figure}[t]
\begin{center}
\includegraphics[width=0.5\textwidth]{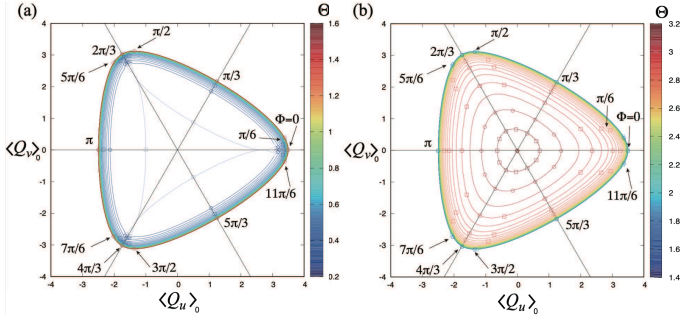}
\end{center}
\vspace{-10pt}
\caption{
$\Phi$-trajectories of 
$\langle \QQ \rangle_0$ for $\Theta \GE \Theta_c$.  
The magnitude $|\langle \QQ \rangle_0|$ 
increases as $\Theta$ increases (a) when $\Theta \LT \pi/2$,  
while decreases when (b) $\pi/2 \LT \Theta \LT \pi$.
}
\vspace{-10pt}
\label{fig6}
\end{figure}

This change in the ground-state wavefunction is clearly reflected 
in the trajectory of quadrupole moment 
$\langle \QQ \rangle_0 (\Theta , \Phi)$$=$
$[\bm{\psi}(\Theta , \Phi)]^\dagger \QQ \bm{\psi}(\Theta , \Phi)$ 
upon varying $\Phi$ from 0 to $2\pi$.  
See Figs.~\ref{fig5} and \ref{fig6}.  
The $\Phi$-trajectory exhibits qualitatively different shapes
in the four regions of $\Theta$: 
(A) $0 \LE  \Theta \LT  \Theta_{c2} 
\Equiv  \cot^{-1} \FRAC{27}{4} \! \simeq \! 0.047\pi$, 
(B) $\Theta_{c2} \LT  \Theta \LT  \Theta_{c}$, 
(C) $\Theta_{c} \LT  \Theta \LT  \FRAC12 \pi$, 
(D) $\FRAC12 \pi \LT  \Theta \LT \pi$.  
We will discuss each region below.  

Let us start from the region A.  
Results for several values of $\Theta\le \Theta_c$ 
are shown in Fig.~\ref{fig5}.  
In the limit $\Theta \! \to \! 0$ (i.e., $|\hh | /E_1 \! \to \! 0$), 
the ground state has no $\Gamma_1$ component, 
and thus can be represented as a pseudospin-1/2 wavefunction. 
This leads to trivial results: 
$|\langle \QQ \rangle_0|=1$ and 
$\langle \QQ \rangle_0 \parallel \hh $ 
for any field direction $\Phi$. 
Thus, the $\Phi$-trajectory is a unit circle. 
As $\Theta$ increases, 
the amplitude of $\langle \QQ \rangle_0$ grows 
and its direction deviates from the field direction $\Phi $. 
As a consequence, the trajectory is no longer circular 
but keeps threefold rotation and mirror symmetries imposed 
by the relations \eqref{eq:Q_trans}.  
This indicates that the hybridization of the $\Gamma_1$ state 
generates an anisotropy in the $\QQ$ space, 
and this is an essentially important aspect of the quadrupole order.  
The three special field directions 
$\Phi \Eq \pi \PLUS n\omega$ ($n \Eq 0, \pm1$) are special, 
and the moment 
is then pinned as 
$ \langle \QQ \rangle_0  (\Theta , \Phi)  \Eq  
(\cos \Phi , \sin \Phi )^{\rm T}$ 
and does not change its amplitude with $\Theta$.  
In their opposite directions $\Phi \Eq n\omega$, 
the moment has a maximal amplitude 
\begin{equation}
\bigl| \langle \QQ \rangle_0 (\Theta , n\omega) \bigr| 
= 
\frac12 
\left[
1 + 
\frac{\cot \Theta \PLUS 36}%
{\sqrt{(\cot \Theta \PLUS 1)^2 + 35}}
\right] , 
\end{equation}
and this value grows monotonically from 1 to $\FRAC72$ 
as $\Theta$ varies from 0 to $\FRAC12 \pi$.  
The upper boundary of the region A is $\Theta_{c2}$, and 
for the three special field directions 
the quadrupole moment points to 
\begin{equation}
\langle \QQ \rangle_0 (\Theta_{c2} , n\omega)  
 = \FRAC{35}{13} (\cos n \omega , \sin n \omega)^{\rm T}
 =: \bar{\QQ}_{p}^{(n)} . 
\end{equation}
These become cusps of the $\Phi$-trajectory, 
and actually the $\Phi$-dependence is singular there: 
for example, 
$\langle Q_v \rangle_0 (\Theta_{c2} , \delta \Phi) \propto 
(\delta \Phi)^5$ for $|\delta \Phi| \! \ll \! 1$.  

In the region B of $\Theta$, the moment 
$\langle \QQ \rangle_0$ changes its direction non-monotonically 
with $\Phi$,  
and the trajectory has three pinch points 
located at $\bar{\QQ}_{p}^{(n)}$.  
The pinch points are reached when 
the field direction is 
$\Phi \Eq  n\omega \! \pm \! \delta$ 
where $ \cos \delta \Eq  \cot \Theta / \cot \Theta_{c2}$.  
It is remarkable that each pinch point $\bar{\QQ}_{p}^{(n)}$ is 
visited \textit{twice} upon varying field direction $\Phi$ 
with $\Theta$ fixed:
once from ``above'' and from ``below'' the other time.  
See the inset of Fig.~\ref{fig5}. 
These pinch points do not move with $\Theta$ in the region B, 
since the ground state there is fixed to 
$\bm{\psi} (\Theta , n\omega \PM \delta) 
 \msm2 \propto \msm2 \mathcal{U}(n\omega) (1,\alpha,0)^{\rm T}$.  

The upper boundary of the region B is $\Theta \Eq  \Theta_{c}$, 
and this case is exceptional.  
The $\Phi$-trajectory is not connected but consists of three 
disconnected parts.  
This is due to the level crossing of the ground state 
at $(\Theta_{c},\Phi \Eq  \pi \!+\! n\omega)$ discussed before.  
The doubly degenerate ground states are 
$\bm{\psi}_1 \!\propto\! \mathcal{U}(n\omega)(2,-\alpha,0)^{\rm T}$ and 
$\bm{\psi}_2 \Eq  \mathcal{U}(n\omega)(0,0,1)^{\rm T}$.  
With approaching one level-crossing point from some direction 
in the $(\Theta,\Phi)$ space, these two states are 
hybridized to form the ground state as 
$\bm{\psi}_1 \cos \xi \PLUS  \bm{\psi}_2 \sin \xi$, and 
the mixing angle $\xi$ is determined by the approaching direction.  
With varying $\xi$ for each crossing point, 
the expectation value $\langle \QQ \rangle_0$ 
traces a fraction of an ellipse.  
For example, 
around the crossing point at $\Phi \Eq  \pi$ (i.e., $n$=0), 
this ellipse is expressed as 
\begin{equation}
\left(
\frac{\langle Q_u \rangle_0 + 26/17}{9/17}
\right)^2 
+ 
\left(
\frac{\langle Q_v \rangle_0 }{\sqrt{105/17}}
\right)^2 
= 1 . 
\end{equation}
One should note that the ``pinned'' point 
$\langle \QQ \rangle _0 \Eq (-1,0)^{\rm T}$ is on the minor axis 
of this ellipse.   
The ellipses for the other level crossing points are 
obtained by rotating this by the angle $\pm \omega$ about the origin.  
When the parameter set passes a level-crossing point 
$(\Theta_c,\Phi \Eq \pi \PLUS n\omega)$, 
$\langle \QQ \rangle_0$ jumps from one point on the corresponding 
ellipse to its opposite point. 
In the case of the $\Phi$-trajectory, this jump takes place 
from a point on the ellipse's \textit{major} axis to its opposite point.  
Thus, the interiors of these ellipses 
are \textit{a forbidden region} of $\langle \QQ \rangle_0$ 
as far as $E_1 \GT 0$, 
and this property is also important 
in the discussion of the quadrupole order.  

%
%
%
%

In the region C of $\Theta$, the $\Phi$-trajectory 
becomes simple again. 
Just above $\Theta_c$, $|\langle \QQ \rangle_0|$ shows jumps to the
larger magnitude as shown in Fig.~\ref{fig6}(a). 
This is what we have just discussed above.  
For example, $|\langle \QQ \rangle_0| (\Phi \Eq \pi)$ jumps from $1$ at 
$\Theta \Eq \Theta_c -0$ 
to $\frac{35}{17}$ at $\Theta_c +0$. 
In this region, the quadrupole moment 
changes its direction monotonically with $\Phi$ 
as shown in Fig.~\ref{fig6}(a).  

Finally, while the region D ($\FRAC12 \pi \LT  \Theta$) is 
not directly relevant to the following sections, 
we also study this region to complete the single-site analysis.  
This corresponds to the situation where the singlet $\Gamma_1$ 
level is below the $\Gamma_3$ doublet.   
This indicates that $|\langle \QQ \rangle _0 |$ 
should shrink to 0 
as $\Theta$ approaches $\pi$. 
This process is shown in Fig.~\ref{fig6}(b). 
The change is smooth with respect to both $\Theta$ and $\Phi$, and  
shows no jump or singularity.  
This is consistent with the energy level analysis
in Fig.~\ref{fig4}(a), where there is no ground-state level crossing.

\section{Mean field approach}
\label{sec:MFresults}

In this section, we employ a mean-field approach to 
determine the phase diagram and investigate ordered phases.  
We will first analyze the limit of $E_1 \To  \infty$.  
This corresponds to a pseudospin-$\FRAC12$ model with no internal anisotropy.  
As shown in Fig.~\ref{fig5}, the trajectory of 
field response is a unit circle, 
despite the anisotropic interaction $K$ exists. 
We will show that only three states appear in that limit: 
two antiferro states with O$_{20}$ or O$_{22}$ order parameter, 
and an isotropic ferro state.  
Then, we will proceed to develop a four-sublattice mean-field theory.  
Its results predict several triple-$\qq$ states. 
We will determine the $J$--$K$ phase diagram and calculate 
the temperature dependence of the order parameters.  
Details of the triple-$\qq$ orders will be also discussed.

%
%
\begin{figure}[t]
\begin{center}
\includegraphics[width=0.43\textwidth]{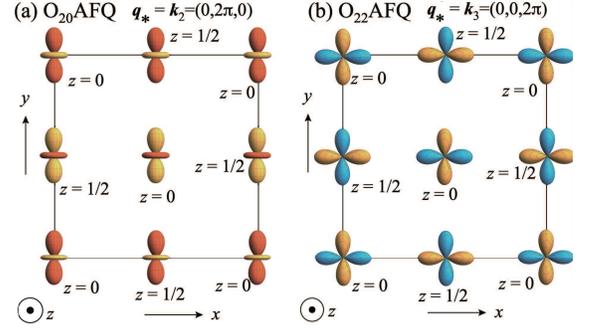}
\end{center}
\vspace{-10pt}
\caption{
Quadrupole configuration in the 
AFQ ordered states viewed from the positive $z$ direction. 
The $z$ coordinate of each site 
is indicated beside the orbital. 
(a) O$_{20}$ AFQ order with $\qq_*=\kk_2$. 
(b) O$_{22}$ AFQ order with $\qq_*=\kk_3$. 
}
\label{fig8}
\end{figure}

%
%
\begin{figure}[t]
\begin{center}
\includegraphics[width=0.45\textwidth]{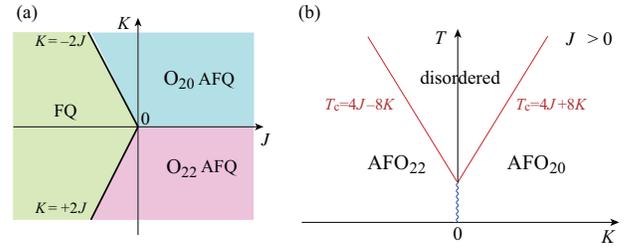}
\end{center}
\vspace{-10pt}
\caption{
Phase diagram in the limit of $E_1\to \infty$   
in (a) the $J$--$K$ and (b) the $K$--$T$ plane.  
In (b) the wavy line represents a first-order transition.
}
\label{fig7}
\end{figure}

\vspace{10pt}
\subsection{Pseudospin-1/2 limit}
\label{sec:spin-onehalf}

In the limit of $E_1 \To  \infty$, we can safely ignore 
the excited $\Gamma_1$ state for both the Hamiltonian (\ref{eq2:HQ}) 
and the quadrupole operators $\QQ (\rr)$.  
Thus $\QQ (\rr)$'s are treated as pseudospin-$\FRAC12$ operators,  
and the mean-field equation for the order parameter  
is easy to analyze \cite{Tsunetsugu2021}.  
For an isolated quadrupole, the response to the mean field $\hh$
has an elementary form 
$\langle \QQ \rangle \Eq  \mathscr{N} (h/T) \, \chi_0 (T) \, \hh$.  
Here, 
$\chi_0 \Eq  1/T$ is the linear susceptibility, and 
the nonlinear correction is
$\mathscr{N} (x) \Equiv  x^{-1} \tanh x$ 
 The latter can be neglected for $|x| \!\ll\! 1$.  

First, 
let us consider the region of ferro order ($2J < |K| < -2J$). 
The eigenmode analysis 
shows that the two eigenvalues of $\mathsf{J}  (\kk)$ are degenerate, 
since the $K$ terms vanish at $\kk_0 \Eq  \bm{0}$.  
This means there is no preference for the ordering direction 
in the $Q_u$--$Q_v$ quadrupole space. 
Recalling that each site has 12 nearest neighbors, one sets 
$\hh \Eq  12 |J |\langle \QQ \rangle$, 
and this leads to 
the condition of the transition temperature 
$12 |J| \chi_0 (T_c) \Eq  1$.  
Its solution and the ground-state energy are 
\vspace{-6pt}
\begin{equation}
T_c = 12 |J|, \quad 
E(\mbox{$T$=0})=6J, \quad (\rm FQ). 
\label{eq:TcFQ}
\end{equation}

For the antiferro case, one chooses one of the ordering vectors 
$\qq_*$ at the X points.  
Let us first discuss the O$_{20}$ type realized for $K \GT  0$. 
See the eigenvector of the exchange coupling $\mathsf{J} (\pp)$ 
in Fig.~\ref{fig3}. 
For the domain with $\qq_* \Eq  \kk_2 \Eq  (0,2\pi,0)$, 
the order parameter is $\mbox{$3y^2$$-$$r^2$}$ type:
$\uuu  ^\pm (\kk_2)$ with 
$\bar{\vartheta} (\kk_2) \Eq \FRAC43 \pi$ or $\FRAC13 \pi$ 
as shown in Eq.~(\ref{eq:eigen_theta}).  
Since this ordering vector corresponds to two-sublattice (A and B) orders, 
one uses a two-sublattice version of the mean-field theory 
with 
$\hh_{\rm A,B} \Eq  
- 4J (\langle \QQ_{\rm A} \rangle \!+\! 2 \langle \QQ_{\rm B} \rangle)
\!\mp\! 4K \mathsf{g} 
( \bdelta _2 ) 
(\langle \QQ_{\rm A} \rangle \!-\! 
 \langle \QQ_{\rm B} \rangle)$. 
Here, $-(+)$ is for $\bm{h}_{\rm A(B)}$. 
The self-consistent equations for $\langle \QQ_{\rm A,B} \rangle$ lead to 
\begin{equation}
T_c=4J+8K,\quad 
E(\mbox{$T$=0})=-2J-4K,\ ({\rm AFO}_{20}). \quad
\label{eq:TcO20}
\end{equation}
This antiferro quadrupole pattern is schematically shown 
in Fig.~\ref{fig8}(a), 
where the  quadrupole moments 
exhibit a ferro alignment on each $zx$ plane, 
while an antiferro alignment along the $y$ direction. 
Similarly, for $K \LT  0$, the antiferro O$_{22}$ quadrupole solution 
leads to 
\begin{equation}
T_c=4J-8K,\quad 
E(\mbox{$T$=0})=-2J+4K,\ ({\rm AFO}_{22}), \quad 
\label{eq:TcO22}
\end{equation}
which are obtained from Eq.~(\ref{eq:TcO20}) 
by just replacing $K \To  -K$. 
As an illustrative example, 
Fig.~\ref{fig8}(b) shows the ordering pattern for O$_{22}$ 
antiferro order with $\qq_* \Eq  \kk_3 \Eq  (0,0,2\pi)$,  
where the orbital type is $\mbox{$x^2$$-$$y^2$}$: 
$\vvv  ^\pm(\kk_3)$ with 
$\theta(\kk_3) \Eq  \pm \FRAC12 \pi$. 
See also Fig.~\ref{fig3}(b) and Eq.~(\ref{eq:eigen_theta}).

The ground state phase diagram determined 
from Eqs.~(\ref{eq:TcFQ})--(\ref{eq:TcO22}) is shown in Fig.~\ref{fig7}(a). 
The phase boundaries agree with those obtained 
by the mode analysis in Sec.~\ref{sec:exJ}. 
Note that the transition from 
the high-temperature disordered phase 
is always second order, 
while that between the different ordered states are first order 
as shown in Fig.~\ref{fig7}(b).

\vspace{10pt}
\subsection{Triple-$\qq$ orders}
\label{sec:tripleqorders}

Now, let us consider realistic situations where $E_1 \LT  \infty$, 
and examine whether the solutions in 
the previous section remain stable or not.  
The discussion above has examined only 
the leading instability due to the exchange coupling $\mathsf{J} (\pp)$. 
A crucial point is missing, and 
that is the anisotropy in the $Q_u$--$Q_v$ space 
that emerges from a hybridization of the excited $\Gamma_1$ state. 
When this anisotropy is taken into account, 
most parts of the ordered antiferro states 
in Fig.~\ref{fig7}(b) are to be replaced by various triple-$\qq$ orders, 
and they have four-sublattice configurations in the real space. 
In this section, 
we concentrate on showing the results of 
the four-sublattice mean-field calculations.  
The mechanism stabilizing the triple-$\qq$ states will be 
discussed in detail in Sec.~\ref{sec:Landau} 
based on a phenomenological Landau theory.

%
%
\begin{figure*}[t]
\begin{center}
\includegraphics[width=\textwidth]{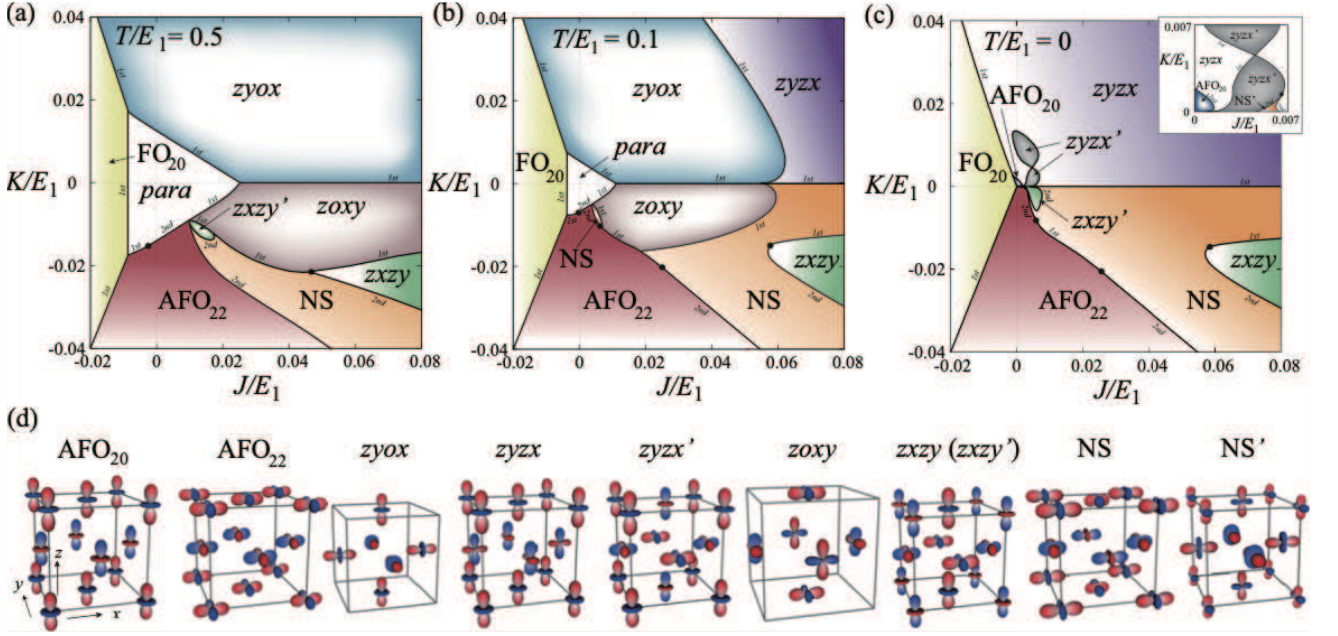}
\end{center}
\caption{ 
(a)--(c) $J$-$K$ phase diagram 
at fixed values of $T$, and 
(d) real-space configuration $\QQ (\rr)$ in each ordered state.
(a) $T/E_1=0.5$, (b) $0.1$, and (c) $0.0$. 
The order of transition is 
indicated for the borders in (a)-(c) as ``1st'' or ''2nd''. 
Tricritical points are marked by filled circles. 
``$para$'' in (a) and (b) denotes 
a disordered phase with no symmetry breaking. 
See also Fig.~\ref{table1} for 
the symmetry and $\QQ$ configuration in each state. 
In (b), a region of the $zxzy$ state exists 
near $(J,K)/E_1\sim (0.005,-0.006)$, 
but it is too small to see. 
In (c), the $zxzy'$ state contains a small $zxzy$ region 
but they are not distinguished since both have the same symmetry.  
Inset in (c) is a zoom up of the part 
near the origin for $K>0$. 
The $zyzx'$ state has two disconnected parts.
In the part at large $J$, 
two of the four sublattice $\QQ$'s are the largest there
(see the $zyzx'$ state in Fig.~\ref{table1}). 
In the part at small $J$,
one $\QQ$ is larger than the others, 
(see the $zyzx$ state in the Fig.~\ref{table1}).
A small region of the NS$'$ state [not shown in the main panel of (c)] 
bridges the $zyzx'$ and $zyzx$ states via two second-order transitions. 
The phase diagram at $K=0$ is discussed in Appendix~\ref{KzeroAnal}.
(d) Schematic quadrupole configuration 
in each ordered state in a cubic unit cell. 
In the $zyzx$ state, 
moments at the corners have a slightly different magnitude 
from those at the face centers.
}
\label{fig9}
\end{figure*}

%
%
\begin{figure}[t]
\begin{center}
\includegraphics[width=0.49\textwidth]{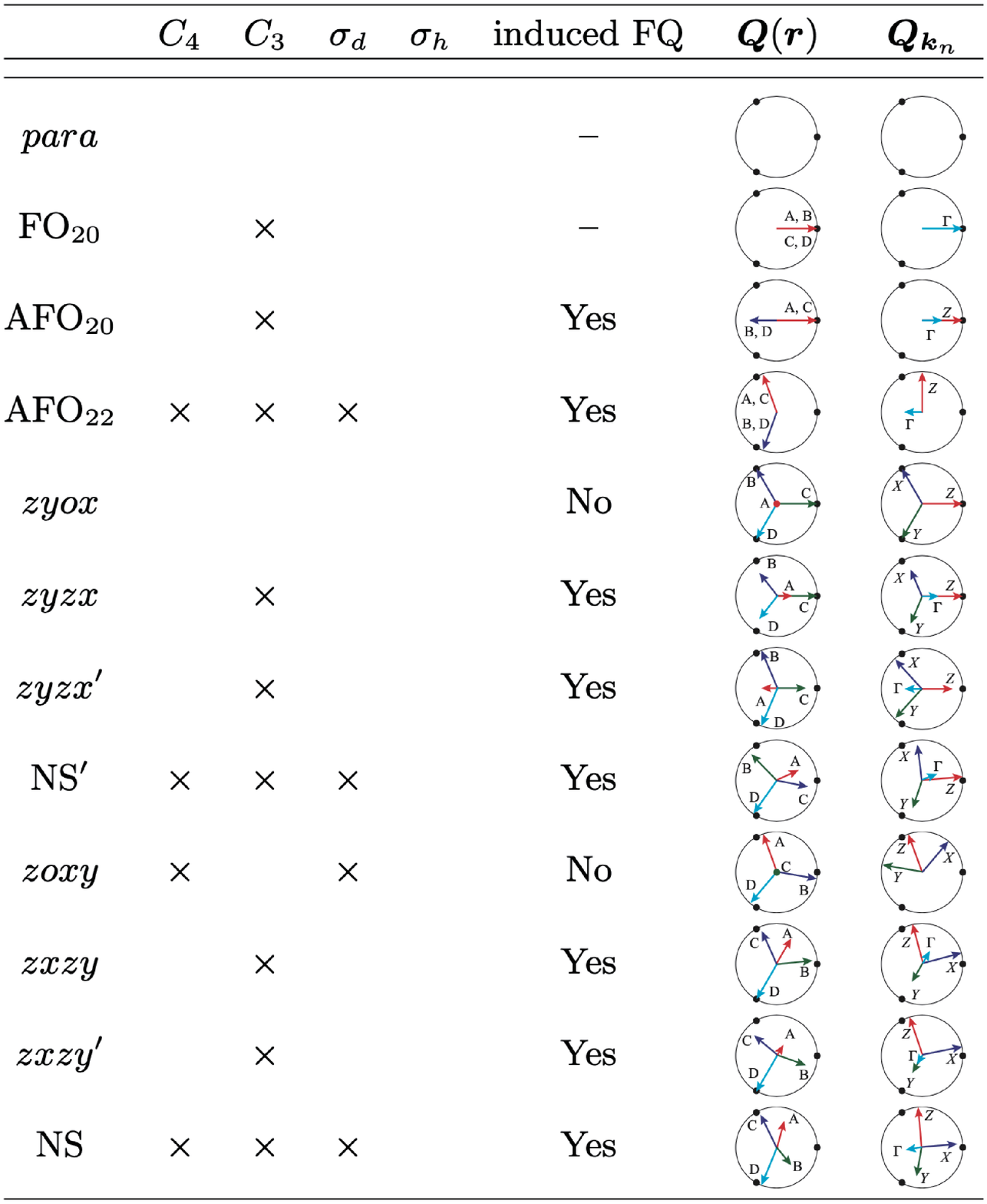}
\end{center}
\caption{ 
Symmetry of the ordered states and 
schematic quadrupole configurations
in real space $\QQ(\rr)$ and in momentum space 
$\QQ_{\kk_n} (n=0,1,2,3)$. 
The first four columns show a list of broken symmetries 
among those of the original point group:
($C_4$) the four-fold rotation about a principal axis, 
($C_3$) the three-fold rotation about [111] direction, 
($\sigma_d$) the mirror with respect to ($1\bar{1}0$) plane, 
($\sigma_h$) the mirror with respect to ($001$) plane. 
Broken ones are indicated by a $\times$ symbol.
$\QQ$'s are shown by
arrows in the two-dimensional quadrupole space 
where the horizontal (vertical) direction corresponds to 
the $Q_u$ ($Q_v$) component. 
The $zyzx$ $(zxzy)$ and $zyzx'$ $(zxzy')$ states
have the same symmetry, but are 
distinguished by the direction of the smallest $\QQ(\rr)$ $(\QQ_{\kk_0})$.
Circles with dots are guide for the eye.
$\QQ(\rr)$'s are calculated using 
Eq.~(\ref{eq:Qloc}) from $\QQ_{\kk_n}$'s, for 
the domain chosen in Figs.~\ref{fig10}--\ref{fig12}.  
} 
\label{table1} 

\end{figure}

For a general configuration of the four-sublattice order, 
the sublattice order parameters $\QQ_{\rm A,B,C,D}$ 
are related to those in the Brillouin zone at the four wavevectors 
$\kk_n$ ($n=0$, 1, 2, 3).  
Labeling the sublattices as shown in Fig.~\ref{fig1}(a), 
the relation is 
\begin{align}
\left[
\begin{array}{@{\hspace{2pt}}c@{\hspace{2pt}}}
\QQ _{\rm A} \\
\QQ _{\rm B} \\
\QQ _{\rm C} \\
\QQ _{\rm D}
\end{array}
\right]
=
\frac{1}{2}
\left[
\begin{array}{@{\hspace{2pt}}c@{\hspace{4pt}}r@{\hspace{4pt}}r@{\hspace{4pt}}r@{\hspace{2pt}}} 
1 & 1 & 1 & 1 \\
1 & 1 & -1 & -1 \\
1 & -1 & -1 & 1 \\
1 & -1 & 1 & -1 
\end{array}
\right]
\left[
\begin{array}{@{\hspace{2pt}}c@{\hspace{2pt}}}
\QQ _{\kk_0} \\
\QQ _{\kk_1} \\
\QQ _{\kk_2} \\
\QQ _{\kk_3} 
\end{array}
\right] , 
\label{eq:Qloc}	
\end{align}
where 1's in the matrix elements are $2 \!\times\! 2$ identity matrices.  
This is evident from the representation of the triple-$\qq$ states 
$
\QQ(\rr)
\Eq 
\frac{1}{2}\sum_{n=0}^3 \cos (\kk_n\cdot \rr)\,  \QQ_{\kk_n} 
$.
We note that these order parameters $\QQ _{\kk _n}$ are real, 
since $-\kk _n$ is equivalent to $+\kk_n$ 
in the Brillouin zone. 

For later purposes, 
we introduce 
the following polar coordinates  
of the quadrupole moments:
\vspace{-6pt}
\begin{align}
&\QQ_{\kk_0} \equiv Q \hee _{\theta_\Gamma}, \ \ 
\QQ_{\kk_1} \equiv X \hee _{\theta_X}, 
\nonumber\\
&\QQ_{\kk_2} \equiv Y \hee _{\theta_Y}, \ \  
\QQ_{\kk_3} \equiv Z \hee _{\theta_Z}. 
\label{eq:PolarQQ} 
\end{align}
Here, $Q,X,Y,Z \! \ge \! 0$.
We sometimes use an alternative notation 
\begin{equation}
\QQ _{\kk _n} = X_n \, \hee (\theta _n) . \label{eq:PolarQQ2}
\end{equation}
We will use these variables to distinguish and identify
various states in the following sections.

%
%
\begin{figure*}[t!h]
\begin{center}
\includegraphics[width=\textwidth]{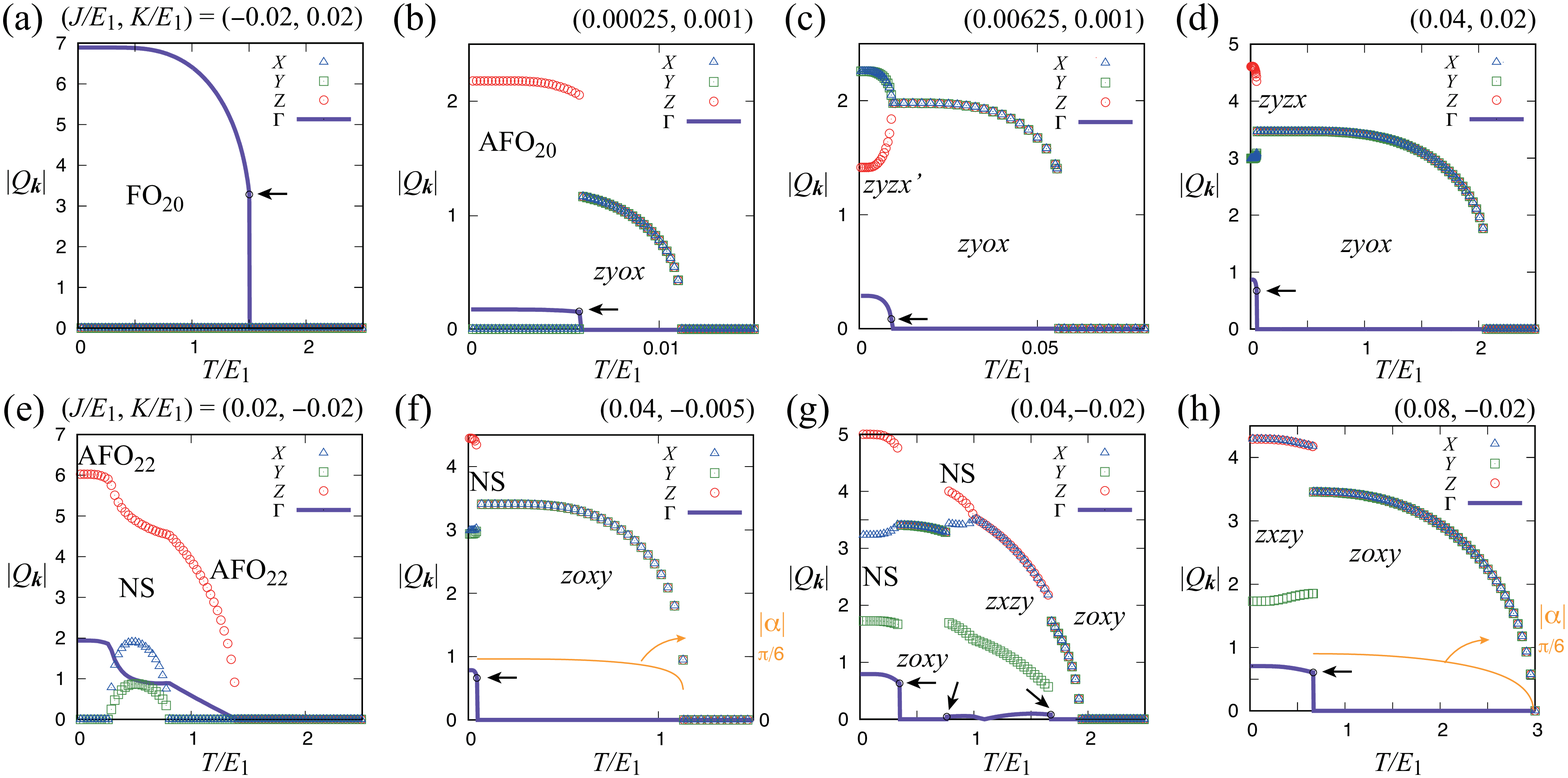}
\end{center}
\vspace{-14pt}
\caption{
Temperature dependence of the moment magnitudes 
$|\QQ_{\kk_{n}}|$ 
for (a)--(d) $K>0$ and for (e)--(h) $K<0$. 
Blue $\triangle$, green $\square$,  red {\large $\circ$}, 
and purple line show 
$|\QQ_{\kk_1}| \Eq X$, $|\QQ_{\kk_2}| \Eq Y$,  
$|\QQ_{\kk_3}| \Eq Z$, $ |\QQ_{\kk_0}| \Eq \Gamma \Eq Q$, 
respectively. 
The parameter set $(J,K)/E_1$ is shown in each panel.
Discontinuous changes in $\Gamma  \Eq Q$ 
are marked by a black arrow. 
In (f) and (h) also plotted is 
$\alpha\equiv (\theta_X \PLUS \theta_Y \PLUS \theta_Z)/3 \MINUS \pi/2$ 
(mod $2\pi$). In (b)--(d) and (f), the transition 
between the para and $zyox$ or $zoxy$ state\Blue{s} 
is discontinuous. 
}
\label{fig10}
\end{figure*}

\vspace{10pt}
\section{Mean-field phase diagram}
\label{sec:phase}

In this section, we will determine the phase diagram of 
the $\Gamma_3$--$\Gamma_1$ model \eqref{eq:defH} 
based on a four-site mean-field approach.  
We will identify various ordered states and 
briefly summarize their transitions before more detailed 
analyses in the later sections. 

First, we show the $J$--$K$ phase diagrams for $T$=$0.5E_1$, $0.1E_1$,
and $0$ in the panels (a), (b), and (c), respectively, 
of Fig.~\ref{fig9}.  
Each of the phases is identified, and 
its symmetry and schematic quadrupole configuration 
are summarized in Fig.~\ref{table1}. 
Simple antiferro (AFO$_{22}$) and ferro (FO$_{20}$) states 
appear in similar regions as those in Fig.~\ref{fig7}(b), 
while the area of the AFO$_{20}$ state is small at low temperatures. 
This is a striking difference from the results of 
the two-sublattice calculations. 
Note that the ferro quadrupole state is named 
FO$_{20}$ rather than FQ, 
since the system favors O$_{20}$ over O$_{22}$ 
due to the anisotropy driven by the hybridization $\alpha$ 
of the excited $\Gamma_1$ state in Eq.~(\ref{eq:QuQv}).   
Other regions in the $J$-$K$ parameter space are occupied 
by various types of the triple-$\qq$ orders or otherwise 
the high-temperature disordered state denoted by \textit{$para$}. 
We name these triple-$\qq$ states 
 by a combination of four letters 
according to the real-space 
quadrupole configurations in the four sublattices \cite{Tsunetsugu2021}. 
These four letters denote the principle axis of 
uniaxial orders in the sublattices A--D. 
For example, $zyox$ means that 
$\QQ _{\rm A}$, $\QQ _{\rm B}$, 
and $\QQ _{\rm D}$ are $z^2$, $y^2$, 
and $x^2$ type, respectively.  
 The C-sublattice is disordered, namely, ``$o$'' represents 
 disordered. Thus 
the $zyox$ and $zoxy$ states are  partially ordered. 
They appear only at high temperatures and do not exist 
in the ground state phase diagram in Fig.~\ref{fig9}(c). 
In these partial-order states, the ferro component 
has zero amplitude, 
$Q=0$, while $Q \!\ne\! 0$ in the fully ordered states. 
The phase with the lowest symmetry in the four-site mean-field calculations 
is named \textit{no symmetry state} (NS and NS$'$). 
They are not invariant for any operation of the cubic lattice symmetry 
or for sublattice exchanges.  

We comment about new aspects of the phase diagrams 
untouched in the first report \cite{Tsunetsugu2021}. 
The first aspect is about the NS$'$ state. 
This appears when $K \GT 0$ in an extremely small region 
near the phase boundary between the $zyzx'$ and $zyzx$ states 
at very low temperatures as shown in the inset of Fig.~\ref{fig9}(c). 
As shown in Fig.~\ref{table1}, the two states $zyzx$ and $zyzx'$ 
have the same symmetry, 
and thus the transition between the two should be first order. 
When the NS$'$ state intervenes into their boundary, 
the new boundary becomes a line of second-order transition 
with either the $zyzx$ or $zyzx'$ state.   
Remember that the two modes 
$\QQ _{\kk_1}$ and $\QQ _{\kk_2}$ 
are both dominant in the $zyzx'$ state (i.e., $X \Eq  Y>Z$), 
while 
only one mode is dominant in the $zyzx$ state: $\QQ _{\kk_1}$ ($X \GT  Y=Z$). 
See Fig.~\ref{table1}. 
Thus, as $J$ increases starting from the $zyzx'$ state, 
we observe that the balance of the two modes is 
eventually lost such that $X>Y$ in the NS$'$ state 
and then $Y$ approaches $Z$  
with approaching the $zyzx$ state. 
This change may happen continuously, and 
this explains the calculated results.  
We note that while the NS and NS$'$ states have the same symmetries, 
they are distinct. Further details are analyzed in Appendix~\ref{KzeroAnal}.

The second new aspect is about a tiny region of the $zxzy'$ state 
near the small island of the NS state 
at $T/E_1=0.1$ in Fig.~\ref{fig9}(b).  
The area of this  
is too small to see there.  
The third aspect is about the order of transitions.  
Some parts of the phase boundaries are of the first-order transition, 
and they are separated from the parts of 
second order transition by tricritical points.  
In Ref.~\cite{Tsunetsugu2021}, these tricritical points are shown only 
in the ground state phase diagram corresponding to Fig.~\ref{fig9}(c).
We have carefully examined the order of 
transitions at finite temperatures, and 
the locations of the tricritical points are 
indicated by filled circles 
also in the panels (a) and (b).
For example, a tricritical point exists 
on the $para$-AFO$_{22}$ phase boundary.  
This was one of our predictions in Ref.~\cite{Hattori2014}, 
where antiferro orders only differentiate two sublattices 
inside the unit cell without breaking the translation symmetry.   
The discussion there is applicable to the present case 
that an instability of $\QQ _{\kk}$ takes place 
at the X points in the Brillouin zone.  

Let us now examine the temperature dependence 
of the order parameters.  
Figure \ref{fig10} shows their amplitudes $|\QQ_{\kk_\ell}| (T)$ 
for the typical ordered states.  
We just show the results for one 
of the degenerate domains.  
The transitions of the $para$-FO$_{20}$ (a), 
$para$-$zyox$ (b), 
AFO$_{20}$-$zyox$ (b), 
and NS-$zoxy$ (f) are clearly all discontinuous, i.e. 
first order.  
The transitions of the 
$zyox$-$zyzx$ ($zyzx'$) [(c) and (d)] and 
$para$-$zoxy$ [(f)--(h)] 
are either first order or continuous as depending on the parameters.  
In contrast, the transitions of the 
$para$-AFO$_{22}$ [(g)], 
NS-AFO$_{22}$ [(e)], and  
NS-$zxzy$ [(g)] are all 
continuous for the parameter sets used.  
We note that there is no direct $para$-AFO$_{20}$ transition.
The high-temperature side above the AFO$_{20}$ state 
is always the $zyox$ state.

\begin{figure}[tb]
\begin{center}
\includegraphics[width=0.5\textwidth]{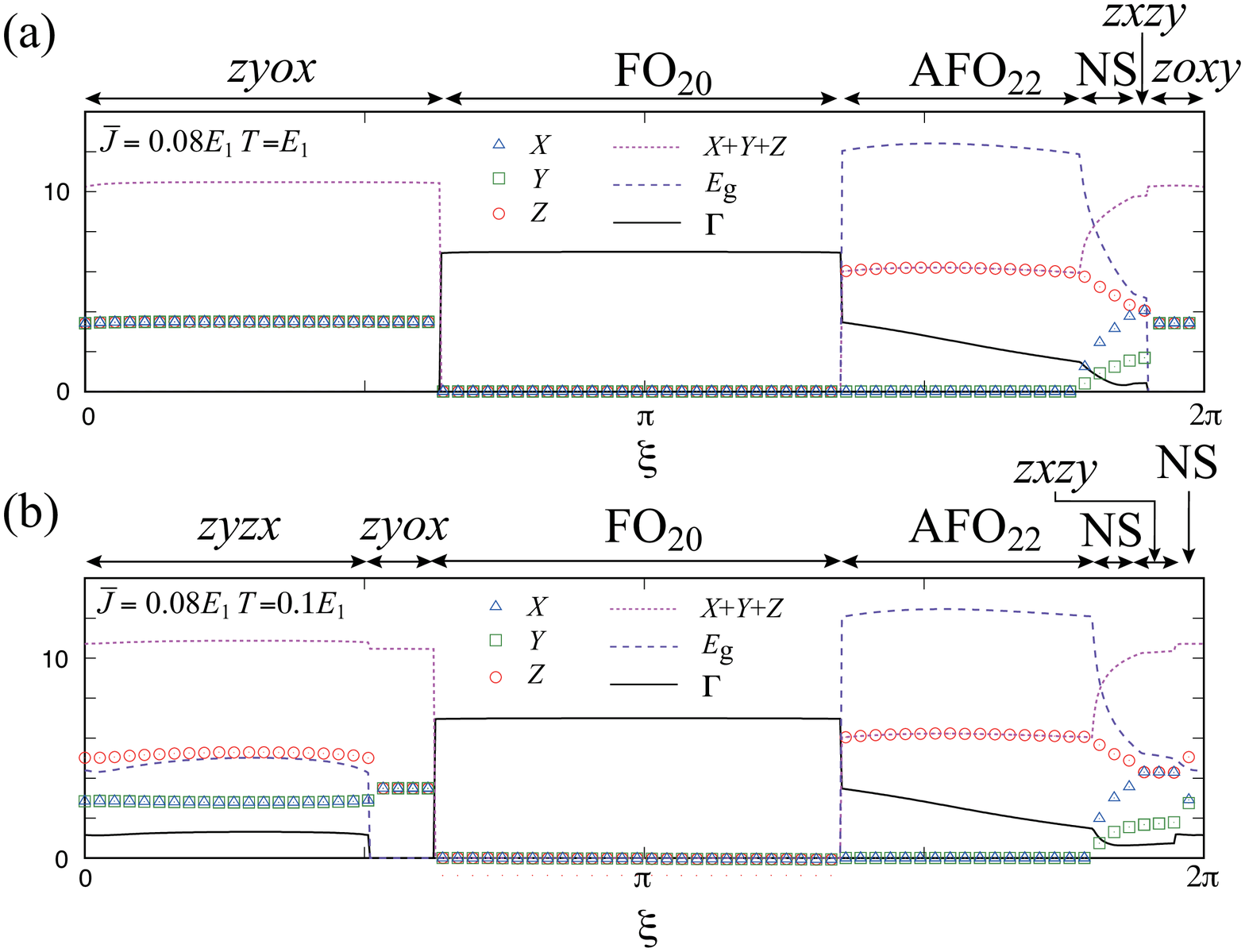}
\end{center}
\caption{
$\xi$ dependence of $|\QQ _{\kk_n|}$. 
$X$, $Y$, $Z$, and $\Gamma$ are those defined in Fig.~\ref{fig10}. 
Dotted and dashed lines represent $Q_{XYZ}$ and $Q_{E_g}$, 
respectively. 
$\bar{J}/E_1=0.08$. 
(a) $T/E_1 =1.0$ and (b) 0.10. 
}
\label{fig11}
\end{figure}

Next, we discuss the order parameter variations 
in the $J$--$K$ space for several fixed values of $T$.
To this end, we parametrize $J$ and $K$ as 
\vspace{4pt}
\begin{equation}
(J,K) \equiv \bar{J} (\cos\xi,\sin\xi),
\vspace{4pt}
\end{equation}
and vary $\xi$ from 0 to $2\pi$.
Figure~\ref{fig11} shows the $\xi$-dependence of $|\QQ_{\kk_\ell}|$ 
for $\bar{J}$=$0.08E_1$ 
at $T$=$E_1$ and $0.1E_1$.
In addition to $|\QQ_{\kk_\ell}|$'s, we introduce 
\begin{subequations}
\begin{align}
Q_{XYZ}&\equiv X+Y+Z,
\\
Q_{E_g,1}
&\equiv 
\FRAC{1}{\sqrt{6}} \msp2 (2Z-X-Y),
\ \ 
Q_{E_g,2}
&\equiv 
\FRAC{1}{\sqrt{2}} \msp2 (X-Y),
\color{black}
\end{align}
with
\begin{align}
Q_{E_g} 
&\equiv 
\bigl( Q_{E_g,1}^2+Q_{E_g,2}^2 \bigr)^{1/2}
\nonumber\\		
&=
\left[ X^2+Y^2+Z^2- \FRAC{1}{3}(X+Y+Z)^2 \right]^{1/2} .
\end{align}%
\end{subequations}
These $Q_{E_g,1}$ and $Q_{E_g,2}$ 
describe the symmetry breaking 
in the triple-$\qq$ states.
For example, both of the $zoxy$ and $zyox$ partial ordered states 
have $X$=$Y$=$Z$, 
and thus $Q_{E_g} \Eq  0$. 
These two differ in the angle variables $\theta_{X,Y,Z}$. 
The transition between AFO$_{22}$ and NS states is about 
the changes in the two modes ($X$ and $Y$ in Fig.~\ref{fig11}).  
They stay zero inside the AFO$_{22}$ state 
and emerge continuously in the NS state.  
 The two states have the same internal 
symmetries in Fig.~\ref{table1}, and 
what breaks is the translation symmetry;  
A two-sublattice order changes to a four-sublattice one.
At the NS--$zxzy$ transition, 
the equality of the largest amplitude modes breaks.
Namely, the equality $Z \Eq  X$ in the $zxzy$ state breaks down 
in the NS state.
The related transition is either first order or continuous 
as shown in Figs.~\ref{fig9} and \ref{fig11}.

%
%

%
%
\begin{figure}[h!t]
\begin{center}
\includegraphics[width=0.5\textwidth]{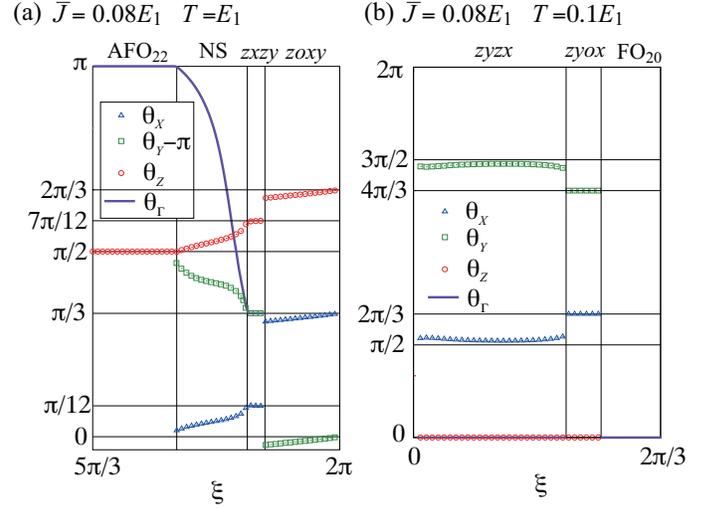}
\end{center}
\caption{
Order parameter directions 
$\theta_{\rm X,Y,Z,\Blue{\Gamma}}$ 
in a specific domain at high and low temperatures for 
$\bar{J}/E_1 =0.08$.
Note that $\theta$ is not shown when its $\QQ _{\kk_n} \Eq \bm{0}$, 
e.g., $\theta_{\Gamma}$ for the $zoxy$ and $zyox$ states.
}
\label{fig12}
\end{figure}
%
%
\begin{figure}[h!t]
\begin{center}
\includegraphics[width=0.5\textwidth]{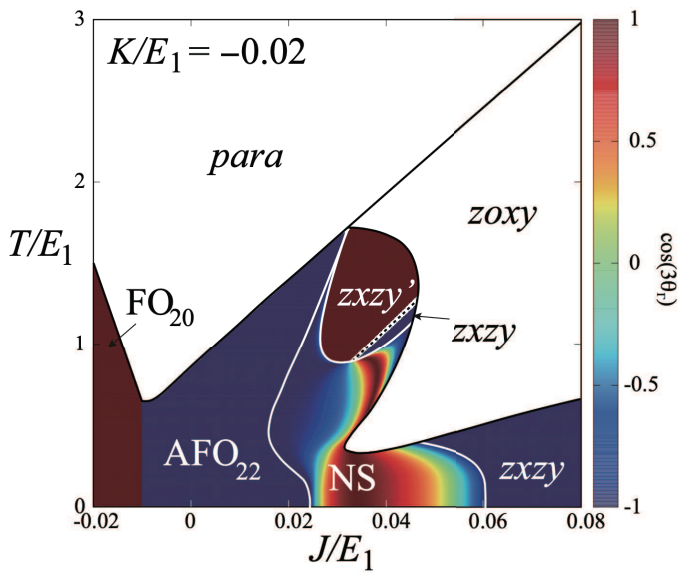}
\end{center}
\caption{
Color plot of $\cos(3\theta_\Gamma)$ in 
the $J$--$T$ plane for $K/E_1=-0.02$.  
Phase boundaries are indicated by black or white lines.  
$\theta_\Gamma=\omega\times(\mbox{integer})$ 
for the FO$_{20}$ and $zxzy'$ states, 
while $\theta_\Gamma={\pi+\omega\times(\mbox{integer})}$ 
for the AFO$_{22}$ and $zxzy$ states.  
Note that $Q$ stays zero in the $zoxy$ state.  
A dotted line represents the boundary 
between the $zxzy$ and $zxzy'$ states, 
but this does not mean a phase transition there.  
This is just the line of $Q=0$.
}
\label{fig13}
\end{figure}

Let us switch to the analysis of the moment directions.  
In the relatively high symmetry states, AFO$_{20}$, AFO$_{22}$, 
and $zyox$, the directions $\theta_{\Gamma,X,Y,Z}$ 
are fixed to the high symmetry axes 
$\FRAC14 \omega \times \mathrm{(integer)}$.  
Figure \ref{fig12} shows typical examples of their change 
as a function of $\xi$.  
In Fig.~\ref{fig12}(a), $\theta_{X}$ and $\theta_{Y}$ seem to 
abruptly appear at the AFO$_{22}$-NS phase boundary.  
However, it does not mean that the transition is of first order, 
since $X \Eq  Y \Eq  0$ inside the AFO$_{22}$ state.  
In the partially ordered $zoxy$ state, 
the quadrupole directions are equally separated 
and satisfy the relations 
$\theta_Y \MINUS  \theta_X \Eq  
 \theta_X \MINUS  \theta_Z \Eq  
 \theta_Z \MINUS  \theta_Y \Eq  
 \omega$, 
by properly shifting the origin of the four-site unit cell, 
since this shift changes two of the three directions by the angle $\pi$. 
See Eq.~(\ref{eq:Qloc}).  
In contrast, the angle average 
$\alpha \Equiv  
(\theta_X \mspace{-2mu} + \mspace{-2mu} 
 \theta_Y \mspace{-2mu} + \mspace{-2mu} 
 \theta_Z)/3$ 
varies continuously inside the $zoxy$ state 
as shown in Figs.~\ref{fig10}(f), \ref{fig10}(h), and \ref{fig12}(a).  
Note that the sign of $\alpha$ distinguishes the two domains.  
The counterpart of this state is $zyox$ in the $K \GT  0$ region. 
There, $\alpha$ is fixed to 0 and $\theta_Z - \theta_Y \Eq  -\omega$, 
while the other $\theta$-relations are unchanged.  
See Fig.~\ref{fig12}(b).  
One should also note that these $zyox$ and $zoxy$ states
have no uniform moment and thus $Q=0$. 

%
%
\begin{figure*}[t!h]
\begin{center}
\includegraphics[width=\textwidth]{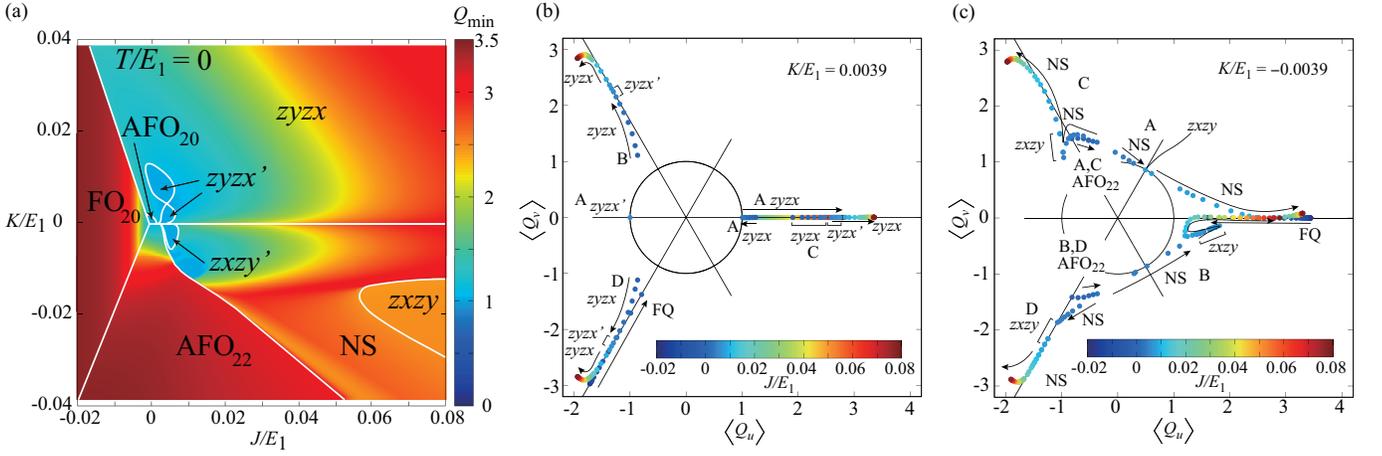}
\end{center}
\vspace{-14pt}
\caption{
(a) Color plot of $Q_{\rm min}$ in the $J$--$K$ plane for $T=0$. 
(b) Trajectories of the sublattice moments 
$\QQ_{\rm A}$--$\QQ_{\rm D}$ as $J$ varies at $K/E_1 \Eq 3.9 \times 10^{-3}$ 
and $T \Eq 0$ fixed. 
(c) Trajectories now for $K/E_1 \Eq -3.9 \times 10^{-3}$. 
Arrows indicate the variation 
upon increasing $J$ and the symbol color 
represents $J$-value. Unit circles and lines indicating directions of $\theta=\omega \times$(integer) 
are guides for the eye.
}
\label{fig14}
\end{figure*}


While the direction of the uniform moment $\theta_\Gamma$  
changes continuously in the NS state, 
it is fixed in the other states to one of the symmetry axes 
$\FRAC12 \omega\times \mathrm{(integer)}$. 
Figure \ref{fig13} plots $\cos (3\theta_\Gamma)$ 
in the $J$--$T$ plane for $K/E_1$=$-0.02$.  
Inside the NS state, it  
varies continuously between $-1$ and 1.  
One should also note that the $para$--$zoxy$ phase boundary 
extends smoothly across the multi-critical point at $J/E_1 \!\sim\! 0.03$ 
to the $para$--AFO$_{22}$ boundary.  
We will discuss this aspect in Sec.~\ref{sec:symtilting}.

We close this section by discussing the consequence of the 
topological transition explained in Sec.~\ref{sec:SSproperty}. 
In the following discussion, we discuss the local quadrupole moments $\QQ_{\rm A-D}$. 
In the mean field theory, 
strong interactions $J$ and $K$ 
enhance the effective mean fields. 
Thus, the local moments increase their amplitude 
correspondingly for most of the cases.  
However, when $\QQ$ points to 
one of the special directions 
$\theta \Eq  n\omega \PLUS \pi$ ($n$: integer), 
its amplitude is pinned to $|\QQ |=1$ and does not grow  
even when the interactions increase.  
This is because 
the system belongs to the region of ``half-integer spin'' where 
$|\langle \QQ \rangle_0 
 (\Phi\Eq  n\omega \PLUS \pi)|$=1 
as shown in Fig.~\ref{fig5}, 
as far as the mean fields 
are not so strong ($\Theta<\Theta_c$).
With further increasing the interactions, 
the parameter $\Theta$ 
increases and finally reaches $\Theta=\Theta_c$, 
where the topological transition occurs. 
For larger interactions exceeding 
the critical value ($\Theta>\Theta_c$), 
the ground state wavefunction is ``integer spin'' type,  
and $|\QQ|$ is no longer pinned and grows continuously 
as shown in Fig.~\ref{fig6}. 
The point is that as long as the local 
mean-field direction is $\Phi\Eq  n\omega \PLUS \pi$, 
$|\QQ|$ cannot change up to $\Theta \Eq \Theta_c$.
Once the direction $\Phi$ 
tilts from $n\omega \PLUS \pi$, the quadrupole 
moments also tilt 
and their amplitude can change.
 
Such a $\QQ$ pinning is  
indeed realized in the $zyzx'$, $zxzy'$, and 
AFO$_{20}$ states.  
Figure \ref{fig14}(a) is a color plot of the minimum amplitude 
$Q_{\rm min} \Equiv  \min |\QQ _{\rm A,B,C,D}|$ for $T=0$.  
It clearly shows that $Q_{\rm min}$=1 in the 
$zyzx'$, $zxzy'$, and AFO$_{20}$ states.  
Apart from these three states, 
there is a regime with  $Q_{\rm min} \!\simeq\! 1$ 
inside the NS state for $K/E_1 \!\sim\! -0.01$ and $J/E_1 \!\sim\! 0.01$.  
In the NS state, the quadrupole moment changes its direction 
away from $\theta \Eq  n\omega \PLUS \pi$, 
and $Q_{\rm min}$ varies continuously.  
In the $zyzx$ state for $K \GT  0$, 
no moments point to $\theta \Eq  n\omega \PLUS \pi$,  
and thus no such pinning effect takes place.  
In contrast, 
in the $zxzy$ state for strong coupling $J$ and $K<0$, 
one moment points to $\theta \Eq  n\omega \PLUS \pi$,  
but this time the system is already in the ``integer-spin'' domain, 
and thus pinning effects are absent. 

Figures \ref{fig14}(b) and \ref{fig14}(c) illustrate 
the variation of the sublattice moments 
$\QQ_{\rm A-D}$ with 
increasing $J$ at $T \Eq 0$. 
The panel (b) is the data for 
$K/E_1 \Eq 3.9 \times 10^{-3}$ and shows that 
$\QQ_{\rm A}$ stays $(-1,0)^{\rm T}$ inside the $zyzx'$ state  
and jumps to $(1,0)^{\rm T}$ 
at the transition to the $zyzx$ state. 
The panel (c) is for 
$K/E_1 = -3.9 \times 10^{-3}$ and shows that 
$|\QQ_{\rm A}|=1$ inside the $zxzy$ state and 
the transition to 
the NS state is continuous. 
This is a clear contrast to the cases for $K>0$, 
and the direction of $\QQ_{\rm A}$ gradually changes 
around the transitions. 
These transitions occur inside the regime 
where the A-sublattice state behaves 
as a ``half-integer spin''. 
We also point out that the variations of 
$\QQ _{\rm A-D}$ 
in the panel (c) is quite complicated. 

%
%
\vspace{10pt}
\section{Landau theory}
\label{sec:Landau}

In this section, we employ a phenomenological Landau analysis   
and interpret the determined mean-field phase diagrams (Fig.~\ref{fig9}). 
Our aim is to explain the stability of various ordered states 
and describe their transitions based on a phenomenological theory.  

The determined phase diagrams in Sec.~\ref{sec:phase} are quite complicated, and 
their complete analysis is beyond the scope of the present paper. 
Since the partially ordered $zyox$ and $zoxy$ states 
are exotic states characteristic to the present model, 
we set them as our main targets and investigate mainly 
their stability mechanism and instability to other states.

In our phenomenological analysis,
we will construct the Landau free energy 
in terms of $\QQ _{\kk _0}$ and $\QQ _{\kk _\ell}$'s 
and analyze various triple-$\qq$ orders in detail.  
An important point is that the ordering vectors $\qq_*$=$\kk _\ell$
allow cubic couplings of the three antiferro modes $\QQ _{\kk _\ell}$,  
which play a crucial role in stabilizing several ordered states 
in this system.  
This analysis succeeds in explaining 
most of the results of the microscopic mean field calculations 
in Sec.~\ref{sec:MFresults}.

\vspace{10pt}
\subsection{Landau-Ginzburg free energy expansion}

Let us first introduce the local free energy $f_{\rm loc} (\rr)$ 
with the quadrupole moment  
$\bphi (\rr ) \Equiv [(\phi_u,\phi_v) (\rr)]^{\rm T}$ 
as 
\begin{align}
f_{\rm loc} (\rr) 
=&
{ \FRAC{1}{2}} a_0 (T) |\bphi(\rr)|^2 
-b \bigl[ \phi_u^3(\rr)-3\phi_u(\rr)\phi_v^2(\rr) \bigr ] 
\nonumber
\\
&+ c |\bphi(\rr)|^4 + O (|\bphi (\rr)|^5 ).  
\label{eq:Floc}
\end{align}
The coefficients $a_0$, $b$, and $c$ are all positive 
constants depending on temperature,  
which may be derived from the local CEF model 
through a Legendre transformation \cite{Hattori2014}. 
It is customary to consider the temperature dependence 
of $a_0 (T)$ alone and neglect the change\Blue{s} in $b$ and $c$.  
One should note that $a_0 (T)$ is the inverse 
of local quadrupole susceptibility and decreases  
monotonically towards zero with decreasing temperature.  
An important characteristic of this system 
is the presence of the $b$-term.  
This third-order term exists 
only for order parameters  
with even parity under both time reversal 
and space inversion operations, 
and our quadrupole moments belong to this category.  
Its explicit form in terms of $\phi_u$ and $\phi_v$ 
is independent of the details of the system, 
and determined from the coefficient related to the reduction 
of the triple product of the $\Gamma_3$ representation 
 to the trivial one, 
$\Gamma_3\otimes\Gamma_3\otimes\Gamma_3 \To  \Gamma_1$.  

Following the conventional procedure, 
we add to Eq.~(\ref{eq:Floc}) the exchange interaction energy, 
which are given by 
the mean-field approximation. 
Thus the total free energy 
density $\bar{f} = F/N$
reads as 
\begin{subequations}
\begin{align}
& 
\bar{f}
\equiv \!\! 
\frac{1}{N}
\sum_{\rr } 
\Bigl[ 
f_{\rm loc} (\rr) 
+\sum_{j =1}^{6} \, 
  \bphi(\rr+\bdelta _j) \cdot 
  \mathsf{J}_{\bdelta _j} \, \bphi(\rr) 
\Bigr]
\mspace{-10mu}
\\
&=
 \frac{1}{N}
\sum_{\pp}
\bphi(-\pp) \cdot 
\bigl[ \FRAC{1}{2}a_0 +\mathsf{J} (\pp) \bigr]
\bphi(\pp)
\nonumber
\\[-4pt]
&
 -\frac{b}{N^{3/2}} 
 {\sum} ' 
 \phi_u (\pp_3) 
\bigl[ 
   \phi_u (\pp _1) \phi_u (\pp _2) 
 -3\phi_v (\pp _1) \phi_v (\pp _2) \bigr] 
\nonumber
\\
&+
\frac{c}{ N^2} 
{\sum} ' 
\bigl[ \bphi(\pp _1) \cdot \bphi(\pp _2) \bigr] 
\bigl[ \bphi (\pp _3) \cdot 
       \bphi (\pp _4 ) 
\bigr] , 
\mspace{-10mu}
\vspace{-6pt}
\label{eq:F}%
\end{align}
\end{subequations}
where $\sum '$ 
denotes the sum over $\pp _i$'s 
under the constraint $\sum_i \pp _i = $(some reciprocal 
lattice vector $\GG$).
The interaction matrix 
$\mathsf{J}_{\bdelta}$ 
is given by Eq.~(\ref{eq:def_J_delta}), 
while its Fourier transform 
$\mathsf{J} (\pp)$ is given by Eq.~(\ref{eq:def_J_p}).  
We have neglected a constant energy, 
and $N$ is the number of the sites. 
The terms for $\GG \ne \bzero$ are Umklapp processes, 
and they will turn out to be important later. 

Since our main concern is 
various antiferro orders of quadrupole, 
we split the total free energy as follows:
\begin{equation}
 \bar{f} = f_{\rm X} + f_{\Gamma} + f_{\Gamma {\rm X}} + f_{\mathrm{other}}, 
\end{equation}
where $f_{\rm X}$ is the antiferro part, 
i.e., contributions of $\bphi (\kk _{\ell})$'s alone. 
$f_{\Gamma}$ is the ferro part contributed by $\bphi (\kk_0)$, 
and $f_{\Gamma {\rm X}}$ is the coupling of $\bphi (\kk _{\ell})$'s 
with $\bphi (\kk_0)$. 
We will show that this coupling modifies 
and eventually destabilizes several antiferro orders.  
Lastly, $f_{\mathrm{other}}$ is the sum of 
all the remaining parts. 

%
%
\begin{figure}[tb]
\begin{center}
\includegraphics[width=0.5\textwidth]{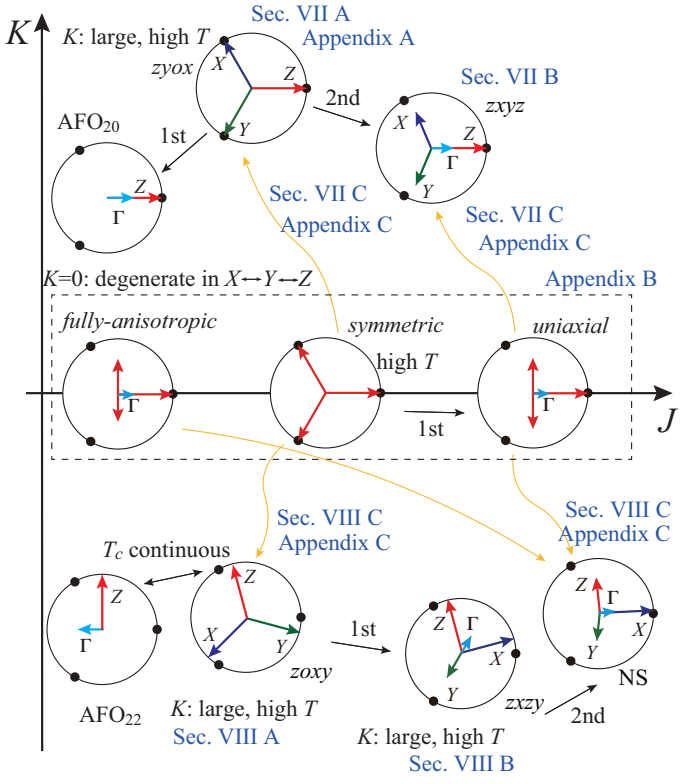}
\end{center}
\caption{
Summary of the order parameter configurations discussed 
on the basis of Landau theory in 
Secs.~\ref{sec:Landau}--\ref{sec:Knegative}. 
We list only the major states 
(NS$'$ not included) 
and also indicate in which section each state is discussed. 
The three configurations enclosed 
by a dashed rectangular are those for $K=0$. 
They are analyzed in Appendix~\ref{A1}. 
The label ``1st'' or ``2nd'' marked for an arrow 
indicates the order of the corresponding transition, but 
one should note that the possibility of a first order transition 
is not excluded for any of them.  
Yellow arrows indicate how each configuration at $K \Neq 0$ 
evolves from one of the degenerate configurations at $K \Eq 0$. 
}
\label{figSummary}
\end{figure}

Figure \ref{figSummary} summarizes
the relations among various states 
in the $J$--$K$ plane
along with the section numbers where they are discussed. 
The detailed discussions about several states are 
summarized in Appendices. 
In particular, the configurations at $K=0$ are 
discussed in detail in Appendix~\ref{A1}. 
The yellow arrows illustrate schematically 
the relation\Blue{s} between the states at $K\ne 0$ 
and those at $K=0$ through small $K$ perturbations.

\vspace{10pt}
\subsection{Free energy of the antiferro modes}

In this section, we will rewrite the antiferro part $f_{\rm X}$ 
into a convenient form for later analyses.  
When $|K| \PLUS 2J \GT  0$, 
the maximally negative eigenvalue of the matrix $\mathsf{J} (\pp)$
is $\Lambda_{\rm min} \Eq -2J-4|K|$ and it is realized 
at $\pp \Eq \kk _{\ell}$'s. 
Thus it is natural to expect a single-$\qq$ order 
with the ordering vector located at one of $\kk_\ell$'s.  
The transition temperature of this order $T \Eq  T_c^0$ 
is determined by the equation 
$\FRAC{1}{2}a_0 (T_c^0) \PLUS  
 \Lambda_{\rm min} \Eq  0$.  
However, the situation is not so simple in this system, 
since the free energy contains the third-order terms 
including $\bphi (\kk _{\ell})$'s.  
The three wavevectors at the X points satisfy the relation 
$\kk_1 \PLUS    
 \kk_2 \PLUS   
 \kk_3 \Eq  (2\pi,2\pi,2\pi) \Eq  \GG$, and thus 
the corresponding third-order coupling is nonvanishing.  
This may lead to a first-order transition to 
a triple-$\qq$ state with a transition temperature 
$T_c \GT  T_c^0$. 

To analyze such triple-$\qq$ states, 
we restrict the degrees of freedom 
to those $\{ \bphi (\kk _{\ell}) \}_{\ell=1,2,3}$ 
and ignore other modes.  
Notice that $\bphi (\kk _{\ell})$'s are real, 
since each $\kk _\ell$ is equivalent to $-\kk _\ell$, 
 and proportional to $\sqrt{N}$ in the antiferro ordered states.  
We thus rescale them as 
$\bpsi ^{({\ell})} \Equiv  \bphi(\kk_{\ell})/\sqrt{N}$.  
The related part of the free energy density 
reads as 
\begin{align}
&f_{\rm X}  = 
\sum_{\ell} 
\bpsi^{({\ell})} \cdot 
 \bigl[ \FRAC{1}{2}a_0 (T) + 
 \mathsf{J}(\kk_{\ell}) \bigr] 
\bpsi ^{({\ell})}
\nonumber\\[-4pt]
&-6b 
\Bigl[ 
\psi_u ^{(1)} \psi_u ^{(2)} \psi_u^{(3)} - 
\sum_{\ell}
\psi_u^{({\ell})} 
\psi_v^{({\ell}+1)} 
\psi_v^{({\ell}+2)} 
\Bigr]
+c  
\sum_{\ell} \, 
\bigl| \bpsi^{({\ell})} \bigr|^4
\nonumber\\[-4pt]
&
+2c' 
\sum_{{\ell}<{\ell}'}  
\bigl| \bpsi^{({\ell})} \bigr|^2 \, 
\bigl| \bpsi^{({\ell}')} \bigr|^2 
+4c\dprime  
\sum_{{\ell}<{\ell}' } 
\bigl[ 
  \bpsi^{({\ell})} \mspace{-4mu} \cdot \bpsi^{({\ell}')} 
\bigr]^2, 
\label{eq:F2}
\end{align}
where one should understand that 
$\bpsi^{({\ell}+3)} \Equiv  \bpsi^{({\ell})}$.  
Here, although the direct calculations provide
$c' \Eq  c\dprime  \Eq  c$, 
we regard these three as independent parameters, 
since each term is separately invariant. 
We rewrite Eq.~\eqref{eq:F2} in terms of 
the polar coordinates defined as 
$\bpsi^{(\ell)} = (\psi^{(\ell)}_u,\psi^{(\ell)}_v)^{\rm T} 
\equiv X_{\ell} 
\hee (\theta_{\ell})$ as in Eqs.~(\ref{eq:PolarQQ}) and (\ref{eq:PolarQQ2}).
We also define 
$\XX \Equiv (X_{1},X_{2},X_{3})$ and 
$\btheta \Equiv (\theta_1,\theta_2,\theta_3)$ 
for later use.  
Substituting these into Eq.~(\ref{eq:F2}), we obtain
\begin{align}
f_{\rm X} 
=& 
\bigl[ \FRAC{1}{2} a_{0}(T)-2J \bigr] R^2 
+cR^4
-4K 
\sum_{\ell}
  X_{\ell}^2 \cos \left( 2\theta_{\ell} \msm2 + \msm2 \ell \omega \right)
\nonumber
\\
&
- 6bX_1X_2X_3 \cos\bar{\theta}
+ 2(c' \msm2 - \msm2 c) 
\sum_{\ell < \ell '} X_{\ell}^2 X_{\ell '}^2 
\nonumber
\\
&
+4c\dprime  
\sum_{\ell < \ell '} 
  X_{\ell}^2 X_{\ell '}^2 \, 
\cos^2 (\theta_{\ell} \msm2 - \msm2 \theta_{\ell'}) ,
\label{eq:F3}
\end{align}
where 
\begin{equation}
R^2 \equiv 
X_1^2+X_2^2+X_3^2,
\quad 
\bar{\theta} \equiv 
\theta_1+\theta_2+\theta_3.
\label{eq:Polar2}
\end{equation}
Further analysis depends on the sign of 
the anisotropic interaction $K$.  
In the following sections, we will discuss the two cases separately.

\vspace{10pt}
\section{Analysis of the $K>0$ part}
\label{sec:LandauC}

In this section, 
we perform phenomenological analyses on the triple-$\qq$ states 
for the $K>0$ part of the mean-field phase diagrams.
We will mainly examine two limiting cases, 
large and small $K$ limits, 
in the following sections.  
We first attempt to find solutions 
that minimize $f_{\rm X}$ in Eq.~(\ref{eq:F3}) 
for the case of $K \GT  0$. 
As discussed in Sec.~\ref{sec:exJ}, 
the eigenvector of the maximally negative eigenvalue of $\mathsf{J}(\kk_{\ell})$: 
$\Lambda_{\mathrm{min}} \Eq \Lambda_{-} (\kk_\ell)$ 
corresponds to 
$\btheta \Eq  (\omega,-\omega,0)$. 
We will fix these angles and find an approximate solution 
in Sec.~\ref{sec:zyox}.  
Then, we will introduce their couplings to the uniform moment 
in Sec.~\ref{sec:inducedQ}.   
This mode coupling induces a finite uniform moment 
and also deforms the angles $\btheta$ from $(\omega,-\omega,0)$.
In Sec.~\ref{sec:smallKpositive}, we will discuss possible triple-$\qq$
 states in the small $K$ limit, starting from the results for $K=0$. 

\vspace{10pt}
\subsection{Large $K$ case: the $zyox$ state}
\label{sec:zyox}

In the limit of $K \To \infty$, 
the free energy $f_{\rm X}$ has a symmetry inherited from the $K$-term.  
The replacement 
$\theta_{\ell} \To  \theta_{\ell} \PLUS \pi$ for any $\ell$ 
does not change the $K$-term, 
and this results in an eightfold degeneracy of $f_{\rm X}$ minimum: 
$ \btheta \Eq  
(\omega,-\omega,0) \PLUS 
(n_1,n_2,n_3) \msp2 \pi$ where $n_j \Eq  0, 1$. 
The third-order term favors four 
out of these eight possibilities. 
Since $b \GT  0$, the favored ones are 
those with $\sum_j n_j \Eq (\mbox{even})$.

We continue the minimization procedure for $f_{\rm X}$ 
now with respect to 
$X_{\ell}$'s.
This is a cumbersome but straightforward calculation, 
and we show only its results. 
There exist two types of solutions, and both are 
controlled by the renormalized second-order 
coupling 
\begin{equation}
a_{\rm X}(T) \Equiv \FRAC12 a_0 (T) \MINUS 2J \MINUS 4K.   \label{eq:def_aXT}
\end{equation}
One type is the solution that only one of 
$X_{\ell}$'s 
is nonvanishing, 
while 
$X_1 \Eq X_2 \Eq X_3$ 
in the other type.  
The former one is the single-$\qq$ order, and the latter one 
is the $zyox$ order. 

A solution for the single-$\qq$ order 
($\msp2 X_{\ell} \Eq \delta_{\ell 3} R_{1\qq}^* \msp2$)
exists when $a_{\rm X} (T) \LT 0$,  
and it is represented as follows
\begin{equation}
R_{1\qq}^* \equiv \left[ \frac{-a_{\rm X} (T)}{2c} \right]^{1/2} , \quad
f_{\rm X}^{1\qq} =  - \frac{a_{\rm X} (T)^2}{4c} .  
\end{equation}
See Eq.~(\ref{eq:A1f_single}) in Appendix~\ref{sec:AppAA1}. 

A $zyox$-type solution 
($\msp2 X_{1} \Eq X_{2} \Eq X_{3} \Eq : R_{zyox}^*/\sqrt3 \msp2 $)
appears as a local minimum, 
and we denote its transition temperature by $T^\star$. 
See Eq.~(\ref{eq:A1fzyox}) for the 
effective free energy as a function of $R_{zyox}^*$.  
We define $R^\star$ as the $R_{zyox}^*$ value 
just below $T^\star$.  
These values are determined as 
\begin{equation}
 R^\star \equiv 
\frac{\sqrt{3}b}{4(c \PLUS c_1/3)} , \quad
a_{\rm X} (T^\star ) =  \frac{\sqrt{3}}{2} b R^\star \equiv   a^\star > 0 , \label{eq:astar}
\end{equation}
where 
\begin{equation}
c_1 \equiv  2(c' \MINUS c) \PLUS  c\dprime . 
\label{eq:c1}
\end{equation}
Then, the temperature dependence of important quantities 
is written as 
\begin{align}
R_{zyox}^* &= 
R^\star 
\bigl[ 1 + \sqrt{1 - a_{\rm X}(T)/a^\star} \, \bigr], 
\\
f_{\rm X}^{zyox} 
&= 
(4\sqrt3)^{-1} b {R^\star}^3  \, 
\mathscr{F}_2 \bigl( 1 - a_{\rm X}(T)/a^\star \msp2 \bigr), 
\end{align}
where $\mathscr{F}_2 (x) = 4 - 3 (1+x)^2 - 8 x^{3/2}$. 

Let us now compare the stability of these two ordered states 
by calculating their free energies.  
One should recall that $a_{\rm X}(T)$ decreases monotonically 
with lowering temperature.  
Four temperatures characterize possible phase transitions: 
\begin{equation}
T^{\times}  < 
T_{1 \qq}^c <  
T_{zyox}^c  < 
T^\star. \label{eq:Tc_relation}
\end{equation}
As defined before, $T^\star$ is the temperature where 
a solution of the $zyox$ order appears. 
However, since its free energy is still higher than 
that for the disordered phase, 
this order is not realized yet.  
The $zyox$ order is stable below $T_{zyox}^c$ 
and this temperature is determined by the condition 
$a_{\rm X} (T_{zyox}^c ) \Eq \FRAC89 a^\star$.  
With further lowering temperature, 
the effective second-order coupling 
$a_{\rm X} (T)$
vanishes eventually and this defines the ``transition temperature'' 
of the single-$\qq$ order $T_{1 \qq}^c$.  
Below this temperature, the single-$\qq$ ordered state is 
more stable than the disordered state. 
However, its free energy 
$f_{\rm X}^{1\qq}$
remains 
higher than $f_{\rm X}^{zyox}$ for a while, 
and thus the single-$\qq$ order does not appear yet.  
With further decreasing temperature, 
$f_{\rm X}^{zyox} (T)$ may cross with $f_{\rm X}^{1\qq} (T)$. 
This determines the final characteristic temperature $T^{\times}$, 
 below which
the single-$\qq$ order is finally realized.  
Equating the two free energies, 
$T^{\times}$ is obtained 
as a solution of the following equation
\begin{align}
\frac{c_1}{8c}
= 
x^{-2} 
\bigl[ 
1 - \FRAC32 x +(1-x)^{3/2} \bigr]
\Big|_{x = a (T^{\times}) /a^\star}. 
\end{align}
One should note that the transitions 
at $T_{zyox}^c$ and $T^{\times}$ 
are both first order.  

We have not examined a possibility of double-$\qq$ orders, 
and there is a reason for that.  
In the free energy density $f_{\rm X}$ in Eq.~\eqref{eq:F3}, 
the third-order term generally has a nonzero coupling $b \ne 0$.  
Thus, when two order parameters are nonvanishing, say 
$X_1, \, X_2 > 0$, their product acts as a field linearly 
coupled to $X_3$, and this induces a nonvanishing amplitude of $X_3$. 
Therefore, any double-$\qq$ order is inevitably converted 
to a triple-$\qq$ order, and 
genuine double-$\qq$ orders do not exist as a stable phase.  

Unfortunately, the above free energy analysis 
does not fully explain the actual phase diagram 
in the $K>0$ part.  
For example, the $zyzx$ state is hardly realized.  
This is because we have not taken account of  
coupling to the uniform component 
$\bpsi_{0} \Equiv   \bphi (\kk _0 )/\sqrt{N}$, 
or its effects on tilting $\btheta$ from the assumed values.  
In addition, the single-$\qq$ state also couples 
with $\bpsi_{0}$ and this lowers the free energy.  
Thus, the above analysis is satisfactory 
only for the $zyox$ state, 
and we need to include those corrections for 
the $zyzx$ and single-$\qq$ states.  
Nonetheless, it remains true that 
the $zyox$ state appears at a temperature higher 
than the transition temperature $T_{1\qq}^c$ 
of the single-$\qq$ state.  
This is one of the main results in this paper. 
 
Finally, let us demonstrate the partially ordered configuration of 
quadrupoles in real space for the $zyox$ state.  
Since 
$\bpsi_0 \Eq \bzero$, 
the direct substitution of 
$X_{\ell} \Eq R_{zyox}^* /\sqrt{3} \Equiv  q^*/2$ and 
$\theta_{\ell} \Eq  \ell \omega$ 
into Eq.~(\ref{eq:Qloc}) leads to 
\begin{equation} 
\QQ _{\rm A}={\bf 0}, 
\ 
\QQ _{\rm B}=q^* \hee _{\omega}, 
\ 
\QQ _{\rm C}=q^* \hee _{0}, 
\ 
\QQ _{\rm D}=q^* \hee _{2\omega}, 
\label{eq:localQ_zyox}
\end{equation}
where 
$\hee _{\theta}$ is defined in Eq.~\eqref{eq:evec}.  
This is precisely what we have obtained 
in the microscopic calculations in Sec.~\ref{sec:phase}. 

\vspace{10pt}
\subsection{Instability to the $zyzx$ state}
\label{sec:inducedQ}

So far, we have only examined the antiferro orders 
of the modes 
of $\bpsi^{({\ell})}\ ({\ell}=1,2,3)$. 
As shown in Sec.~\ref{sec:MFresults}, 
they couple to the uniform moment $\bpsi^{(0)}$ in the 
single-$\qq$ and $zyzx$ ($zyzx'$) states. 
In the latter state, 
the order parameters consequently 
tilt from the directions $\theta_{\ell} \Eq \ell\omega$. 
We will study this tilting in this section.  
To avoid complication in a full analysis, 
we employ an alternative approach based 
on a perturbation analysis of the mode coupling.  
We focus on the part of $K \GT  -2J$.  
Thus, we consider the case where 
the order parameters are modified only slightly 
from the previous solutions in Sec.~\ref{sec:zyox}.
We restrict ourselves to the second-order stability analysis 
and show possible types of instability 
in each of the two states.
We parametrize its small deformations as follows: 
\begin{alignat}{2}
&\XX 
- R \msp2 \bar{\ee}_0 &
&\sim 
{\FRAC{1}{\sqrt3}}  
R 
\left(
- 
\msp2 d_2 \msp2 \bar{\ee}_1
+ d_1 \msp2 \bar{\ee}_2 
\right), 
\label{eq:ParamXYZ}
\\
&\btheta
- \omega (1, 2 , 0) &
&\sim 
  \eta_0 \msp2 \bar{\ee}_0 
+ \eta_1 \msp2 \bar{\ee}_1
+ \eta_2 \msp2 \bar{\ee}_2 . 
\end{alignat}
Here, the deformations 
$|d_j|$'s and $|\eta_j|$'s are all assumed to be small, 
and 
$\bar{\ee}_0 \Eq \FRAC{1}{\sqrt3} (1,1,1)$, 
$\bar{\ee}_1 \Eq \FRAC{1}{\sqrt2} (1,-1,0)$, and 
$\bar{\ee}_2 \Eq \FRAC{1}{\sqrt6} (1,1,-2)$.  

The above small deviations from the $zyox$ state couple with 
 the uniform moment $\bpsi_0 \equiv (\psi_{0u},\psi_{0v})^{\rm T}\equiv Q\hee _{\theta_\Gamma}$, 
and we also assume $Q \ll 1$. 
See Eqs.~(\ref{eq:evec}) and (\ref{eq:PolarQQ}). 
The $\bpsi_0$ contributes 
to the free energy starting from the second-order term  
as $a_{\Gamma} (T) \bpsi_0^2 \Eq a_{\Gamma} (T) Q^2$ 
with 
\begin{equation}
a_\Gamma (T) \equiv 
\FRAC{1}{2} a_0 (T) + 
J \gamma_0 (\bzero  )
= 
\FRAC{1}{2} a_0 (T) +  6J 
> a_{\rm X}(T) .  
\end{equation}
The couplings between $\bpsi_0$ and the deformation 
of the $zyox$ order parameters $|d_j|$'s and $|\eta_j|$'s arise from 
the following mode coupling terms $f_{\Gamma {\rm X}} \Eq  
 f_{\Gamma {\rm X}}^{(3)} \PLUS  
 f_{\Gamma {\rm X}}^{(4)}$, where 
\begin{align}	
&f_{\Gamma {\rm X}}^{(3)}
=
-\lambda \sum_{\ell=1}^3
\bigl[ 
\psi_{0u} 
\bigl( \psi_u^{(\ell)} {}^2 - \psi_v^{(\ell)} {}^2 \bigr)
-2 
\psi_{0v} 
\psi_u^{(\ell)} \psi_v^{(\ell)}  
\bigr]
\nonumber
\\
&\hs{5mm}
=
-\lambda Q 
\sum_{\ell=1}^3
  X_{\ell}^2 \cos (2\theta_{\ell}+\theta_\Gamma) , 
\label{eq:F0X_revised}\\
&f_{\Gamma {\rm X}}^{(4)}
=
\lambda' \sum_{\ell < \ell '} \, 
[\bpsi ^{(\ell)} \cdot \bpsi ^{(\ell ')} ] 
[\bpsi ^{(6-\ell - \ell')} \cdot 
\bpsi_{0}
 ] 
\nonumber\\
&= 
\lambda' X_1 X_2 X_3 Q 
\sum_{\ell < \ell'}
\cos(\theta_{\ell} \MINUS \theta_{\ell'}) 
\cos(\bar{\theta} \MINUS \theta_{\ell} \MINUS \theta_{\ell'} 
      \MINUS \theta_\Gamma) 
\nonumber\\[-2pt]
&=
\lambda' X_1 X_2 X_3 Q 
\sum_{\ell=1}^3
\cos( 2\theta_{\ell} \PLUS \theta_\Gamma \MINUS \bar{\theta} ),
\label{eq:F0X4_revised} 
\end{align}
where 
$\lambda \Equiv  3b$ and 
$\lambda' \Equiv  8c$. 
Collecting the terms within the second order in the 
deviations $d_{1,2}$, $\eta_{0,1,2}$, and $\bpsi_0$ in 
the free energy Eqs.~(\ref{eq:F3}), 
(\ref{eq:F0X_revised}), and (\ref{eq:F0X4_revised}), 
one obtains the change in the free energy $\delta f$ as 
\begin{subequations} 
\begin{align}
\delta f\simeq& \  
\FRAC12 \alpha_0 \msp2  \eta_0^2
+\FRAC12 \alpha_1 \msp2  |\bm{\psi}_0|^2 
+\FRAC12 \alpha_2 \msp2 |\bm{d}|^2
+\FRAC12 \alpha_3 \msp2 |\bm{\eta}_\perp|^2
\nonumber\\
&- g_1 \msp2 \bm{d} \cdot \bm{\eta}_{\perp} 
 - g_2 \msp2 {\bm{\psi}_0 } \cdot \bm{\eta}_{\perp}
 + g_3 \msp2 {\bm{\psi}_0 } \cdot \bm{d}.  \hs{0.5cm}
\label{eq:deltaf}
\end{align}
Here, $\bm{d}\equiv (d_1,d_2)^{\rm T}$ and $\bm{\eta}_\perp\equiv (\eta_1,\eta_2)^{\rm T}$. 
The coefficients are given as 
\vspace{-5pt}
\begin{align}
\alpha_0
&=
  \FRAC{16}{3} \msp2 KR^2 
+ 2\sqrt{3} \msp2 bR^3, 
\ \ \ \ 
\alpha_1
=
2a_\Gamma (T) , 
\label{eq:a0a1}
\\
\alpha_2
&=
\FRAC{2}{3} a_{\rm X} (T) R^2+
  \FRAC{2}{\sqrt{3}} \msp2 bR^3 
+ \FRAC43 
\msp2 c  R^4,  
\label{eq:a2}
\\
\alpha_3
&=
  \FRAC{16}{3} \msp2 K R^2 
+ \FRAC43 \msp2 c\dprime  R^4 ,  
\ \ \ \ 
g_1
= 
\FRAC43 \msp2 c\dprime  R^4 , 
\label{eq:a3g1}
\\
g_2
&=
\FRAC{\sqrt{2}}{3} \msp2 R^2
(\sqrt{3} \lambda - \lambda' R ), 
\ \ \ \ 
g_3
=
\sqrt{\FRAC{2}{3}} \msp2 \lambda R^2, 
\label{eq:g2g3}
\end{align}
\label{eqs:deltaf}
\end{subequations}

We first note that the part of $\eta_0$ is decoupled 
in Eq.~\eqref{eq:deltaf} from the others, 
and the minimization leads to $\eta_0 \Eq 0$  
since $\alpha_0 \GT 0$. 
For discussing other diagonal coefficients $\alpha$'s, let us assume
$c \Eq c' \Eq c''>0$ as in the original form (\ref{eq:F}). 
One notices 
that $\alpha_2$ may change its sign with 
lowering temperature.
The remaining $\alpha_1$ and $\alpha_3$ are 
positive, since we consider the situation 
of $a_{\Gamma}(T) \GT 0$ and $K \GT 0$. 
We denote by $T_{\alpha_2}$ the
temperature at which $\alpha_2 \Eq 0$. 
Then it is obtained by evaluating 
$a_{\rm X}(T_{\alpha_2}) \Eq 
-R^*_{zyox} (\sqrt{3} b \PLUS 2 c R^*_{zyox})
< 0 \Eq a_{\rm X}(T_{1\qq}^c)$, 
which shows $T_{\alpha_2} \LT T_{1\qq}^c$. 
This value is $a_{\rm X}(T_{\alpha_2})=-63b^2/c$.
Thus, ignoring the couplings $g$'s, one sees 
that the $zyox$ state is a locally stable solution for $T>T_{\alpha_2}$.  
We also note that $g_2$ [Eq.~\eqref{eq:g2g3}] 
may change its sign to negative as $R$ increases.

The pure $zyox$ state becomes unstable, once 
$\delta \msm2 f$ in Eq.~\eqref{eq:deltaf} can take a negative value. 
A new stable configuration then acquires 
a nonvanishing value of one or some of 
$\bm{\psi}_0$, 
$\bm{d}$, and $\bm{\eta}_{\perp}$, 
which deform the original $zyox$ order.  
This instability is signaled by the appearance of a negative 
eigenvalue in the coefficient matrix.  
That is, the instability takes place 
at the position where the determinant $D_3$ of 
the block-diagonal $3\times 3$ matrix changes 
from positive to negative.  This determinant is given by 
\vspace{-7pt}
\begin{equation}
D_3 = 
\alpha_1 \alpha_2 \alpha_3 + 2 g_1 g_2 g_3 - 
\sum_{i=1}^3 \alpha_i g_i^2  .  \label{eq:D3}
\end{equation}
Since Eq.~(\ref{eq:D3}) includes the effects of  
the coupling between 
$\bpsi^{(\ell)}$ and $\bpsi^{(0)}$, 
the condition of $D_3 \Eq 0$ determines 
a transition temperature which is 
not necessarily the same as $T_{\alpha_2}$.

Let us discuss this instability of the $zyox$ state 
but do it only qualitatively 
in order to simplify discussions. 
For large $K(>0)$, it is 
natural to set $\bm{\eta}_{\perp} \Eq \bm{0}$ 
in the zeroth-order approximation, 
since its coefficient $\alpha_3$ is a large positive value.  
Analyzing the variations with 
$\bm{\psi}_0$ 
and $\bm{d}$, 
we find that the $zyox$ state is unstable 
when 
$g_3^2 \GT  \alpha_1 \alpha_2$. 
Then, a nonvanishing deformation is spontaneously induced 
and it has the amplitude 
$\bpsi _0 \Eq -C \dd$ and 
$\bm{\eta}_{\perp} \Eq C' \dd$ 
with 
$C \Sim g_3 /\alpha_1 \GT 0$ and 
$C' \Sim (g_1 \MINUS C g_2 )/\alpha_3$.  

Suppose the induced ferro component 
$\bpsi _0$ points to the direction $\theta_{\Gamma} \Eq 0$. 
This implies that 
the other induced deformations have the form of 
$d_1 \LT 0$ and $d_2 \Eq \eta_2 \Eq 0$.  
The sign of $\eta_1$ is also negative if $C g_2 \LT g_1$ 
or positive otherwise.  
In our microscopic calculations, 
we have observed a wide region of the $zyzx$ state, 
which corresponds to the solution 
with deformation $\eta_1 \LT 0$.  
This is a consequence of the decrease of $g_2$ 
due to increasing $R$ associated with lowering temperature. 
Although the transition between $zyox$ and $zyzx$ states 
in Fig.~\ref{fig9}(b) in the microscopic mean-field results 
is first order, the deformation is consistent 
with this analysis.

If $\bpsi_0$ is induced in the opposite direction 
$\theta_{\Gamma} \Eq \pi$,
the other induced deformation also changes their signs. 
$d_1 \GT 0$ 
with $\eta_1>0$ for $Cg_2 \LT g_1$. 
This is the $zyzx'$ state as shown in Fig.~\ref{table1}. 
See the second-order behavior in Fig.~\ref{fig10}(c).
Since the free energy (\ref{eq:deltaf}) fixes 
only the relative directions among 
$\bpsi_0$, $\bm{\eta}_{\perp}$, and $\dd$, 
we cannot discuss the \NSpm state. 
The stability of the \NSpm state will be discussed 
in Sec.~\ref{sec:smallKpositive}.

\vspace{10mm}

\subsection{Small $K$ limit}
\label{sec:smallKpositive}
So far, we have analyzed mostly the large-$K$ region of 
the $K > 0$ part. 
The order parameters $\bpsi ^{(\ell)}$ 
then point to the directions 
close to $\ell \omega$.  
We now analyze the opposite limit to see which types of 
orders are stabilized.

When $K$ is small, the quadratic part of the free energy is 
nearly isotropic, and thus the quadrupole moments tend to 
rotate freely. 
This indicates that the three $\theta_{\ell}$'s can point 
to arbitrary directions
under the constraint of $\cos \bar{\theta}\Eq  1$ 
in the zeroth-order approximation.  
The details of this straightforward 
but lengthy analysis are explained 
for $K \Eq  0$ in Appendix~\ref{A1}.  
As shown in Fig.~\ref{figSummary}, three triple-$\qq$ states 
exist when $K=0$.
A set of three arrows schematically represent the antiferro moments 
$\{ \bpsi ^{(\ell)} \}_{\ell=1,2,3}$ in each state. 
In addition to the single-$\qq$ order, 
the phase diagram at $K=0$ (see Fig.~\ref{fig:PD-g34}) has 
three regions of the triple-$\qq$ states: 
(i) symmetric triple-$\qq$ with 
$X_1 \Eq X_2  \Eq X_3$, 
(ii) uniaxial triple-$\qq$ 
with $X_1 \Eq X_2 \LT X_3$, 
and (iii) fully-anisotropic triple-$\qq$ with 
$X_{\ell}$'s all different.  
In each triple-$\qq$ state, 
only the relative directions of the order parameters 
$\theta_{\ell} \MINUS \theta_{\ell'}$ are fixed. 
This is because the eigenvalue of the 
exchange interaction for $K \Eq 0$ 
in Eq.~(\ref{eq:EigenValueLambda}) is degenerate,  
$\Lambda_{+}(\pp)=\Lambda_{-}(\pp)$ in Eq.~(\ref{eq:EigenValueLambda}),
which means that the direction of the quadrupole moment $\bpsi^{(\ell)}$ 
can be arbitrary concerning 
the quadratic terms in the free energy. 
For example, a representative state for the symmetric 
triple-$\qq$ state (i) has 
$\btheta \Eq (\omega,2\omega,0)$. 
By modifying this with all the permutations of $\theta_{\ell}$'s, 
one obtains other five states, which have the same 
energy as the original's. 
For the above (i), (ii), and (iii), 
the third and the fourth order terms determine 
the most stable one of the three.  
In the following, we will discuss 
how switching on $K$ lifts this degeneracy.

First, let us examine the symmetric triple-$\qq$ 
states (i) for $K=0$, 
which have the highest transition temperature 
and 24 domains. 
See Figs.~\ref{fig:PD-g34} and \ref{fig:AppK0TJ}. 
One of the domains corresponds to 
the $zyox$ state for $K \GT  0$ with 
$\btheta \Eq  (\omega,2\omega,0)$, 
and this was discussed in Sec.~\ref{sec:zyox}. 
Note that this solution has four-fold degeneracy 
corresponding to different domains 
or equivalently disordered sublattice. 
For finite $K$, it is natural that the free energy of 
these four states becomes lower than those of the other 
states in the symmetric triple-$\qq$ states.

As for the $zyzx$ state, 
its solution at $K \Eq  0$ has 
a uniaxial anisotropy $X_1 \Eq X_2 \ne X_3$. 
As demonstrated in Appendix \ref{A1}, the symmetric triple-$\qq$ state  
becomes unstable and is replaced by the uniaxial state 
at smaller values of $a$, which correspond to lower 
temperatures (Fig.~\ref{fig:AppK0TJ}).  
This has a configuration of the type 
$\btheta \Eq  (\FRAC12 \pi,- \FRAC12 \pi ,0)$, 
and we concentrate on this particular domain.  
For $K>0$, the energy of the $K$-term is 
lowered for the configurations with 
$\theta_\ell \sim  \ell \omega$. 
Because of $\theta_3=0$ in the above domain, 
one expects that 
the direction of $\bpsi^{(3)}$ does not tilt 
even for $K>0$. In contrast, $\bpsi^{(1,2)}$ 
with $\theta_{1,2}=\pm \pi/2$ can tilt to lower 
the energy of the $K$-term for $K>0$ 
with keeping 
$\bar{\theta}=0$,  
$\theta_1 \Eq -\theta_2 \Sim 
 \FRAC12 \pi \PLUS \delta \Sim 
 \FRAC23 \pi$ 
with $\delta \GT 0$.  
See a detailed analysis in Appendix~\ref{sec:AppC}.  

For the \NSpm state, one can perform a similar analysis 
starting from the fully-anisotropic state.  
However, since this reveals no new aspect, 
and so we omit its analysis here.  
We will return to these discussions about the anisotropic $zyzx$ 
and \NSpm states, 
when we analyze the NS state for $K \LT 0$ in Sec.~\ref{sec:C3breaking}.

\vspace{20pt}

\section{Analysis of the $K<0$ part}
\label{sec:Knegative}
In the $K<0$ part,
the situation is more complicated, 
and we now perform its phenomenological analysis. 
This complication is due to the competition of 
the second- and third-order terms in the free energy (\ref{eq:F3}).  
The minimum eigenvalue
$\Lambda_{\rm min}$ in Eq.~(\ref{eq:Lambda_min}) corresponds to 
the eigenvector, for example, $\vvv ^+ (\kk_\ell)$ and 
$\btheta \Eq \pi ( \FRAC76 , \FRAC{11}{6},\FRAC12)$ 
as listed in Eq.~(\ref{eq:eigen_theta}).
For the single-$\qq$ orders, 
the transition temperature is determined by $\tilde{a}_{\rm X}(T)=0$, 
where 
\begin{equation}
\tilde{a}_{\rm X}(T) \equiv 
\FRAC12 a_0(T)-2J+4K. 
\label{eq:a_tilde_pm}
\end{equation}
See Eq.~(\ref{eq:F3}).
This satisfies $\cos\bar{\theta}=0$ 
in Eq.~\eqref{eq:F3}, and this means no energy gain 
 in the $b$-term.  
Therefore, one needs a full minimization to 
determine the order parameters to identify a stable state, 
and this requires solving coupled nonlinear equations 
with six variables.
Instead of this elaborate work, we make a simple analysis
in this section 
to explore an essential mechanism 
stabilizing the triple-$\qq$ orders for $K \LT 0$.

Recall that the $b$-term stabilizes 
the triple-$\qq$ states when $K \GT 0$.  
Thus we first examine whether the same mechanism works for $K \LT 0$.  
Assume that 
$\bpsi _{\ell}$'s tilt slightly 
from the directions 
determined by minimizing the second-order term: 
\begin{equation}
\btheta = 
\btheta _0 
+ ( \delta_1, \delta_2, \delta_3 ) , 
\ \ \ 
\btheta _0 \equiv 
\pi \bigl( \FRAC76 , \FRAC{11}{6}, \FRAC12 \bigr) . 
\label{eq:thetaXYZ}
\end{equation}
Nonvanishing values of 
$\delta_{\ell}$'s 
do not minimize the second-order terms in the free energy, 
but some distorted triple-$\qq$ order 
may have a chance to lower the 
total free energy than that 
for the single-$\qq$.  
This type of order with the three equivalent $X_{\ell}$'s 
will be discussed in the section A below. 
After that, in the following section B, 
we will also examine a configuration 
which can gain both of the $b$- and $K$-terms 
by introducing amplitude modulations in $X_{\ell}$'s.

\vspace{10pt}
\subsection{Case of $X_1 = X_2 = X_3$: the $zoxy$ state}
\label{sec:symtilting}
It is natural to consider that a solution 
with the highest symmetry among the 
triple-$\qq$ states for $K<0$ has a common amplitude of 
$\bpsi^{(\ell)}$'s  ($X_1=X_2=X_3$) 
similarly to the $zyox$ state for $K>0$.  
This also leads to the constraint that 
$\delta_{\ell}$'s should be 
all the same in Eq.~\eqref{eq:thetaXYZ}. 
Let us first examine this simplest case:  
$\delta_{\ell} \Eq \alpha$ and $X_{\ell} \Eq R/\sqrt{3}$ 
for all $\ell$'s.  
This is the $zoxy$ state, and one of the four sublattices is disordered 
as in the $zyox$ state. See also Eq.~(\ref{eq:localQ_zoxy}).  
Its free energy is given as 
\begin{equation}
f_{\rm X}
=
\bigl[
\tilde{a}_{\rm X} (T) \MINUS 8K \sin^2 \alpha
\bigr]
R^2 
\MINUS \FRAC{2}{\sqrt{3}} b R^3  \sin 3 \alpha 
\PLUS \FRAC{1}{4} c_2 R^4, 
\label{eq:fXKnegative} 	
\end{equation}
with $c_2 \Equiv 4(c\dprime \PLUS 2c' \PLUS  c)/3$. 
This has the symmetry $f_{\rm X} (\pi -\alpha) \Eq f_{\rm X} (\alpha )$.  
Minimizing this with respect to $\alpha$ 
leads to the following two types of solutions: 
\begin{subequations} 
\begin{align}
\alpha_{\pm}^{(\mathrm{i})}
&=
\FRAC{1}{2} \pi \pm 
\cos^{-1} 
\left[ 
\sqrt{ \FRAC14 \PLUS \bar{K} ^2 \msp2} \PLUS \bar{K} \right],  \ \ 
\mbox{ for } K \LE 0 , 
\vspace{-10pt}
\label{eq:conditionzoxy}
\\[-6pt] 
\alpha^{(\mathrm{ii})} 
&=
- \FRAC{1}{2} \pi , \ \ \ 
\mbox{ for }  0 \LE K, 
\end{align}
\end{subequations}
\color{black}
where 
$\bar{K} \Equiv K/(\sqrt{3} \msp2 bR)$ 
for $K < 0$ and this is a dimensionless 
parameter which controls the $\alpha$-dependence, 
and $\bar{K} < 0$ for $K < 0$.  
The solution 
$\alpha^{(\mathrm{ii})}$ corresponds to the $zyox$ state with 
$\btheta \Eq (\omega , 2 \omega ,0)$ 
 as discussed in Sec.~\ref{sec:zyox}.
However, since our concern is the region of $K \LT 0$ 
in this section, 
we do not consider that solution further.  

The type (i) solutions exist for any $\bar{K} \LE 0$. 
Their range is $0 \LE \alpha_{-}^{(\mathrm{i})} \LE \FRAC16 \pi$, 
and  
$\alpha_{+}^{(\mathrm{i})} \Eq \pi \MINUS \alpha_{-}^{(\mathrm{i})}$. 
The two solutions $\alpha_{\pm}^{(i)}$ have the same free energy, 
and we consider $\alpha_-^{(i)}$ for the moment.  
Its asymptotic form is 
$\alpha_{-}^{(\mathrm{i})} \msm2 \sim \msm2 (\sqrt{3}/8) |K|^{-1} b R$ 
for small $R/|K|$. 
Inserting this into Eq.~(\ref{eq:fXKnegative}), 
the corresponding free energy is obtained as 
\begin{align}
&f^{\mathrm{(i)}}_{\rm X}
= \tilde{a}_{\rm X}(T)
 R^2
+ 
\FRAC{1}{4} 
\bigl( 
   c_2 \PLUS \FRAC{3}{2} K^{-1} b^2 \bigr) R^4 
+ \cdots. 
\label{eq:fX_ii}
\end{align}
One should note that this has no $R^3$-term.   
Therefore, if $c_2 \PLUS \FRAC32 K^{-1} b^2 \GT 0$, 
a possible transition must be continuous, 
and its transition temperature 
is given by the same expression 
$\tilde{a}_{\rm X}(T)=0$ [Eq.~(\ref{eq:a_tilde_pm})] 
as that for the single-$\qq$ order.  
One can see this in Fig.~\ref{fig13} 
as a straight phase boundary  
with the \textit{disordered} phase  as $J$ varies.  
Thus the fourth-order term determines 
which of the $zoxy$ or single-$\qq$ state appears.  
The free energy of the single-$\qq$ order is
\begin{equation}
f_{\rm X}^{1\qq}
=
\tilde{a}_{\rm X}(T)
R^2
+ c R^4. 
\label{eq:fsingle}
\end{equation}
Comparing Eqs.~\eqref{eq:fX_ii} and \eqref{eq:fsingle}, 
one obtains 
the appearance condition of the triple-$\qq$ state 
\begin{equation}
4c > 
c_2 \PLUS \FRAC{3}{2} K^{-1 }b^2 
= \FRAC43 (c \PLUS 2c' \PLUS c\dprime ) 
\MINUS \FRAC32 (-K)^{-1} b^2 . 
\label{eq:cond_state2}
\end{equation}
For the parameters estimated 
from the local potential ($c \Eq c' \Eq c\dprime $), 
this condition reads as 
\begin{equation}
-\frac{9b^2}{8c} < K   <  0. 
\end{equation}
This is consistent with the results of microscopic calculations 
in Fig.~\ref{fig9}, 
which show the single-$\qq$ AFO$_{22}$ state 
for larger $(-K)$. 
The transition can be first order, 
when the fourth-order coefficient 
$[ c_2 - 3b^2/(-2K)]/4$ 
turns into negative for large $b$.  
However, its quantitative analysis needs to include 
fifth- and sixth-order terms in Eq.~(\ref{eq:F}).  
Using the expression (\ref{eq:conditionzoxy}) 
in $f_{\rm X}$, one sees that 
the free energy has a local minimum at $R \Eq R_\circ$ with 
\begin{subequations}
\begin{align}
\hs{-4pt}
R_\circ &= 
c_2^{-1} 
\Bigl\{
g + \sqrt{ g^2 - 
4 \msp2 c_2^2 \msp2 \tilde{a}_{\rm X}(T)\big[ a_0(T)-2J\big]  
}
\Bigr\}^{1/2} , 
\label{eq:Q2tilting}
\\
g &\equiv 
6b^2 - 2 c_2 \msp2 [ \tilde{a}_{\rm X} (T)-2K ].
\end{align} 
\end{subequations}
For $g \LT 0$, a second-order transition occurs 
at $\tilde{a}_{\rm X}(T) \Eq 0$.  
The condition $g \LT 0$ is indeed equivalent to the previous one 
$c_2 \PLUS \FRAC32 K^{-1} b^2 \GT 0$ 
derived from Eq.~(\ref{eq:fX_ii}).  
For $g \GT 0$, 
the transition is generally first order.  
It occurs at the position 
where $f_{\rm X}(R \Eq R_1) \Eq 0$,  
but we do not show the details.  
As discussed in Appendix \ref{A1}, 
the case of $K \Eq 0$ turns out to be easier to analyze, 
and the results are much simpler.   
Indeed, those results are consistent with 
the microscopic mean-field calculations.  
See discontinuous behavior in Fig.~\ref{fig10}(f), while continuous one 
in Fig.~\ref{fig10}(h).

One can apply the above analysis to other cases with $\pi$ shifts, 
which correspond to different domains.
There are four different but equivalent domains 
in the same state, and their $\btheta$ values are given as
\begin{equation}
\begin{split}
\btheta = &\btheta _0 + \alpha (1,1,1) \\
&+ \bigl\{ (0,0,0),\ 
(0, \pi , \pi ), \ 
(\pi , 0, \pi ), \ 
(\pi , \pi , 0) \bigr\}. \label{eq:formersetalpha}
\end{split}
\end{equation}
Here, $\alpha$ is the solution $\alpha_-^{(i)}$ 
in Eq.~\eqref{eq:conditionzoxy}
and $0 \le \alpha \le \FRAC{\pi}{6}$. 
Starting from any one of them, the other three 
can be obtained by translations as discussed 
in Sec.~\ref{sec:tripleqorders}.  
As for the solution $\alpha_{+}^{(i)}=\pi+\alpha=\pi-\alpha_-^{(i)}$ 
with $-\FRAC{\pi}{6}\le \alpha\le 0$ in Eq.~(\ref{eq:conditionzoxy}), 
one obtains
\begin{equation}
\begin{split}
\btheta = &\btheta _0 + \alpha (1,1,1) 
\\
&+
\bigl\{
(\pi, \pi , \pi ), \ 
(\pi, 0 , 0 ), \ 
(0 , \pi , 0 ), \ 
(0 ,  0, \pi ) \bigr\}. 
\end{split}\label{eq:lattersetalpha}
\end{equation}
The free energy for the latter sets (\ref{eq:lattersetalpha}) 
is identical to that for 
the former sets (\ref{eq:formersetalpha}).  
Note the different ranges of the solution $\alpha$ 
 for these two sets. 
In total, the $zoxy$ state has $8=4\times 2$ domains; 
4 corresponds to the trivial translations and 
2 arises from the degeneracy related to 
the mirror operations with respect to one 
of $Z_3$ axes in the $\QQ$ space, 
e.g., $x\leftrightarrow y$, and $z\to z$. 
Note that the mirror symmetry is broken 
in the $zoxy$ states 
as shown in Fig.~\ref{fig9}(d) 
and Fig.~\ref{table1}. 
Seeing Eq.~(\ref{eq:conditionzoxy}), one expects that 
$|\alpha|$ increases with lowering temperature, 
since $R$ usually increases. 
This agrees qualitatively with 
the results of microscopic calculations 
shown in Figs.~\ref{fig10}(f) and \ref{fig10}
(h).  
As $|\alpha|$ approaches $ \FRAC16 \pi$, 
the order 
parameters $\bpsi _{\ell}$'s tilt their directions 
toward $\omega \times \mbox{(integer)}$. 
This is similar to the $zyox$ state realized for $K \GT 0$, 
in which 
$\btheta \Eq (\omega ,2\omega ,0)$.  
In contrast, for the $zoxy$ state, 
the directions $\btheta$ approaches $(2\omega ,0,\omega)$, but
this is not equivalent to {\it any} of the domains in the $zyox$ states. 
To see this, it is useful to check 
the real space configuration of the quadrupole moments.  
Since these configurations have no coupling to $\bpsi_{0}$
[Eqs.~(\ref{eq:F0X_revised}) and (\ref{eq:F0X4_revised})], 
the uniform component vanishes $\bpsi_{0} \Eq \bzero$.  
Then, substituting 
$X_{\ell} \Eq R/\sqrt{3} \Equiv \bar{q}/2$ for all $\ell$'s 
and 
$\btheta \Eq (2\omega ,0, \omega )$ 
into Eq.~(\ref{eq:Qloc}), one obtains
\begin{align}
\QQ _{\rm A}= \bm{0}, \ \ 
\QQ _{\rm B}=\bar{q} \msp2 \hee _{2\omega}, \ \ 
\QQ _{\rm C}=\bar{q} \msp2 \hee _{\omega}, \ \ 
\QQ _{\rm D}=\bar{q} \msp2 \hee _{0}. 
\label{eq:localQ_zoxy}
\end{align}
This configuration indeed manifests a partial order.  
Compare this with Eq.~(\ref{eq:localQ_zyox}). 
This agrees with the results of 
the microscopic calculations in Sec.~\ref{sec:phase}.  
See also Fig.~\ref{table1} and Fig.~\ref{fig9}(d).  
Note that this is the configuration corresponding to 
$\alpha \Eq \FRAC16 \pi$.  
For general values of $\alpha$, O$_{22}$-type components 
also mix as shown in Fig.~\ref{fig9}. This is apparent since 
$\hee_{\pi/2+\alpha}=\hee_{\omega}\cos(\frac{\pi}{6}
-\alpha)-\hee_{\omega/4}\sin(\frac{\pi}{6}-\alpha)$. 
Here, $\hee_\omega$ represents 
the O$_{20}$-type ($2x^2-y^2-z^2$), 
while $\hee_{\omega/4}$ does the O$_{22}$-type 
($\sqrt{3}(z^2-y^2)$).

\vspace{10pt}
\subsection{Case of $X_1 = X_2 \ne X_3$: $zxzy$ state} 
\label{sec:analysis_zxzy}

Now, we study the states with an ``$xxz$''-type anisotropy for $K<0$. 
In addition to the high-symmetry solution with $X_1=X_2=X_3$ discussed 
in the previous section A, another simple way 
lowers the energy 
in both the $K$- and $b$-terms of $f_{\rm X}$. 
This requires ``uniaxial'' modulations 
in the $\bpsi _{\ell}$'s magnitude of quadrupole moments 
such as $X_1 \Eq X_2 \GT X_3  \GT 0$. 
Here, the term ``uniaxial'' means 
that there is only one axis corresponding 
to the direction of $\bpsi^{(3)}$ 
around which the triple-$\qq$ configuration is 
symmetric in the quadrupole space as will be
 explained below. 
The directions $\btheta$ 
are determined as follows. 
First, one expects that 
the free energy of  
the part of $\bpsi _1$ and $\bpsi _2$ in 
the $K$-term for $K<0$ is lowered by setting 
$\theta_1 =-\theta_2 \Sim \FRAC76 \pi$.
Second, it is possible to lower $f_{\rm X}$ through the $b$ term 
by setting $\theta_3 = 0$, since this satisfies $\bar{\theta}=0$. 
An expectation is that the energy cost of the $K$-term is not large 
due to $\bpsi _3 $'s small magnitude.
This is indeed the $zxzy$ state 
obtained in Sec.~\ref{sec:tripleqorders}. 
Note that the trend is 
$\theta_1 \To \FRAC76 \pi$ as $K \To -\infty$. 
This choice of $\btheta$ can lower 
both $K$- and $b$-terms of the free energy.

Let us concentrate on the case of large $(-K)$ in the following.  
The free energy reads
\begin{align}
f_{\rm X}^{zxzy}=& 
2\tilde{a}_{\rm X} (T) X^2 + 
a_{\rm X} (T) Z^2 - 
6 b X^2 Z
\nonumber\\
&+ 
c_3 X^4 + 
c_4 X^2 Z^2 + 
c Z^4, 
\label{eq:Fzxzy}
\end{align}
where $X \Equiv X_1 \Eq X_2$, $Z \Equiv X_3$, 
$c_3 \Eq 2 (c \PLUS c') \PLUS c\dprime$, and 
$c_4 \Eq 4c' \PLUS 6c\dprime$. 
The coefficients of the quadratic terms 
$a_{\rm X}$ and $\tilde{a}_{\rm X}$ were defined in 
Eqs.~\eqref{eq:def_aXT} and \eqref{eq:a_tilde_pm}, respectively. 
Minimizing Eq.~(\ref{eq:Fzxzy}) with respect to $X$, 
we obtain a stationary value of $X$ and the result is 
\begin{align}
X = 
\left[ 
\frac{-2\tilde{a}_{\rm X} (T)
  \PLUS 6 b Z \MINUS c_4 Z^2}{2c_3} 
\right]^{1/2}
\equiv 
X_{*} (Z) . 
\end{align}
Substituting this to Eq.~(\ref{eq:Fzxzy}), we obtain 
\begin{align}
f_{\rm X}^{zxzy}=& \, 
c_3^{-1} 
\Bigl\{
\bigl[ 
  c_3 a_{\rm X}(T) \MINUS 
  c_4 \tilde{a}_{\rm X} (T) \MINUS 
  9b^2
\bigr] Z^2
-\tilde{a}_{\rm X}^2 (T) 
\nonumber\\
& 
\PLUS 
6 \tilde{a}_{\rm X} (T) b Z \PLUS 
3 c_4 b Z^3 
+
\bigl( 
  c_3 c \MINUS 
  \FRAC14 c_4^2 
\bigr) Z^4
\Bigr\} . 
\end{align}
The stationary value of $Z$ is calculated via 
$\partial f^{zxzy}_{\rm X}/\partial Z=0$ with the 
constraint $X_*^2(Z)\ge 0$. 
Since its analytic solution is not so simple, 
we do not discuss the detail here. 
Nevertheless, 
it is certain that the transition between the $zyox$ 
and the $zxzy$ states is first-order, 
since it is determined by the crossing of their free energy values.  
One can also examine the stability of the $zxzy$ states 
in comparison with the lower symmetry NS state 
as has been done in Sec.~\ref{sec:inducedQ}, but 
we do not show them here. 
Note that the above analysis neglects the ferro component $\bpsi _0$. 
This is induced in the $zxzy$ state, 
and Eq.~(\ref{eq:Fzxzy}) is valid only for large $(-K)$. 
When $(-K)$ is very large, 
the moment directions are fixed to 
$\btheta \Eq \FRAC76 \pi (1,-1,0)$ 
or one of the equivalent directions.  
However, at $K \Eq 0$, as shown in  Appendix~\ref{A1}, 
the favored configurations are 
uniaxial ones, e.g.,  
$\btheta \Eq  \FRAC32 \pi (1,-1,0)$,  
or its equivalent ones. 
This indicates that with decreasing $(-K)$, 
$|\theta_{1}|$ and $|\theta_{2}|$ decrease 
with keeping the symmetry of the $zxzy$ state.  
In the next section C, 
we will discuss the question 
whether such a $zxzy$ state survives for 
smaller $(-K)$.  

The $zxzy$ state can break its symmetry down 
to that for the NS state with $X_1\ne X_2\ne X_3$. 
The phase transition is either first or second order. 
Once the three modes are inequivalent, 
the relation $|\theta_1|=|\theta_2|$ no longer holds. 
The discussion can be done in a similar way to that 
in Sec.~\ref{sec:inducedQ}, 
but we omit it for simplicity.  See the discussions in 
the next section C.

\vspace{10pt}
\subsection{Fully asymmetric case: NS states}
\label{sec:C3breaking} 

Lastly, we will discuss the no symmetry (NS) state. 
Instead of carrying out the stability analysis 
as has been done in the last sections A and B, 
we will take an alternative approach and study the limit of small $(-K)$.  
As discussed in Appendix~\ref{A1}, 
some solutions at $K \Eq 0$ have 
anisotropic configurations.  
These degenerate anisotropic states 
have different values of $\btheta$.  
Below, we will discuss how this degeneracy is lifted for $K \LT 0$.

Among the solutions at $K \Eq 0$, 
the isotropic configuration shown in Fig.~\ref{figSummary} 
obviously appears for $K \LT 0$ 
in a state connected to the $zoxy$ state  
discussed in Sec.~\ref{sec:symtilting}.  
At low temperatures, a wide range of 
the parameter space is 
covered by a phase with 
``$xxz$''-type uniaxial configurations and
unbalanced magnitudes $X_1 \Eq X_2 \Neq X_3$. 
This is denoted by uniaxial triple-$\qq$ 
in Fig.~\ref{figSummary}.  
See detailed discussions 
in Appendices~\ref{sec:AppC} and \ref{KzeroAnal}. 
They have the configurations of either 
$\XX \Eq (p,q,q)$ with 
$\btheta \Eq \FRAC12 \pi (0,1,-1)$ 
or $\XX \Eq (q,p,q)$ with 
$\btheta \Eq \FRAC12 \pi (-1,0,1)$. 
We will investigate the possibility of tilting for the first case.  
To simplify our discussion, 
we assume that $p$ and $q$ are fixed.
Representing the three modes of the tilting as
$\delta \btheta \Eq 
\alpha (1,1,1) \PLUS 
\delta (2,-1,-1) \PLUS 
\eta (0,1,-1)$, 
the change in the $K$-term of the free energy is calculated 
up to the linear order as 
\begin{align}
\delta f_{2K}^{pqq} 
( \delta \btheta ) 
&\sim 
4 \sqrt{3}  K 
\bigl(
p^2 \delta \theta_X + q^2 \delta \theta_Y
\bigr) 
\nonumber
\\
&= 
4 \sqrt{3}  K 
\bigl[
  (p^2 \PLUS q^2) \alpha
\PLUS (2 p^2 \MINUS q^2 ) \delta
\PLUS q^2 \eta 
\bigr] . 
\label{eq:df2Kpqq}
\end{align}
Since the linear-order coefficients 
are nonvanishing above,  
these three modes are all induced, 
but the direction $\theta_3$ 
is pinned to $- \FRAC12 \pi$ in this order. 
By taking account of other terms such as the cubic $b$ terms, 
$\theta_3$ may eventually tilt.  
Thus the uniaxial state becomes unstable for $K \LT 0$, 
and is replaced by the NS state.  

This result partly explains the isolated island of the $zxzy$ state 
in the phase diagram shown in Fig.~\ref{fig9}. 
The stability of the $zxzy$ state
for large $(-K)$ depends on the free energy 
of the NS [Eq.~(\ref{eq:a_tilde_pm})] and 
the single-$\qq$ [Eq.~(\ref{eq:fsingle})] states. 
We do not try further analysis in this paper.

We close this section with a 
comment on a very small region 
where the configurations are 
$X_1\ne X_2\ne X_3$ and 
$\btheta=\FRAC12 \pi (0,1,-1)$.  
This is denoted in Figs.~\ref{figSummary} and \ref{fig:AppK0TJ}  
as the fully-anisotropic triple-$\qq$ 
for $K \Eq 0$ and for small $J$.  
Since the three magnitudes differ to each other, 
the directions $\btheta$ 
tilt for finite $K$ away from the directions
$\btheta \Eq \FRAC12 \pi (0,1,-1)$, 
and this leads also to the NS (NS$'$) state. 
Indeed, such changes have been observed in the microscopic calculations 
shown in Fig.~\ref{fig:AppK0}(b).

\vspace{10pt}
\section{Discussion}
\label{sec:Discussion}

In this section, we 
discuss the implications of the present theory for 
the related materials 
including PrMgNi$_4$ and 5$d^1$ double perovskites.   
We also briefly comment on further implications of 
the multiple-$\qq$ physics of multipoles in other systems. 
It should be noted that the microscopic mean-field results 
in Sec.~\ref{sec:phase} based on the 
localized model are supported 
by the Landau analysis in Sec.~\ref{sec:Landau}. 
This indicates that 
the discussions about the triple-$\qq$ physics 
in this paper are also applicable to metallic systems. 
One should understand that the ordering wavevectors 
are determined with taking into account 
the effects of the conduction electrons. 

\vspace{10pt}
\subsection{$\Gamma_3$ quadrupole moments in real systems}

The rare earth compound PrMgNi$_4$ has a structure in which 
Pr$^{3+}$ ions form an fcc sublattice \cite{Kusanose2019}, 
and the CEF ground state of the Pr$^{3+}$ ion has been identified 
as the non-Kramers doublet $\Gamma_3$. 
This material is metallic but shows no indication 
of the quadrupolar Kondo effects. 
For discussing the phase transition in this material, 
it is useful to compare it with the 
results obtained for the localized model. 
A detailed quantitative analysis needs more 
elaborate calculations and it is one of our future studies.
The excited states of Pr$^{3+}$ ion are 
the $\Gamma_4$ triplet at 1.16 meV, 
the $\Gamma_1$ singlet at 2.78 meV, 
and the $\Gamma_5$ triplet at 11.6 meV, 
and these excitation energies have been determined 
by the inelastic neutron scattering experiments \cite{Kusanose2022}. 
Here, the excitation gap to $\Gamma_1$ corresponds to the parameter 
$E_1$ used in Eq.~\eqref{eq:defH}. 
Since the other states 
have no quadrupole matrix elements with the 
ground states $\Gamma_3$, 
we have not taken them into account. 
The unidentified inelastic peaks at 2.5 and 5.9 meV 
suggest that the cubic lattice symmetry is weakly broken. 
However, thermodynamic experiments such as 
specific heat and magnetization measurements have shown no 
signature of phase transitions down to the temperature $\sim 0.1$ K. This broken lattice symmetry is now considered as an extrinsic 
effect of lattice imperfections or excess Mg atoms.
They mask the intrinsic quadrupole ordering discussed in this paper, 
and thus further experimental studies using single crystals 
are necessary to identify the type of quadrupole order 
realized in this system. To explore exotic quadrupolar physics, 
it is important to find other materials related to PrMgNi$_4$.  
Mg or Ni may be replaced by nearby elements in the periodic table 
with similar chemical properties.  

In a recent study, PrCdNi$_4$ was synthesized 
and found to show a clear phase transition \cite{KusanoseOnimaru2022}. 
The estimated entropy $S$ at the transition temperature $T_q\simeq 1$ K is 
less than $\sim 0.5$ln~2, and
 the ordering degrees of freedom have not been 
 identified. A broad peak at $T\simeq 5$ K is reported in its specific heat above $T_q$. 
Although this might be a Schottky peak due to the CEF excited states,
its origin remains unclear, 
since $S(T\simeq 5~{\rm K})$ is less than ln~2 and too small
 to conclude that this anomaly 
 is due to the CEF excitations. 
In this respect, it is interesting to apply the present theory 
and explore a possibility of triple-$\qq$ ordered states above $T_q$. 
We have no information on
the detailed bond dependence of the quadrupole exchange interactions, 
except the point that 
the total magnitude is about 1 K. 
In addition to the nearest-neighbor interactions, 
some further-neighbor ones may also be large and 
have non-negligible effects.  
For determining their values, 
it is useful to observe the spin-orbital wave 
in the ordered phase, 
and such experiments are highly desirable. 
We also expect
that further experimental studies 
clarify the nature of the low temperature phase in PrCdNi$_4$.

Other interesting materials related to the present theory 
are the family of double-perovskites containing an fcc sublattice of  
ions with 5$d^1$ electron configuration \cite{Erickson2007,%
De_Vries2010,Lu2017,Gao2020,Hirai2020,Arima2022-vh}. 
A characteristic point is that  
those ions have a quartet ground state with the effective 
total angular momentum $J_{\rm eff} \Eq \FRAC32$ 
due to strong spin-orbit coupling in $5d$ orbitals.  
Various nontrivial orders have been proposed for this system  
\cite{Chen2010,Churchill2022,Svoboda2021}.  
Under the cubic CEF, 
this quartet can be regarded as a product state made of 
a spin-$\FRAC12$ doublet and an orbital $\Gamma_3$ ($E_g$) doublet.  
Thus, in the temperature range where their spins remain disordered, 
we may expect that their orbital degrees of freedom are 
described by the present theory.  

One member of this family is Ba$_2$MgReO$_6$. 
Hirai et al.~studied it by synchrotron x-ray-diffraction measurement 
and observed a phase transition 
of the AFO$_{22}$-type quadrupole order at $T_q \Eq 33$ K above 
the magnetic ordering temperature $T_m \Eq 18$ K \cite{Hirai2020}. They also found a ferro O$_{20}$ component below $T_q$, 
and it has been explained by considering the electron-lattice anharmonic coupling or lattice anharmonicity 
\cite{Hirai2020, Svoboda2021,Iwahara2022-mt}. 
We propose to apply the present theory to this system and explain a ferro component as an induced moment due to the third-order coupling. 

From our point of view, it is worthwhile to examine 
the effects of CEF excited states above 
 the $J_{\rm eff}=\FRAC32$ 
multiplet on the magnitude of the observed ferro orbital moments.  
 Important excited states are orbital singlet (spin doublet) states. 
They realize a situation of the orbital degrees of freedom 
 similar to those studied in this paper, 
 where the third-order couplings of quadrupoles take effects \cite{Kubo2023-sl}.

\vspace{10pt}
\subsection{Other applications}

We have demonstrated in this paper that 
the triple-$\qq$ quadrupole orders emerge generically, if not always,  
in the fcc lattice with nearest-neighbor interactions, 
and that some of them are partially ordered states. 
One of the main results is that the partial-order 
state for $K \GT 0$ 
(named $zoxy$) has a higher transition temperature 
than the single-$\qq$ quadrupole order.  
This is a consequence of the cooperation of 
the anisotropic interaction $K$ and the third-order $b$-term 
 of the local potential in the free energy.  
It is also important that the $b$ term couples the modes 
at all the three X points,  
and this causes several triple-$\qq$ order patterns of quadrupoles.  
The translation symmetry imposes the important matching condition 
$\kk_1 \PLUS \kk_2 \PLUS \kk_3 \Eq \GG $. 
Some three-dimensional systems have a set of 
high-symmetric $\kk$-points in the Brillouin zone 
satisfying this condition. 
For example, $\kk_1=(0,\pi,\pi)$, $\kk_2=(\pi,0,\pi)$, 
and $\kk_3=(\pi,\pi,0)$ in a simple cubic lattice 
satisfy $\kk_1+\kk_2+\kk_3=\bm{G}$. 
In two dimensions, this condition is easily satisfied particularly in 
systems with a hexagonal symmetry, 
since the three vectors are confined in the same plane \cite{Okubo2012}.

Multipoles have such third-order couplings, 
if their parity is even under both time reversal and 
spatial inversion operations.  
One can expect similar triple-$\qq$ orders in some other systems.  
For example, promising candidates are the  system of 
electron $t_{\rm 2g}$ orbitals 
($d_{xy}$, $d_{yz}$, and $d_{zx}$) 
in cubic materials. 
Other candidates are those of the $e_{\rm g}$ orbitals 
in two- and three-dimensional systems.
In the $t_{\rm 2g}$ systems, a possible third-order coupling  
has a form of O$_{xy}$O$_{yz}$O$_{zx}$  \cite{Nikolaev1999}, 
and this is similar to the $b$ term discussed in this paper.

We also note that 
such a third-order coupling also exists for 
composite degrees of freedom.
For example, most natural candidates are the  
systems with both active dipole and quadrupole moments.  
This case was studied using the $R$Zn compounds 
($R$: rare earth element such as Tm or Nd) \cite{Morin1978-ov,Amara1995} or  
actinide monopnictides \cite{Lander1995}, which have 
the CsCl-type crystal structure, i.e., an fcc structure.  
Recently, two of the present authors 
discussed that UNi$_4$B is also categorized 
to this type of materials \cite{Ishitobi2023UNi4B}. 
It was motivated 
by the experiments pointing out the importance of quadrupole 
degrees of freedom in this system \cite{Yanagisawa2021}. 
A triple-$\qq$ charge-density-wave (CDW) order 
has also been discussed for 
the kagome-lattice superconductors 
$A$V$_3$Sb$_5$ ($A=$K, Rb, Cs) 
\cite{Ortiz2019,Jiang2021,Denner2021}. 
Its free energy includes a cubic term similar to 
ours, and this also leads to triple-$\qq$ CDW orders in this system. 
This suggests an interesting possibility of 
superconductivity mediated by fluctuations in a triple-$\qq$ order, 
but this is not an issue of this paper and we do not discuss it further.  

When a leading instability occurs at an incommensurate 
wavevector $\qq$ with small $|\qq|$, 
this leads to several large-scale structures such as mosaic, 
(half-)vortex, or skyrmion, 
and this corresponds to triple-$\qq$ orders 
in magnetic systems \cite{Okubo2012}.  
We once again emphasize that those exotic configurations are stable 
at high temperatures as is the $zoxy$ state in this paper.  
These fascinating possibilities will be examined
in our future studies \cite{Hattori2022}.


\vspace{10pt}
\section{Summary}
\label{sec:Summary}

In this paper, we have studied $\Gamma_3$ quadrupole orders 
in an fcc lattice.  
We have employed a four sublattice mean-field theory, 
and shown the presence of various triple-$\qq$ states, 
which include partially ordered states with disordered sites 
at high temperatures.  
We have discussed the stability of these states,  
based on the phenomenological Landau theory 
and shown that its results can well explain 
those of the microscopic mean-field calculations.  
The third-order coupling of quadrupoles in the free energy plays a  
crucial role for stabilizing the triple-$\qq$ states 
with the ordering vectors located at the zone boundary X points.  
This unique mechanism for the triple-$\qq$ orders 
is quite ubiquitous 
in the systems with the time-reversal even parity 
including electric multipoles, 
and it  also works in many other systems.  
We believe that our work stimulates further theoretical studies 
and experiments on exotic quadrupole or other multipole orders in future.

\section*{Acknowledgment}
This work was supported by JSPS KAKENHI 
(Grant Nos. JP16H04017, JP18K03522, and JP21H01031).

\appendix

\section{Detail Analysis on the Large $K$ limit for $K>0$}
\label{sec:AppAA1}

We start with minimizing the free energy $f_{\rm X}$ in 
Eq.~\eqref{eq:F3} with respect to the following two angle variables 
$\Theta$ and $\Phi$ with $0\le \Theta,\Phi\le \pi/2$ in two steps:
\begin{equation}
(X,Y,Z) = R ( 
\sin\Theta\cos\Phi, \, 
\sin\Theta\sin\Phi, \, 
\cos\Theta ).
\end{equation}
We will see that an important dimensionless parameter is 
\begin{equation}
  y \equiv  \frac{3b}{c_1 \,R } >  0 . 
\end{equation}  

The first step is the minimization with respect to $\Phi$. 
Calculation of $\partial f_{\rm X} / \partial \Phi$ 
leads to the condition 
$\sin ^4 \Theta \cos 2 \Phi \, 
[ \sin 2 \Phi - 
\sigma (\Theta ) / \sigma (\Theta_c )
] \Eq  0$ 
with $\sigma (x) \Equiv  \cos x/\sin ^2 x$. 
Here, $\Theta_c = \Theta_c (y)$ is defined as 
\begin{equation}
\Theta_c (y) = 
\cos^{-1} 
\bigl[ 
\sqrt{1+ y^2} - y \bigr]. 
\end{equation} 
Note that 
$\Theta_c(y)$ 
is a monotonically increasing function, 
and $\Theta_c (0) \Eq  0$ and 
$\Theta_c ( y \To  \infty) \Eq  \pi/2$.  
An important relation is 
$\cos \Theta_c (y _{\star}) \Eq  y_{\star}$ 
at the special value  
$y_{\star} \Eq  1/\sqrt{3}$.  
Judging also from the corresponding values of second-order derivative, 
the minima are located at 
\begin{equation}
\Phi^* = \left\{ 
\begin{array}{@{}ll@{}}
\pi/4, & 
\mbox{(for $0$$\le$$\Theta$$\le$$\Theta_c$)}
\\[2pt]
\Phi_2, \  
\pi/2 \MINUS  \Phi_2, & 
\mbox{(for $\Theta_c$$\le$$\Theta$$\le$$\pi/2$)}
\end{array}
\right. 
\label{eqA:Phistar}
\end{equation}
with
\begin{equation}
\Phi_2 (\Theta ) = 
\frac12 \sin^{-1} 
\left[ 
\frac{ \sigma (\Theta ) }{ \sigma (\Theta_c ) } \right]. 
\end{equation} 
The case of $\Theta \Eq  0$ corresponds 
to the single-$\qq$ state, which is not under consideration here.  
For these obtained $\Phi^*$ values in Eq.~\eqref{eqA:Phistar}, 
we further examine 
the extremum and minimum conditions with respect to $\Theta$.  

For $\Phi \Eq  \pi/4$, the extremum condition is 
$0 = \partial f_{\rm X}/\partial \Theta 
\propto 
\sin \Theta ( 3 \cos^2 \Theta \!-\! 1 ) 
(\cos \Theta \!-\! t )$. 
Therefore, there is at most one minimum and 
it is located at 
$\Theta_{\star} \Eq \cos^{-1} y_{\star}$ for  
$y \GT  y_{\star}$.  
The energy of this local minimum is 
\begin{equation}
f_{\rm X} \left(\Theta_{\star} , \mbox{$\Phi^*$=} \frac{\pi}{4}\right) 
=
\left( y^{-1}-\FRAC{2}{\sqrt{3}}\right) bR^3, 
\end{equation}
and this is negative when 
$y \GT  \sqrt{3}/2$.  
This solution 
corresponds to the symmetric triple-$\qq$ state with 
$X \Eq  Y \Eq  Z$.  The case of $0 \LT  y \LT y_{\star}$ 
has no minimum in this $\Theta$-region.   

For $\Theta_c \LT  \Theta \LT  \pi/2$, 
the minimum in the $\Phi$-direction 
at $\Phi^* \Eq  \pi/4$ splits into two minima 
located at $\Phi_2 (\Theta )$ and 
$\pi/2 \!-\! \Phi_2 (\Theta )$.  
The extremum condition with respect to $\Theta$ is 
\begin{equation}
\left.
\frac{\partial f_{\rm X}}{\partial \Theta} 
\right|_{\Phi \Eq  \Phi_2(\Theta )} 
\propto 
\sin (2 \Theta) \, [ t^2 + \cos (2 \Theta) ], 
\end{equation}
for both of the two new positions.  
Therefore a local minimum exists only when 
$y \LT  1$ and 
its position is 
the symmetric point $\Theta^* \Eq  \pi/2$.  
The other extremum points $(\Theta^*,\Phi^*)$ 
are located at 
$\Theta^* \Eq  (1/2) \cos^{-1} (-y^2)$
or 
$\pi \MINUS  (1/2) \cos^{-1} (-y^2)$
and $\Phi^* \Eq  \Phi_2 (\Theta ^*)$ or 
$\pi/4 \MINUS  \Phi_2 (\Theta ^*)$, 
but they are all saddle points, 
because they are local maxima in the $\Theta$-direction. 

Let us summarize the results for the local minima 
of $f_{\rm X}(\Theta,\Phi)$. 
Their locations depend on the value of $t$ and 
\begin{equation}
(\Phi^*, \Theta^* )   
= 
\left\{
\begin{array}{l}
\left( \Phi^*, 0 \right), \ 
\left( 0, \frac{\pi}{2} \right),  \ 
\left( \frac{\pi}{2}, \frac{\pi}{2} \right), \ \ \ 
\mbox{for $0 \LT  t \LT  t_{\star}$,}
\\[4pt]
\left( \frac{\pi}{4}, \, \cos^{-1} t_{\star} \right), \ \ \ 
\mbox{for $t \LT  t^\star \LT 1$}
\end{array}
\right. 
\label{eqA1:min2}
\end{equation}
Note that $\Phi ^*$ is arbitrary when $\Theta ^* =0$.

There are two classes of stationary solutions. 
One is single-$\qq$ configurations. 
The other is symmetric triple-$\qq$ ($zyox$) state with 
$X \Eq Y \Eq Z \Eq R/\sqrt{3}$ . 
The free energy for the single-$\qq$ reads
\begin{align}
f_{\rm X}^{1\qq}=a_{\rm X}(T)R^2+cR^4,
\label{eq:A1f_single}
\end{align}
while that for the $zyox$ configuration is 
\begin{align}
f_{\rm X}^{zyox} = 
a_{\rm X}(T)R^2
-\frac{2b}{\sqrt{3}}R^3
+\left(c+\frac{1}{3}c_1\right)R^4.
\label{eq:A1fzyox}
\end{align}
Here, $a_{\rm X}(T)$ is defined in 
Eq.~(\ref{eq:def_aXT}). 
Equation \eqref{eq:A1fzyox} shows 
that the $zyox$ state is stabilized by the third-order term, 
while the magnitude of the fourth-order term depends on the anisotropic 
coupling $c'$ and $c''$ 
according to the definition of $c_1$ in Eq.~\eqref{eq:c1}.

\section{Analysis of the $K=0$ case: 
effects of the local free energy}
\label{A1}

We analyze in this Appendix the triple-$\qq$ orders at 
$K \Eq  0$. 
The isotropic $J$ term 
alone contributes to the 
inter-site interaction 
part of the free energy $f_{\rm X}$ 
in Eq.~(\ref{eq:F3}) as $(\FRAC12 a_0-2J)R^2$.  
This analysis is qualitatively the same as that for the local free energy, 
since the cubic and the fourth-order terms in the free energy 
arise from the local CEF potential and the form of the quadratic part is 
isotropic. 
Thus, the stable states  
at $K \Eq  0$ can be regarded 
as those favored by the single-ion potential.  
For complete analysis, it is necessary to 
take into account the effect of 
anisotropic inter-site interactions, 
i.e., the $K$-term. 
Nevertheless it is very useful to analyze 
the properties of the states favored by the single-ion potential. 
By using this knowledge as a starting point, 
we perform a stability analysis 
at $K \neq 0$ 
in Secs.~\ref{sec:LandauC} and \ref{sec:Knegative}. 

\subsection{Minimization with respect to $\theta_i$'s}

First, we minimize $f_{\rm X}$ in Eq.~\eqref{eq:F3} 
with respect to the order parameter directions $\theta_{X,Y,Z}$.  
Their amplitudes are assumed to be known.  
Ignoring the quadratic $K$-term proportional to $K$, 
one can write down the free energy 
as a sum of the following four terms.  
\begin{subequations} 
\begin{equation}
f_{\rm X}
=
f_{\rm X}^{24} + 
f_{\rm X}^{3} + 
f_{\rm X}^{\mathrm{4A}} + 
f_{\rm X}^{\mathrm{4B}}.
\label{eq:appB1}
\end{equation}
With the notation of 
$\XX \Eq (X_1,X_2,X_3) \Eq  (X,Y,Z)$ and 
$\btheta \Eq (\theta_1,\theta_2,\theta_3) \Eq  
(\theta_X,\theta_Y,\theta_Z)$, 
two of the four terms do not dependent on $\theta_i$'s: 
$f_{\rm X}^{24} = a(T) \, R^2 + c \, R^4$ and 
$f_{\rm X}^{\mathrm{4A}} =
2 \varDelta c \msm2 \sum_{i \LT  j} X_i^2 X_j^2$, 
where $a(T) \Eq  \FRAC12 a_0 (T) \MINUS  2J$ and 
$\varDelta c \Eq  c' \MINUS  c$, and 
$R^2 \Eq  \sum_i X_i^2$ as before.  
The other two terms do depend as 
\begin{align}
&f_{\rm X}^3
=
-6b \, X_1 X_2 X_3 \cos \bar{\theta},
\ \ 
\Bigl( \bar{\theta} \Eq  \sum_i \theta_i \Bigr), 
\label{eq:AppB3}
\\
&f_{\rm X}^{\mathrm{4B}}
=
4c\dprime \msm2 
\sum_{i \LT  j}  
X_i^2 X_j^2 
\cos^2 \theta_{ij} , 
\ \ 
\bigl( \theta_{ij} \Equiv  
\theta_i \MINUS  \theta_j  \bigr). 
\label{eq:AppB4}
\end{align}
\end{subequations}
We can assume without any loss of generality 
the relation 
$X_1 \mspace{-2mu}\ge \mspace{-2mu}
 X_2 \mspace{-2mu}\ge \mspace{-2mu}
 X_3$, and we will examine this case.  
As for the order parameter directions, the free energy depends on 
their three combinations, 
$\bar{\theta}$, $\theta_{13}$ and $\theta_{23}$.  
Note that they constitute a complete set of the directions, 
since $\theta_{12} \Eq  \theta_{13} \MINUS  \theta_{23}$.  
The $\bar{\theta}$-dependence immediately shows the minimum 
is located at $\bar{\theta}^* \Eq  0$.

Therefore, in order to minimize with respect to $\theta_{ij}$'s, 
it is convenient to consider the function 
$
 \bar{f} \Equiv f_{\rm X}^{\mathrm{4B}}/(4c\dprime X_1^2 X_2^2) 
 \Eq  A \cos^2 \theta_{23} \PLUS 
 B \cos^2 \theta_{13} \PLUS 
 \cos^2 (\theta_{23} \MINUS  \theta_{13} )
$, 
where the coefficients are 
\begin{equation}
 A \equiv 
\Bigl( \frac{X_3}{X_1} \Bigr)^2 \le
 B \equiv 
\Bigl( \frac{X_3}{X_2} \Bigr)^2 \le 1, 
\end{equation}
and their ratio is denoted as 
\begin{equation}
\kappa \equiv 
\frac{A}{B}  = 
\Bigl( \frac{X_2}{X_1} \Bigr)^2 \le 1. 
\end{equation}
See Fig.~\ref{figB:region}(a).   
Because of the symmetries 
$\bar{f}(\theta_{23},\theta_{13}) \Eq  
\bar{f}(\theta_{23} \PLUS \pi,\theta_{13}) \Eq  
\bar{f}(\theta_{23} ,\theta_{13}\PLUS \pi) \Eq  
\bar{f}(-\theta_{23} ,
        -\theta_{13})$, 
it suffices to consider the fundamental region,  
$(\theta_{23}, \theta_{12}) \Sp{\in} 
(-\pi/2 ,\pi/2] \times  [0, \pi/2]$.  
The symmetric points 
$(\theta_{23},\theta_{13}) \Eq  \frac{\pi}{2} (n_2,n_1)$ 
are local extrema for any integers 
$n_2$ and $n_1$ 
irrespective of the values of $A$ and $B$, 
but most of them are maxima or otherwise saddle points. 
In the fundamental region, 
$(0,\pi/2)$ is the only point among them 
which has a chance of being minimum.   
We calculated the $\bar{f}$'s Hessian and found 
that this local minimum is stable 
as far as $A \Sp{\le} B/(1 \PLUS  B) \Equiv  A_c (B)$.  
Note that the upper bound $A_c$ cannot exceed 1/2.  
When $A$ increases beyond $A_c (B)$ with $B$ fixed, 
the local minimum at $(0,\pi/2)$ becomes unstable and starts 
to move towards the direction proportional to $-(B, A_c (B))$.  
The minimum position $(\theta_{23}^*,\theta_{13}^*)$ 
is determined by solving the extremum 
conditions
\begin{equation}
  \sin [2 (\theta_{23}^* - \theta_{13}^*)] 
= -A \sin (2 \theta_{23}^*) 
=  B \sin (2 \theta_{13}^*).  
\end{equation}
If $A_c (B) \Sp{<} A \Sp{\le} B$, 
one and only one solution 
exists \textit{inside} the fundamental region, 
and this is a minimum position of $\bar{f}$.  
Its explicit expression reads as
\begin{subequations} 
\begin{align}
&\theta_{23}^* = 
 -\FRAC12 \pi 
 +\FRAC12 \sin^{-1} 
  \bigl[ \mathscr{S} /( 2\kappa A ) \bigr], 
\\
&\theta_{13}^* =  \phantom{-}
  \FRAC12 \pi 
 -\FRAC12 \sin^{-1} 
  \bigl[ \mathscr{S} / ( 2\kappa B ) \bigr], 
\\
&\mathscr{S} \equiv 
\left\{ 
\bigl[ 1 \MINUS  (\kappa \MINUS  A)^2 \bigr]
\left[ \left( A / A_c (B) \right)^2 \MINUS  1 \right]
\right\}^{1/2}. 
\end{align}
\label{eqB3:solution}
\end{subequations}
\!\!Since 
$(\kappa \MINUS A)^2 \Sp{<} 1$, 
this result once again manifests that this nontrivial solution exists 
only when $A \Sp{>} A_c(B)$.  
For evaluating the free energy, we need the values of 
$\cos^2 \theta_{ij}$'s.   
We have calculated them from Eq.~\eqref{eqB3:solution} and found 
that their expressions are particularly simple in terms of $X_j$': 
\begin{equation}
\cos^2 \theta_{ij}^* 
= 
\bigl[ \msp2 R^2 / (2X_i^2) \MINUS  1 \msp2 \bigr] 
\bigl[ \msp2 R^2 / (2X_j^2) \MINUS  1 \msp2 \bigr] , 
\end{equation}
where $R^2 \Eq  \sum_i X_i^2$ as defined before.  
Thus, the minimum value 
$f_{\rm X}^{\mathrm{4B}}{}^* 
 = 4c\dprime X_1^2 X_2^2 \bar{f} (\{\theta_{ij}^*\})$ 
is immediately calculated with these values, and the result is 
\begin{equation}
f_{\rm X}^{\mathrm{4B}} {}^*
= \left\{
\begin{array}{@{}ll@{}}
\displaystyle
4c\dprime  X_2^2 X_3^2 , 
& \mbox{when } R/\sqrt{2} \Sp{<} X_1 
\\[2pt]
\displaystyle 
c\dprime  
\Bigl( R^4  - 2\sum_i X_i^4 \msp2 \Bigr),  & \mbox{otherwise}. 
\end{array}
\right. 
\label{eqB:f41result}
\end{equation}
Here, $X_1$ and $X_3$ should be understood as 
the largest and smallest respectively 
of $\{ X, Y, Z\}$ in general cases, while $X_2$ 
is the remaining one.  
The condition for the upper case is equal to the previous one 
$A \Sp{<} A_c (B)$.  

Let us now determine the values of $\theta_i$'s 
from the above results.  
In order to obtain all the possibilities, 
one should also consider the solutions not limited 
to the fundamental region  
$\Sp{\pm} (\theta_{23}^*,\theta_{13}^*) \PLUS  \pi 
 (n_2 , n_1)$.   
Combining the result for $\bar{\theta}$, one find 
the minimum positions are represented as 
$\btheta^* \Eq (\theta_1^*,\theta_2^*,\theta_3^*) \Eq 
n_0 \omega \bm{d}_0 \PLUS 
\FRAC23 
[ (\pm \theta_{13}^* \PLUS  n_1 \pi ) \bm{d}_1 \PLUS  
  (\pm \theta_{23}^* \PLUS  n_2 \pi ) \bm{d}_2 ]$ 
where 
the two plus-minus signs should take an identical value.  
The vectors 
$\bm{d}_0$, $\bm{d}_1$, and $\bm{d}_2$ 
are $(1,1,1)$, $(1,-\FRAC12,-\FRAC12)$, and $(-\FRAC12,1,-\FRAC12)$, 
respectively. 
The representative value $(\theta_{23}^*,\theta_{13}^*)$ is 
the symmetric point $(0,\pi/2)$ for $A \Sp{<} A_c (B)$, or 
the nontrivial solution \eqref{eqB3:solution} for $A \Sp{>} A_c (B)$.  
Counting independent combinations, one finds 24 different 
sets of $(\theta_1^*,\theta_2^*,\theta_3^*)$ for a general value 
of $(\theta_{23}^*, \theta_{13}^*)$, 
but all of them can be generated from one representative 
$\btheta ^{*\circ}$ 
using three types of symmetry operations.  
They are 
\begin{subequations} 
\begin{alignat}{2}
&\mbox{(i) inversion:} \quad & 
& \theta_i ^*  \to 
\FRAC23 \bar{\theta} \MINUS  \theta_i ^* , \ 
\mbox{for } {}^\forall i, 
\\
&\mbox{(ii) mirror:} \quad & 
& \theta_i ^* \to \theta_i ^* , \ 
\theta_j ^* \to \theta_j ^* \PLUS \pi , \ 
\mbox{for } j \Sp{\ne}i , 
\\ 
&\mbox{(iii) rotation:} \quad & 
& \theta_i ^* \to \theta_i ^* \Sp{\pm} \omega ,  \ 
\mbox{for } {}^\forall i . 
\end{alignat}
\end{subequations}
The total number of the combinations is indeed 
$(1 \PLUS  1) \Sp{\times} (1 \PLUS  3 ) \Sp{\times} (1 \PLUS  2) \Eq  24$.  
For the trivial minimum point 
$(\theta_{23}^*,\theta_{13}^*) \Eq  (0,\pi/2)$ for $A \Sp{\le} A_c (B)$, 
the number is reduced to 12, since the half of the operations 
duplicate the points.  
One may choose 
$\btheta^{*\circ}  \Eq  (\pi/2) (0,1,-1)$ 
as a representative for this case.  
One should note that for the half of the 12 sets 
two $\theta_i$'s are identical.  
This is a consequence of one of the mirror operations.  
When $\bar{f}$ has no anisotropy ($A \Eq  B \Eq  1$), 
the value of nontrivial solution is 
$(\theta_{23}^*,\theta_{13}^*) \Eq  \omega (-\FRAC12 ,\FRAC12)$, and 
there also exist 24 minimum points of 
$\btheta ^*$.  
Six of them have a symmetric $120^{\circ}$-configuration 
$\{\theta_1^*,\theta_2^*,\theta_3^* \} \Eq  \{ -\omega, 0, \omega \}$, 
while the others have an umbrella configuration 
$\{\theta_1^*,\theta_2^*,\theta_3^* \} \Eq  
\{ (n\MINUS \FRAC12 )\omega, n\omega, (n\PLUS  \FRAC12 ) \omega\}$.

\subsection{Minimization with respect $X_i$'s} \label{AppB2}

%
%
\begin{figure}[tb]
\begin{center}
\includegraphics[width=8cm]{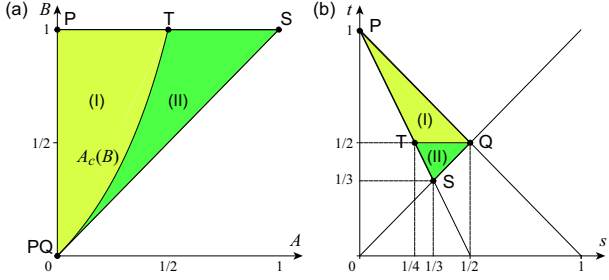}
\end{center}
\vspace{-10pt}
\caption{
(a) Parameters in the renormalized free energy $\bar{f}$. 
Special points in the panel (b) are also shown.  
Note that the point P there corresponds to the edge 
connecting $(A,B)=(0,0)$ and $(0,1)$. 
The colored two parts are the region under consideration.  
They are separated by the boundary $A=A_c (B)$, where 
the minimum position of $\bar{f}$ starts to 
shift from $(\theta_{23},\theta_{13}) \Eq (0,\FRAC{\pi}{2})$. 
(b) Domain of 
the dimensionless free energy $\bar{g}^{\msp2 34}(t,s)$. 
The fundamental region ($X_3 \Sp{\le} X_2 \Sp{\le} X_1$) 
is colored.  
}
\label{figB:region}
\end{figure}

Now that $f_{\rm X}$ has been minimized with respect to 
the order parameter directions 
$\{ \theta_i \}$, the next minimization procedure is about 
their amplitudes $\{ X_i \}$.   
As before, we do this under the constraint of 
$\sum_i X_i^2 \Eq  R^2$ being fixed.  
The simplest way of imposing this constraint is 
the use of the parametrization 
$X_1^2 \Eq  R^2 t$,  
$X_2^2 \Eq  R^2 s$, and  
$X_3^2 \Eq  R^2 ( 1 \MINUS  s \MINUS  t)$, 
and we search a minimum point in the $(s,t)$-space.  
Since the free energy is invariant upon any permutation of 
$\{ X_i \}$, it suffices to consider the fundamental region 
illustrated in Fig.~\ref{figB:region} (b), 
which corresponds to the part of $X_3 \Sp{\le} X_2 \Sp{\le} X_1$.  

For minimization with respect to $s$ and $t$, 
it suffices to consider the following dimensionless 
function $\bar{g}^{\msp2 34}$, 
since $f_{\rm X}^{24}$ is independent of $s$ and $t$: 
\begin{subequations} 
\begin{align}
\hs{-2mm}
\bar{g}^{\msp2 34} 
&(t,s) \equiv 
\frac{
f_{\rm X}^{3} \! +\! 
f_{\rm X}^{\mathrm{4A}} \! +\! 
f_{\rm X}^{\mathrm{4B}}{}^*}
{bR^3}
= 
\bar{g}^{\msp2 3} \!+\! 
\bar{c}_{\mathrm{A}} \msp2 \bar{g}^{\msp2 \mathrm{4A}} \msm6 +\! 
\bar{c}_{\mathrm{B}} \msp2 \bar{g}^{\msp2 \mathrm{4B}} , 
\\
&\bar{g}^{\msp2 3} 
= 
-6 \msp2 
[ t  s  (1-s-t)]^{1/2} , 
\\
&\bar{g}^{\msp2 \mathrm{4A}} 
= 
2 \msp2 
[ t + s - (t +s )^2  + ts ]
=: \mathscr{W}(t,s),
\\
&\bar{g}^{\msp2 \mathrm{4B}} 
= 
\left\{
\begin{array}{ll}
4 s (1-s-t) \ & 
\mbox{for $t \GE \FRAC12$,}
\\[4pt]
2 \mathscr{W}(t,s) - 1 & 
\mbox{for $t \LT \FRAC12$,}
\end{array}
\right. 
\end{align}
with the coefficients scaled as 
\begin{equation}
\bar{c}_{\mathrm{A}} \equiv \varDelta c R/b, \quad 
\bar{c}_{\mathrm{B}} \equiv    c\dprime R /b.  \hs{1cm} 
\label{eq:CA_CB}%
\end{equation}
\end{subequations}
Note that its domain is a narrow region shown in 
Fig.~\ref{figB:region} (b), 
and its control parameters are only 
$\bar{c}_{\mathrm{A}}$ and $\bar{c}_{\mathrm{B}}$. 
As for 
$\bar{g}^{\msp2 \mathrm{4A}}$, 
its minimum and maximum locate at the P and S point, respectively:  
$\bar{g}^{\msp2 \mathrm{4A}}(\mathrm{P}) \Eq 0$ and 
$\bar{g}^{\msp2 \mathrm{4A}}(\mathrm{S}) \Eq \FRAC23$. 
As for 
$\bar{g}^{\msp2 \mathrm{4B}}$, 
the local maximum value is 
$\bar{g}^{\msp2 \mathrm{4B}}(\mathrm{T}) \Eq \FRAC14$ 
and 
$\bar{g}^{\msp2 \mathrm{4B}}(\mathrm{S}) \Eq \FRAC13$ 
in the region I and II, respectively.  
Its minimum is degenerate and 
$\bar{g}_{\textrm{min}}^{\msp2 \mathrm{4A}}\Eq 0$ 
at all the points on the edge PQ.  

%
%
\begin{figure}[tb]
\begin{center}
\includegraphics[width=0.37\textwidth]{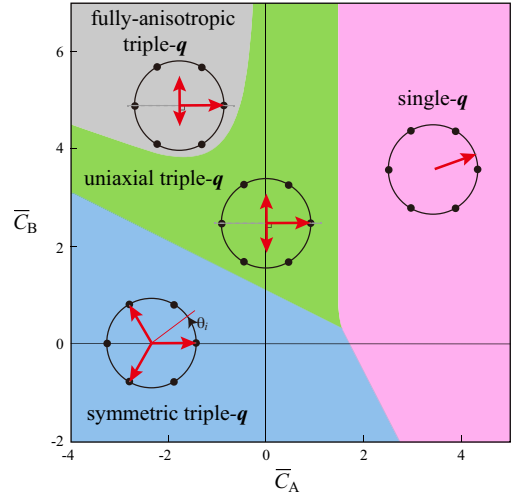}
\end{center}
\vspace{-10pt}
\caption{
Configurations of the order parameters at the X points 
in the parameter space of 
$\bar{c}_{\rm A}$ and $\bar{c}_{\rm B}$ defined in Eq.~(\ref{eq:CA_CB}) for $K \Eq 0$.   
Arrows represent three vectors $\bpsi _{\kk_\ell}$.  
The fully-anisotropic state is not realized 
at $\bar{c}_{\rm A} \Eq 0$ ($c=c'=c\dprime$). 
In the single-$\qq$ state, the arrow can freely rotate, 
since the free energy has no anisotropy with respect to its 
direction in Eq.~(\ref{eq:appB1}). 
}
\label{fig:PD-g34}
\end{figure}

Therefore, numerical minimization of $\bar{g}^{\msp2 34}$ 
is easy to perform with searching the entire domain, 
and we have determined the phase diagram covering the main part of 
the parameter space.   
The result is shown in Fig.~\ref{fig:PD-g34}, and 
the minimum position $(t_{\sharp},s_{\sharp})$ 
differs among the four parts marked by different colors.   

First, the red part (large\Blue{-}$\bar{c}_{\mathrm{A}}$ region) 
is the single-$\qq$ state, 
since the minimum locates at the point $(t_{\sharp},s_{\sharp}) \Eq (1,0)$  
corresponding to $\XX \Eq (R,0,0)$.

Secondly, 
the blue part 
(region of negatively large $\bar{c}_{\mathrm{A}}$) 
is the symmetric triple-$q$ state, and 
the minimum locates at the symmetric position 
$(t_{\sharp},s_{\sharp}) \Eq (\FRAC13 ,\FRAC13 )$  
corresponding to $X_1 \Eq X_2 \Eq X_3 \Eq \FRAC{R}{\sqrt3}$.  
The minimum position does not move within each of these two states.  
Therefore, the transition between them is first order, 
and the border is given by the line 
$2\bar{c}_{\mathrm{A}} \PLUS 
  \bar{c}_{\mathrm{B}} \Eq 2\sqrt{3}$.  

Thirdly, in the green-color part 
the minimum point locates on the edge PT in the domain, 
and thus the solution has an uniaxial symmetry 
$(X_2 \Eq X_3 \msm2 \LT 
 \FRAC{R}{2} \LT 
  \FRAC{R}{\sqrt2}  \LT \msm2 X_1)$. 
As $\bar{c}_{\mathrm{A}}$ decreases, 
the single-$\qq$ state (red part)
becomes unstable and continuously turns into this state.  
This phase boundary is determined by a breakdown 
of the stability condition of the minimum at the P point, 
and this gives the vertical line $\bar{c}_{\mathrm{A}} \Eq \FRAC32$.  
However, this continuous transition terminates at 
the tricritical point 
$(\bar{c}_{\mathrm{A}}^\star, \bar{c}_{\mathrm{B}}^\star)
 \Eq ( \FRAC32 ,\FRAC34 )$, 
and the transition becomes first order for 
$\bar{c}_{\mathrm{B}} \LT \bar{c}_{\mathrm{B}}^\star$.  
The first order transition line slightly winds and 
connects to the end point 
of the boundary between the single-$\qq$ and symmetric triple-$\qq$ states 
$(\bar{c}_{\mathrm{A}}^{\star \star}, \bar{c}_{\mathrm{B}}^{\star \star}) 
 \Eq (1.567,0.330)$.  
The symmetric triple-$\qq$ state (blue part) 
also becomes unstable and turns into the uniaxial state, 
as $\bar{c}_{\mathrm{B}}$ increases.  
The minimum at the S point becomes unstable 
on the line 
$\bar{c}_{\mathrm{A}} 
 \PLUS 2\bar{c}_{\mathrm{B}} \Eq 3\sqrt{3}/2$.  
However, a first-order transition takes place before that 
and the minimum jumps to a point with $t_{\sharp} \GT \FRAC12$.  
This determines the boundary of the symmetric and uniaxial 
triple-$\qq$ states. 

Lastly in the gray part, the minimum of $\bar{g}^{\msp2 34}$
locates \textit{inside} the triangle PTQ, 
and thus the solution has no symmetry 
corresponding to fully-anisotropic triple-$\qq$ state. 
This also means that the transition to the uniaxial triple-$\qq$ 
state is continuous.  

 This phase diagram of $\bar{g}^{\msp2 34}$ 
in Fig.~\ref{fig:PD-g34} is actually very useful, 
and we can make many predictions based on it 
for possible phase transitions 
in the $f_{\rm X}$ system upon lowering temperature.  
In any ordered state, the order parameter amplitude $R$ 
is nonvanishing and varies with $T$.  
Usually, $R(T)$ grows as $T$ decreases.  
When the disordered phase changes to an ordered state,  
$R(T)$ varies continuously starting from 0 if the transition is continuous, 
while jumps to a finite value otherwise.  
In any case, the two parameters 
$\bar{c}_{\mathrm{A}}$ and $\bar{c}_{\mathrm{B}}$ vary 
with $T$ according to Eq.~(\ref{eq:CA_CB}) with 
$R \Eq R(T)$, but 
\textit{they are confined on a ray} 
starting from the origin, 
i.e., $\bar{c}_{\mathrm{B}}/\bar{c}_{\mathrm{A}}$ is 
independent of $T$. 
Therefore, we can predict which ordered states may appear 
upon temperature control 
by looking at the changes on the ray in Fig.~\ref{fig:PD-g34}. 
By repeating this procedure with varying the ray's direction 
$u \Eq \tan^{-1} \bar{c}_{\mathrm{B}}/\bar{c}_{\mathrm{A}}$, 
we can determine the phase diagram.  

Let us write down explicitly the above procedure.  
Suppose a set of parameters $a$, $b$, and $c$'s is given, and consider 
possible phase transitions upon lowering temperature.  
The first step is the construction of the following 
function: 
\begin{equation}
\bar{g}_u (R) \equiv 
\bar{g}^{\msp2 34} (t_{\sharp},s_{\sharp}) 
\Bigr|_{\bar{c}_{\mathrm{A}} =  R C \cos u, \ 
        \bar{c}_{\mathrm{B}} =  R C \sin u }, 
\end{equation}
where 
the two dimensional vector $b^{-1}(2 \varDelta c , c\dprime)$ 
is parameterized by its modulus $C$ and angle $u$.  
Here, $(t_{\sharp},s_{\sharp})$ denotes the minimum position 
for the given value of $R$.  
The total free energy density is then given as 
\begin{align}
\tilde{f}_{\rm X} = 
a(T) R^2 + c R^4 + bR^3 \bar{g}_u (R) .  
\end{align}
Now, the Landau free energy functional has been minimized with respect to 
all the degrees of freedom except for $R$. 
Therefore, the minimization with respect to $R$ is the last task. 
The minimum is determined by the stationary condition 
$d\tilde{f}_{\rm X} / dR \Eq 0$, 
but it always has the trivial solution $R \Eq 0$. 
A nontrivial solution is the one satisfying the 
following equation:
\begin{equation}
 -2a(T) - 4c R^2 = 
3 b R \bar{g}_u (R) + b R^2 \frac{d\bar{g}_u(R)}{dR}
=: \mathscr{Y}_u (R^2).  
\end{equation}
One can solve this graphically:
plot $\mathscr{Y}_u$ 
as a function $\rho \equiv R^2$ and find 
its crossing with the straight line 
$-2a(T) \msp2 - \msp2 4c \rho$.  
If it crosses from below as $\rho$ increases, 
its crossing point $\rho_{\star}(T)$ 
determines the minimum position 
 as $R \Eq \sqrt{\rho_{\star} (T)}$. 
If there are multiple crossing points of this kind, 
the one with the lowest $\tilde{f}_{\rm X}$ is the global minimum.  
Then, the stable state at the temperature $T$ 
is that 
in Fig.~\ref{fig:PD-g34} 
at the position 
$(\bar{c}_{\mathrm{A}},\bar{c}_{\mathrm{B}}) 
 \Eq \sqrt{\rho_\star (T)} \msp2 C (\cos u , \sin u)$.

\section{Detail of small $K$ analysis}
\label{sec:AppC}

In this Appendix, we examine 
how a finite value of $K$ affects 
the anisotropic configurations at $K=0$.  
Using the previous notation 
$\XX \Eq (X_1, X_2, X_3)$ for $(X,Y,Z)$ and 
$\btheta \Eq (\theta_1 , \theta_2 , \theta_3 )$ 
for $(\theta_X , \theta_Y , \theta_Z )$, 
the $K$-term  in the free energy $f_{\rm X}$ in Eq.~\eqref{eq:F3}
reads as 
\begin{align}
&f_{2K}^{X_1 X_2 X_3 } (\theta_1, \theta_2,\theta_3 ) 
=
-4K \sum_{j=1}^3
  X_j^2 \cos (2\theta_j \PLUS j \omega )  , 
\label{eq:f2Kmini}
\end{align} 
where $\omega \Equiv \FRAC23 \pi$ as 
before.  
Let us start discussing from the limit of $K \Eq 0$. 
There are three equivalent configurations: 
$(\XX  , \btheta )$ $\Eq$
$(q,q,p,\FRAC{\pi}{2},- \FRAC{\pi}{2},0)$ 
and its two equivalents 
$(p,q,q,0,\FRAC{\pi}{2},-\FRAC{\pi}{2})$, 
and $(q,p,q,-\FRAC{\pi}{2} ,0,\FRAC{\pi}{2})$.  
They are indeed realized at low temperatures at $K \Eq 0$ 
as discussed in Appendix~\ref{AppB2}.  
Upon switching on $K$, the degeneracy of 
these three configurations is lifted as is evident from 
the factor $\cos(2\theta_j \PLUS j\omega)$ 
appearing in Eq.~(\ref{eq:f2Kmini}). 
This will result in different behavior in their stability.  

Before discussing the effects of $f_{2K}^{X_1 X_2 X_3}$, 
we first check the stability of the $K \Eq 0$ solution 
against rotating the order parameters.  
Let us consider small variations in the order parameter directions 
such that 
$\pm \FRAC{\pi}{2} \To \pm \FRAC{\pi}{2} +\eta_\pm$
and 
$0 \To \eta_{0} $.  
Then, the corresponding change in the free energy $\delta f_{\rm X}$ is 
calculated by evaluating Eq.~(\ref{eq:F3}) 
\begin{align}
\delta f_{\rm X} \big|_{K =0} \sim
&
q^2 \Bigl[
 3 b \msp2 p (\eta_{+} \PLUS \eta_{-} \PLUS \eta_{0}) ^2  
\PLUS
 2 c\dprime \msp2 p^2 (\eta_{+} \PLUS \eta_{-} \MINUS 2 \eta_{0} )^2 
\nonumber\\
&\hs{5mm}\PLUS
 2 c\dprime (p^2 \MINUS 2 q^2) (\eta_{+} \MINUS \eta_{-})^2 
\Bigr] .
\label{eqC:deltafX}
\end{align}
The leading term in this change 
is quadratic with respect to the three sets of linear combinations of 
$\eta_{\pm}$ and $\eta_{0}$. 
They are the eigenmodes within this harmonic approximation, 
and their energies are all positive, 
since $p^2 \GE 2q^2$ for these states 
and $b$ is assumed positive. 
This concludes that the solution 
at $K \Eq 0$ is stable against small variations in 
$\btheta$.

Now we calculate $f_{2K}^{X_1 X_2 X_3 }$'s for the aforementioned three configurations.  
The results are 
\begin{subequations}
\begin{align}
f_{2K}^{qqp} (\FRAC{\pi}{2},-\FRAC{\pi}{2},0)
&=
-4K(p^2+q^2) , 
\label{eq:f2K0positive}
\\
f_{2K}^{pqq} (0,\FRAC{\pi}{2},-\FRAC{\pi}{2})
&=
f^{qpq}_{2K} (-\FRAC{\pi}{2},0,\FRAC{\pi}{2})
=
2K(p^2+q^2) . 
\label{eq:f2K0negative}
\end{align}
\end{subequations}
Thus, for $K \GT 0$, the configuration 
$\btheta \Eq 
(\FRAC12 \pi,- \FRAC12 \pi ,0)$ is the most stable, 
while those with $(0, \FRAC12 \pi, - \FRAC12 \pi)$ or 
$(-\FRAC12 \pi , 0, \FRAC12 \pi )$ 
are stabilized for $K \LT 0$. 
To simplify the discussion, 
we restrict ourselves to the analysis of $f_{2K}^{X_1 X_2 X_3 }$ 
with fixing $X_1 \Eq X_2 \Eq q$ and $X_3 \Eq p$, 
and consider a free energy change $\delta f_{2K}^{qqp}$
associated with small variations in $\btheta$.  
For $K \GT 0$, straightforward calculation gives 
\begin{subequations}
\begin{align}
& \delta f_{2K}^{qqp} 
(\FRAC{\pi}{2}+\eta_{+},-\FRAC{\pi}{2}+\eta_{-},\eta_{0})
\nonumber
\\
&
\sim 
4K \bigl[
2 p^2 \eta_{0}^2 
\PLUS 
q^2 (\eta_{+}^2 \PLUS \eta_{-}^2 ) 
\MINUS 
\sqrt3 \msp2 
q^2 (\eta_{+} \MINUS \eta_{-} ) \bigr].   
\end{align}
The stability of this configuration is examined by 
minimizing the sum of this $\delta f_{2K}^{qqp}$ and $\delta f_{\rm X} |_{K=0}$ 
in Eq.~\eqref{eqC:deltafX}. 
It is important that this $\delta f_{2K}$ has 
a linear term of $\eta_{+} \MINUS \eta_{-}$. 
Therefore, 
its nonvanishing amplitude is induced as
$
\eta_{+} \MINUS \eta_{-} 
\msm2 \sim \msm2 \sqrt3 K / [ c\dprime (p^2 \MINUS 2q^2 ) ]
$, 
while the other two eigenmodes remain zero in their amplitudes.  
This indicates that the two directions 
$\theta_1$ and $\theta_2$ 
tilt from $\pm \FRAC12 \pi$ 
but the relation $\theta_1 \Eq -\theta_2$ continues to hold.

For $K \LT 0$, similar analyses show that a linear term appears  
once again in $\delta f_{2K}^{X_1 X_2 X_3 }$
\begin{align}
& 
\delta f_{2K}^{pqq} 
(\eta_0 ,\FRAC{\pi}{2} \PLUS \eta_{+},-\FRAC{\pi}{2} \PLUS \eta_{-})
\nonumber\\
&
\hs{-6pt}
\sim 
4|K| \bigl[
p^2 \eta_{0}^2 
\PLUS 
q^2 (2 \eta_{-}^2 \MINUS \eta_{+}^2 ) 
\MINUS 
\sqrt3 \msp2 
(q^2 \eta_{+} \PLUS p^2 \eta_0 ) \bigr], 
\label{eq:fpqq}
\\[6pt]
& 
\delta f_{2K}^{qpq} 
(-\FRAC{\pi}{2} \PLUS \eta_{-},\eta_{0},\FRAC{\pi}{2} \PLUS \eta_{+})
\nonumber\\
&
\hs{-6pt}
\sim 
4|K| \bigl[
p^2 \eta_0^2 
\PLUS 
q^2 (2 \eta_{+}^2 \MINUS \eta_{-}^2 ) 
\PLUS
\sqrt3 \msp2
(q^2 \eta_{-} \PLUS p^2 \eta_0 ) \bigr]. 
\label{eq:fqpq}%
\end{align}
\end{subequations}
The linear term is proportional to $q^2 \eta_{\pm} \PLUS p^2 \eta_{0}$, 
and this combination contains at least two or generally 
all the three eigenmodes in Eq.~\eqref{eqC:deltafX}.  
Therefore, 
these eigenmodes acquire nonvanishing values 
in the configuration minimizing the total $\delta f_{\rm X}$. 
The values of $\eta_{\pm}$ and $\eta_{0}$ are thus 
nonvanishing, and generally they have no symmetry.  
This further induces inequivalent changes 
in the magnitudes $\{ X_j \}$.  
Therefore these configurations with no symmetry 
correspond to NS states for $K \LT 0$.
A final remark is about a negative coefficient of 
one of $\eta_{\pm}^2$ in Eqs.~\eqref{eq:fpqq} and \eqref{eq:fqpq}. 
Since this is proportional $|K|$, 
one can neglect their 
effects, as in the case of $K \GT 0$, as far as $|K|$ is small.  

\black 

\section{Phase changes across the $K \Eq 0$ line}
\label{KzeroAnal}

In this Appendix, we study in detail 
how various symmetry broken phases change 
near the $K \Eq 0$ line in the $(J,K)$ parameter space 
based on the results of the microscopic mean-field 
approximation in Sec.~\ref{sec:MFresults}.  
In particular, we focus on the region of very low temperature.  
As shown in the inset of Fig.~\ref{fig9} (c), 
the \NSpm state exists near the $K \Eq 0$ line.  
Its phase boundary with the $zyzx$ state 
touches the $K \Eq 0$ line around $J/E_1 \sim 0.007$, 
and 
indeed the touching point is $J_c/E_1\simeq 0.0067$ 
at $T \Eq 0$.  
The ground state for $J \GT J_c$ is 
the uniaxial triple-$\qq$ state shown in Fig.~\ref{fig:PD-g34}, 
while it is the fully-anisotropic state 
for $0 \LT J \LT J_c$.  
For $J \LT 0$, the ground state is always the ferro 
 state, 
but this is not our interest in this paper.  
In this Appendix, we use the notations 
$\XX \Eq (X,Y,Z)$ and $\btheta \Eq (\theta_X, \theta_Y, \theta_Z)$.

%
%
\begin{figure}[t]
\begin{center}
\includegraphics[width=0.5\textwidth]{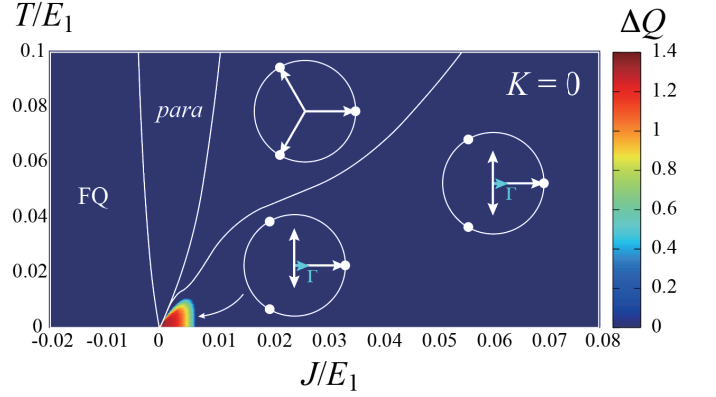}
\end{center}
\caption{
Color plot of $\Delta Q$ in $J$-$T$ plane for $K=0$.  
$\Delta Q$ is finite only in the fully-anisotropic state for $0<J<J_c$.  
The phase boundaries are indicated by the white lines 
and for the triple-$\qq$ states, 
the quadrupole configurations in the wavenumber space are 
schematically shown.
}
\label{fig:AppK0TJ}
\end{figure}

%
%
\begin{figure}[t!]
\begin{center}
\includegraphics[width=0.5\textwidth]{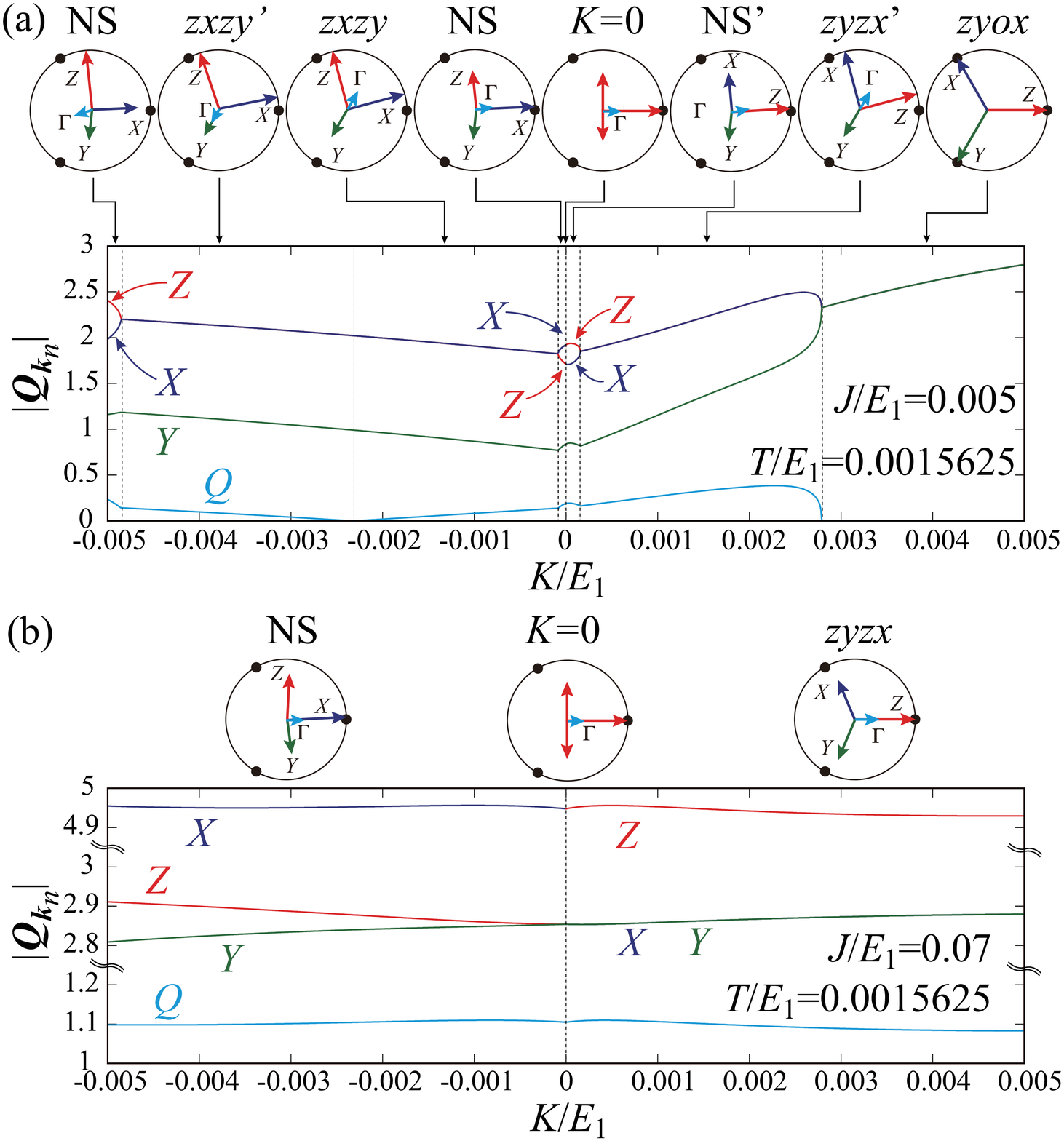}
\end{center}
\caption{
$|\QQ_{\kk_n}|$ as a function of $K$ near $K=0$.  
(a) $J/E_1=0.005$, where the ground state for $K=0$ is 
the fully-anisotropic state with $X\ne Y\ne Z$.  
(b) $J/E_1=0.07$, where the ground state for $K=0$ is 
the uniaxial state with $X=Y\ne Z$.  
The order parameters for $K\to 0+$ and $K\to 0-$ are 
different but they belong to the  equivalent domains 
in the order parameters for $K=0$.  
In each of the figure, schematic configurations $\{\QQ_{\kk_n}\}$ 
are drawn for the intuitive understanding about the changes 
in the order parameters.
}
\label{fig:AppK0}
\end{figure}

Figure \ref{fig:AppK0TJ} shows the $J$--$T$ phase diagram 
at $K \Eq 0$.  
The color map represents the magnitude 
difference $\Delta Q$ 
between the smallest and the second smallest ones among $X$, $Y$, and $Z$.  
This is one type of the order parameter 
identifying the fully-anisotropic state, 
where the three magnitudes 
are all different.  
The region with finite $\Delta Q$ is 
limited to a small part of $J \GT 0$ at low temperatures.  
In the part of $J \GT 0$,  
the phase with the highest transition temperature 
is the symmetric triple-$\qq$ state as discussed in Appendix~\ref{A1}.   
Its lower $T$ side 
is covered by the uniaxial triple-$\qq$ state with ``$xxz$'' anisotropy.  

At $K \Eq 0$, 
the ordered states have many equivalent domains.
Once $K$ becomes finite, these degeneracies are lifted, 
and some of them are stabilized.  
Figure \ref{fig:AppK0} shows the change of 
$|\QQ _{k_0}|$ and 
$\{ |\QQ _{k_\ell}| \}_{\ell=1}^3 \Eq \{X,Y,Z \}$ 
upon varying $K$ with $J$ fixed.  
 The temperature is set to $T \Eq 0.0015625$, 
practically equivalent to $T \Eq 0$.  
The two panels correspond to the results for different $J$'s:
$J/E_1 \Eq $ (a) 0.005 and (b) 0.007. 
The phase at $K \Eq 0$ is the fully-anisotropic state in 
the panel (a), 
and the uniaxial state in the panel (b).

Let us start with discussing the $K \Eq 0$ case in the panel (a) 
of Fig.~\ref{fig:AppK0}. The three magnitudes 
$X$, $Y$, and $Z$ are all different. 
The configuration of one typical domain there is 
schematically illustrated: 
the magnitude is the largest for that with $\theta=0$, 
and the smallest for $\theta=-\pi/2$. 
This state is degenerate in its configuration, 
and this degeneracy corresponds to 36 different domains 
except for trivial translations. 
First, there are 3! ways of assigning 
$\QQ_{{\bm k}_\ell}$'s to these three vectors, 
and the corresponding 6 permutations constitute a first class.
There are two other classes of 
operations generating degenerate domains. 
One type of operation is 
the direction exchange 
$\pi/2\leftrightarrow -\pi/2$ 
while the direction for the largest moment is fixed at $\theta \Eq 0$. 
The other type is the rotation 
$\theta\to \theta\pm\omega$ for all the moments. 
Combining these three types of operations yield
in total $3! \times (1+1) \times (1+2) =36$ domains,  
and they are degenerate in the fully-anisotropic 
triple-$\qq$ state at $K=0$. 

The degeneracy of these 36 domains is lifted when $K\ne 0$. 
In Fig.~\ref{fig:AppK0}(a), 
we show the configuration 
where the smallest magnitude is 
$Y$ and its direction is 
$\theta_Y \sim -\pi/2$.
Upon switching on $K>0$, 
the configuration with $Z\ge X>Y$ is stabilized, 
since $\theta_Z\sim 0$ ensures the {\it maximum gain} 
in the quadratic terms in $f_{\rm X}$. 
At the same time, 
$\theta_X \Eq \pi/2$ and $\theta_Y \Eq -\pi/2$, 
and they are quite close to $\omega$ and $-\omega$, 
respectively, which also lowers $f_{\rm X}$. 
In contrast, for $K<0$, 
the moment with the two larger magnitudes $X$ and $Z$    
point to the direction  $\theta_X \sim \pi/6$ and $\theta_Z\sim \pi/2$. 
 This \textit{maximizes the energy gain}  
in the quadratic terms in $f_{\rm X}$. 
As for the smallest one $Y$, it points to the direction $\theta_Y=-\omega$, 
which \textit{minimizes the energy cost} in $f_{\rm X}$, while  
\textit{maximizes the energy gain} in the $b$-term. 
See the discussion in Sec.~\ref{sec:analysis_zxzy}  
Irrespective of the sign of $K$, 
the phase changes with $K$ as 
$\mbox{NS} \to zxzy \to \mbox{NS}$ 
and $\mbox{NS'} \to zyzx' \to zyox$.  
These changes are accompanied by rotations of the moments, 
which are smooth but quite complicated.
It should be noted that the appearance of the $zyzx'$ and 
$zxzy'(zxzy)$ states is related to the 
topological transition of the local mean-field 
state as discussed in Sec.~\ref{sec:SSproperty}. 
One of the quadrupole moments, say $\QQ_{\rm A}$, is 
pinned to $\QQ_{\rm A}=(-1,0)$ in the ``spin-1/2'' regime. 
Without symmetry breaking to NS or NS$'$, 
$\QQ_{\rm A}$ cannot vary continuously within the ``half-integer spin'' regime. 
See also Fig.~\ref{fig14}. 

Let us switch to the panel (b) of Fig.~\ref{fig:AppK0}.
The uniaxial triple-$\qq$ state is realized at $K=0$.
The configurations shown are for 
the domain in which the moment for 
$\theta \Eq 0$ has the largest magnitude 
among the 36 degenerate domains.  

The stability of the uniaxial triple-$\qq$ state has been analyzed 
in Secs.~\ref{sec:smallKpositive}, 
\ref{sec:analysis_zxzy}, and \ref{sec:C3breaking}, 
and in  Appendix~\ref{sec:AppC}.
There we have demonstrated that the NS$'$ 
state does not appear for $K>0$, 
while the NS state does appears for $K<0$. 
One can see this 
in the mean-field calculations in Fig.~\ref{fig:AppK0}(b). 
Note that, when considering the instability of 
the fully-anisotropic triple-$\qq$ state,
the NS$'$ state can appear as shown 
in the inset of Fig.~\ref{fig9}(c). 
The configurations shown are those 
with $Y$ being the smallest magnitude. 
For $K>0$, one of the previously degenerate domains at $K=0$ 
gradually transforms into the $zyzx$ state. 
During this process, the moment with 
the largest magnitude $Z$ keeps the directions 
$\theta_Z \Eq 0$, while the other two with the same magnitude $X \Eq Y$ 
tilt their directions $\theta_X$ and $\theta_Y$. 
For the parameter space shown in the panel (b), 
the largest moment is that for $X$ for $K<0$
owing to the quadratic terms $f_{\rm X}$. 
Note that we consider the situation with 
$Y$ being the smallest. 
The two smaller magnitude $Z$ and $Y$ start to vary differently; 
$Z$ increases while $Y$ decreases with lowering $K(<0)$. 
This is the NS state and the angles 
$\theta_{X,Y,Z}$ also tilt from the directions at $K \Eq 0$.

%

\end{document}